\documentclass[onecolumn, trackchanges]{aastex63}

\usepackage{multirow}
\usepackage{booktabs}
\usepackage{dirtree}
\usepackage{listings}
\usepackage{rotating}
\usepackage{graphicx}
\usepackage{tasks}
\usepackage[normalem]{ulem}
\usepackage{capt-of}
\usepackage{xspace}

\newcommand{\cmps}{\ensuremath{\text{cm s}^{-1}}\xspace}
\newcommand{\mps}{\ensuremath{\text{m s}^{-1}}\xspace}
\newcommand{\kmps}{\ensuremath{\text{km s}^{-1}}\xspace}

\newcommand{\teff}{$T_{\rm eff}$}

\newcommand{\DARKDARK}{\textsc{dark\_dark}\xspace}
\newcommand{\DARKDARKINT}{\textsc{dark\_dark\_int}\xspace}
\newcommand{\DARKDARKTEL}{\textsc{dark\_dark\_tel}\xspace}
\newcommand{\DARKFLAT}{\textsc{dark\_flat}\xspace}
\newcommand{\DARKFP}{\textsc{dark\_fp}\xspace}
\newcommand{\FLATDARK}{\textsc{flat\_dark}\xspace}
\newcommand{\FLATFLAT}{\textsc{flat\_flat}\xspace}
\newcommand{\FPFP}{\hbox{\textsc{fp\_fp}}\xspace}
\newcommand{\HCHC}{\hbox{\textsc{hcone\_hcone}\xspace}}
\newcommand{\FPHC}{\textsc{fp\_hcone}\xspace}
\newcommand{\HCFP}{\textsc{hcone\_fp}\xspace}
\newcommand{\LFCLFC}{\textsc{lfc\_lfc}\xspace}
\newcommand{\LFCFP}{\textsc{lfc\_fp}\xspace}
\newcommand{\OBJFP}{\textsc{obj\_fp}\xspace}
\newcommand{\OBJDARK}{\textsc{obj\_dark}\xspace}
\newcommand{\POLFP}{\textsc{polar\_fp}\xspace}
\newcommand{\POLDARK}{\textsc{polar\_dark}\xspace}
\newcommand{\BADPIX}{\textsc{badpix}\xspace}
\newcommand{\BACKMAP}{\textsc{backmap}\xspace}
\newcommand{\ORDERP}{\textsc{order\_profile}\xspace}
\newcommand{\LOCO}{\textsc{loco}\xspace}
\newcommand{\REFFP}{\textsc{ref\_fp}\xspace}
\newcommand{\SHAPEX}{\textsc{shape\_x}\xspace}
\newcommand{\SHAPEY}{\textsc{shape\_y}\xspace}
\newcommand{\SHAPELOCAL}{\textsc{shape\_local}\xspace}
\newcommand{\FLAT}{\textsc{flat}\xspace}
\newcommand{\BLAZE}{\textsc{blaze}\xspace}
\newcommand{\THERMALI}{\textsc{thermal\_int}\xspace}
\newcommand{\THERMALT}{\textsc{thermal\_tel}\xspace}
\newcommand{\REFLEAK}{\textsc{ref\_leak}\xspace}
\newcommand{\REFWAVE}{\textsc{ref\_wave}\xspace}
\newcommand{\SONED}{\texorpdfstring{\textsc{s1d}\xspace}{S1D}}
\newcommand{\ETDS}{\textsc{e2ds}\xspace}
\newcommand{\ETDSFF}{\textsc{e2dsff}\xspace}
\newcommand{\ETDSLL}{\textsc{e2dsll}\xspace}

\newcommand{\Apreprocessing}{\lstinline{apero_preprocess}\xspace}
\newcommand{\Adarkref}{\lstinline{apero_dark_ref}\xspace}
\newcommand{\Abadpixel}{\lstinline{apero_badpix}\xspace}
\newcommand{\Alocalisation}{\lstinline{apero_loc_spirou}\xspace}
\newcommand{\Ashaperef}{\lstinline{apero_shape_ref}\xspace}
\newcommand{\Ashapenight}{\lstinline{apero_shape}\xspace}
\newcommand{\Aflatblaze}{\lstinline{apero_flat}\xspace}
\newcommand{\Athermal}{\lstinline{apero_thermal}\xspace}
\newcommand{\Aleakref}{\lstinline{apero_leak_ref}\xspace}
\newcommand{\Awaveref}{\lstinline{apero_wave_ref}\xspace}
\newcommand{\Awavenight}{\lstinline{apero_wave_night}\xspace}
\newcommand{\Aextract}{\lstinline{apero_extract}\xspace}
\newcommand{\Amktellu}{\texorpdfstring{\lstinline{apero_mk_tellu}\xspace}{apero\_mk\_tellu}}
\newcommand{\Amkmodel}{\texorpdfstring{\lstinline{apero_mk_model}\xspace}{apero\_mk\_model}}
\newcommand{\Aftellu}{\texorpdfstring{\lstinline{apero_fit_tellu}\xspace}{apero\_fit\_tellu}}
\newcommand{\Amtemp}{\texorpdfstring{\lstinline{apero_mk_template}\xspace}{apero\_mk\_template}}
\newcommand{\Accf}{\lstinline{apero_ccf}\xspace}
\newcommand{\Apolar}{\lstinline{apero_polar}\xspace}
\newcommand{\Aprocessing}{\lstinline{apero_processing}\xspace}

\newcommand{\OBJECTNAME}{\textsc{drsobjn}\xspace}
\newcommand{\OBSTYPE}{\textsc{obstype}\xspace}
\newcommand{\MJSTART}{\textsc{mjdate}\xspace}
\newcommand{\MJDEND}{\textsc{mjdend}\xspace}
\newcommand{\EXPTIME}{\textsc{exptime}\xspace}
\newcommand{\DPRTYPE}{\textsc{dprtype}\xspace}
\newcommand{\DRSMODE}{\textsc{drsmode}\xspace}
\newcommand{\MIDEXPOSURE}{\textsc{mjdmid}\xspace}

\newcommand{\GAIN}{\textsc{gain}\xspace}
\newcommand{\FRAMETIME}{\textsc{frmtime}\xspace}
\newcommand{\SATURATION}{\textsc{saturate}\xspace}
\newcommand{\RDNOISE}{\textsc{rdnoise}\xspace}

\newcommand{\modAstropy}{\textsc{astropy}\xspace}
\newcommand{\modAstroquery}{\textsc{astroquery}\xspace}
\newcommand{\modBarycorrpy}{\textsc{barycorrpy}\xspace}
\newcommand{\modNumpy}{\textsc{numpy}\xspace}
\newcommand{\modMatplotlib}{\textsc{matplotlib}\xspace}
\newcommand{\modPandas}{\textsc{pandas}\xspace}
\newcommand{\modscipy}{\textsc{scipy}\xspace}
\newcommand{\modscikit}{\textsc{scikit-image}\xspace}

\newcommand{\APERO}{\texorpdfstring{\textsc{apero}\xspace}{apero }}
\newcommand{\spirou}{\texorpdfstring{\textsc{SPIRou}\xspace}{SPIRou }}
\newcommand{\nirps}{\textsc{NIRPS}\xspace}
\newcommand{\nirpshe}{\textsc{NIRPS\_HE}\xspace}
\newcommand{\nirpsha}{\textsc{NIRPS\_HA}\xspace}
\newcommand{\NAN}{\textsc{nan}\xspace}
\newcommand{\FITSTORAMP}{\lstinline{fits2ramp.py}\xspace}
\newcommand{\ARRIJ}{the $i^{th}$ row $j^{th}$ column\xspace}
\newcommand{\latestversion}{v0.7.256\xspace}
\newcommand{\oneoverf}{\texorpdfstring{$1/f$\xspace}{1/f}}
\newcommand{\PRV}{pRV\xspace}

\newcolumntype{J}[1]{>{\raggedright\let\newline\\\arraybackslash\hspace{0pt}}m{#1}}

\received{September 6, 2022}
\revised{October 20, 2022}
\accepted{October 28, 2022}

\submitjournal{PASP}

\shorttitle{APERO\,--Demonstration with SPIRou}
\shortauthors{Cook et al.}
\graphicspath{{./}{figures/}}

\begin{document}

\title{APERO: A PipelinE to Reduce Observations - Demonstration with \spirou}
\correspondingauthor{Neil James Cook}
\email{neil.cook@umontreal.ca}

\author[0000-0003-4166-4121]{Neil James Cook}
\affiliation{Institute for Research on Exoplanets, Université de Montréal, Département de Physique, C.P. 6128 Succ. Centre-ville, Montréal, QC H3C 3J7, Canada}

\author[0000-0003-3506-5667]{\'Etienne Artigau}
\affiliation{Institute for Research on Exoplanets, Université de Montréal, Département de Physique, C.P. 6128 Succ. Centre-ville, Montréal, QC H3C 3J7, Canada}
\affiliation{Observatoire du Mont-Mégantic, Université de Montréal, Département de Physique, C.P. 6128 Succ. Centre-ville, Montréal, QC H3C 3J7, Canada}

\author[0000-0001-5485-4675]{René Doyon}
\affiliation{Institute for Research on Exoplanets, Université de Montréal, Département de Physique, C.P. 6128 Succ. Centre-ville, Montréal, QC H3C 3J7, Canada}
\affiliation{Observatoire du Mont-Mégantic, Université de Montréal, Département de Physique, C.P. 6128 Succ. Centre-ville, Montréal, QC H3C 3J7, Canada}

\author[0000-0002-5945-7975]{Melissa Hobson}
\affiliation{Max-Planck-Institut für Astronomie, Königstuhl 17, 69117 Heidelberg, Germany}
\affiliation{Millennium Institute for Astrophysics, Chile}

\author[0000-0002-5084-168X]{Eder Martioli}
\affiliation{Laboratório Nacional de Astrofísica, Rua Estados Unidos 154, Itajubá, MG 37504-364, Brazil}
\affiliation{Sorbonne Université, CNRS, UMR 7095, Institut d’Astrophysique de Paris, 98 bis bd Arago, 75014 Paris, France}

\author[0000-0002-7613-393X]{Fran\c{c}ois Bouchy}
\affiliation{Observatoire astronomique de l'Universit\'e de Gen\`eve, 51 Chemin des Maillettes, CH-1290 Sauverny, Switzerland}

\author[0000-0002-2842-3924]{Claire Moutou}
\affiliation{Univ. de Toulouse, CNRS, IRAP, 14 Avenue Belin, 31400 Toulouse, France}

\author[0000-0003-2471-1299]{Andres Carmona}
\affiliation{Univ. Grenoble Alpes, CNRS, IPAG, 38000 Grenoble, France}

\author[0000-0002-5232-2883]{Chris Usher}
\affiliation{Canada-France-Hawaii Telescope, CNRS, Kamuela, HI 96743, USA}

\author[0000-0002-1436-7351]{Pascal Fouqué}
\affiliation{Univ. de Toulouse, CNRS, IRAP, 14 Avenue Belin, 31400 Toulouse, France}
\affiliation{Canada-France-Hawaii Telescope, CNRS, Kamuela, HI 96743, USA}

\author[0000-0002-0111-1234]{Luc Arnold}
\affiliation{Canada-France-Hawaii Telescope, CNRS, Kamuela, HI 96743, USA}

\author[0000-0001-5099-7978]{Xavier Delfosse}
\affiliation{Univ. Grenoble Alpes, CNRS, IPAG, 38000 Grenoble, France}

\author[0000-0001-8388-8399]{Isabelle Boisse}
\affiliation{Aix Marseille Univ, CNRS, CNES, LAM, Marseille, France}

\author[0000-0001-9291-5555]{Charles Cadieux}
\affiliation{Institute for Research on Exoplanets, Université de Montréal, Département de Physique, C.P. 6128 Succ. Centre-ville, Montréal, QC H3C 3J7, Canada}

\author[0000-0002-5922-8267]{Thomas Vandal}
\affiliation{Institute for Research on Exoplanets, Université de Montréal, Département de Physique, C.P. 6128 Succ. Centre-ville, Montréal, QC H3C 3J7, Canada}
\author[0000-0001-5541-2887]{Jean-François Donati}
\affiliation{Univ. de Toulouse, CNRS, IRAP, 14 Avenue Belin, 31400 Toulouse, France}

\author[0000-0001-9907-5042]{Ariane Deslières}
\affiliation{Institute for Research on Exoplanets, Université de Montréal, Département de Physique, C.P. 6128 Succ. Centre-ville, Montréal, QC H3C 3J7, Canada}

\begin{abstract}

With the maturation of near-infrared high-resolution spectroscopy, especially when used for precision radial velocity, data reduction has faced unprecedented challenges in terms of how one goes from raw data to calibrated, extracted, and corrected data with required precisions of thousandths of a pixel. Here we present \APERO (A PipelinE to Reduce Observations), specifically focused on \spirou, the near-infrared spectropolarimeter on the Canada--France--Hawaii Telescope (SPectropolarimètre InfraROUge, CFHT). In this paper, we give an overview of \APERO and detail the reduction procedure for \spirou. \APERO delivers telluric-corrected 2D and 1D spectra as well as polarimetry products. \APERO enables precise stable radial velocity measurements on sky (via the LBL algorithm), good to at least $\sim2$ \mps over the current 5-year lifetime of \spirou. 

\end{abstract}


\section{Introduction}
\label{sec:intro}

Astrophysical data, like the echelle data taken with \spirou ({\it Spectro Polarimètre Infra ROUge}, \citealt{donati_spirou_2018}), require data reduction pipelines and data analysis software to produce scientific results. A data pipeline is software that takes data from the origin (in our case, the raw data provided by a specific telescope at an observatory) to a destination (in our case, servers hosting raw, intermediate, and output files accessible by principal investigators).

What exactly is part of the data reduction pipeline and what is part of the data analysis pipeline can be vague. In most cases, there are complex connections and feedback required from data reduction to the observatory (influencing the telescope, the instrument, and thus the raw data) and between the reduction and the analysis software. For the purposes of this paper we define the reduction pipeline as the software that processes raw data in an autonomous way, and specifically per scientific observation, to a point where it can be used by the wider scientific community without prior expertise in the workings of the telescope, instrument or observational procedure (e.g., calibrations, combining of frames, etc). We define a data analysis pipeline as any steps after the reduction pipeline (i.e. after \APERO) where one, many, or all scientific observations (of a specific astrophysical object of scientific interest or, indeed, multiple such objects) may be required to further gain scientific insight.

In the latter half of the 20$^{\rm th}$ Century, astronomy made the transition from analog to digital \citep{McCray2004, McCray2014, Borgman2021}. One could argue that with this digital revolution, data reduction software became a necessity, allowing processing in a more autonomous, uniform manner not possible with a more manual approach. Most early reduction pipelines focused on image processing, but tools such as the Image Reduction and Analysis Facility (IRAF, \citealt{Tody1986-IRAF}) became general-purpose software allowing developers to produce reduction pipelines for imaging and spectroscopy. More recently tools such as \texttt{AstroImageJ} (AIJ, \citealt{Collins2017-AIJ}, an astronomical image analysis software package based on \texttt{ImageJ}, \citealt{Rasband2011-IJ}) have provided astronomers with general-purpose tool kits for reducing data in a uniform manner. However, although generic software is suitable for data sets that have most or all their characteristics in common, such as imaging time series in the optical for photometry which rely on a common set of calibrations (flats, darks, biases), more complex observations, such as multi-object and/or cross-dispersed spectrographs cannot simply have a one-size-fits-all tool that will work on every instrument. As a result many reduction pipelines exist today (see table \ref{tab:echelle_pipelines} for a few examples for some recent echelle spectrographs\footnote{Partially constructed from \url{https://carmenes.caha.es/ext/instrument/index.html}}) and some generalized tools have been developed (e.g., \texttt{PyReduce}; \citealt{pyreduce2021}, \texttt{specutils}; \citealt{specutils2022} and \texttt{HiFLEx}; \citealt{Ermann2020}).

\begin{table*}
    \centering
    \footnotesize
    \begin{tabular}{lllr}
         \hline \hline
         Instrument & Facility & Pipeline & References \\         
         \hline
         CAFE & Calar Alto & CAFExtractor & \citealt{Cafe2013, Cafextractor2020} \\
         CARMENES & Calar Alto  & DRS$^{*}$ & \citealt{carmenes2014, carmenes_pipeline2016} \\
         CRIRES+ & ESO & ESO CPL & \citealt{crires2014, crires_wave_2014} \\
         ESPRESSO & ESO & ESO CPL & \citealt{espresso2021} \\
         ESPaDOnS & CFHT & LIBRE-ESPRIT & \citealt{Donati2003, Donati1997} \\
         GIANO-B & La Palma & IRAF & \citealt{giano2006} \\
         GRACES & Gemini North & OPERA or DRAGraces & \citealt{Martioli2012, Chene2021} \\
         HARPS & ESO & DRS$^{*}$ & \citealt{HARPS2003, HARPS2004} \\
         HARPS-North & La Palma & DRS$^{*}$ & \citealt{HARPS_N_2012} \\
         HIRES & Keck & MAKEE, HIRES Redux & \citealt{Keck_hires_1994, keck_drs_2016} \\
         HPF & McDonald & HPF Pipeline & \citealt{Mahadevan2010, Mahadevan2012} \\ 
         iSHELL & IRTF & Spextools (V5) & \citealt{Cushing2004} \\ 
         iRD & Subaru & DRS$^{*}$ & \citealt{Kotani2018} \\ 
         MAROON-X & Gemini North & DRAGONS & \citealt{Seifahrt2020, Labrie2019} \\ 
         NIRPS & ESO & ESO CPL and APERO & \citealt{NIRPS2017} \\
         SPIRou & CFHT & APERO & \citealt{donati_spirou_2018, donati_spirou_2020} \\
         SOPHIE & OHP & DRS$^{*}$ & \citealt{sophie2009, SOPHIE2011} \\
         X-SHOOTER & ESO & ESO CPL & \citealt{XSHOOTER_2011, XSHOOTER_DRS_2006, XSHOOTER_DRS_2010} \\
         \hline
    \end{tabular}
    \caption{Some recent echelle spectrographs and their pipelines. The asterisk denotes unnamed data reduction software. The ESO Common Pipeline Library (CPL) is a collection of pipelines for ESO instruments where each instrument has specific scripts and some shared functionality \citep{ESO_CPL_2004}.}
    \label{tab:echelle_pipelines}
\end{table*}

The inner workings of a data pipeline should be accessible to the user, meaning it should be open-source (where possible). Open-source software is important as it allows inclusiveness with international partners and collaboration from a wide range of people. Using a free, commonly used language like Python allows developers and users quick access to code and science algorithms. This enables deep exploration when problems arise and results are not understood. Likewise, one wants very specific scripts that handle all the complex details of the data, but also have the ability to crunch tens of thousands of individual observations in a reasonably short amount of time and without human intervention in a way that is uniform through time.

This paper presents \APERO \footnote{Available on GitHub: \url{https://github.com/njcuk9999/apero-drs} \label{footnote:apero_git}}, an open-source Python pipeline to reduce observations, demonstrated with \spirou. Section \ref{sec:overview} gives an overview of the pipeline, Section \ref{sec:spirou} introduces \APERO as the official pipeline for \spirou, Sections \ref{sec:preprocessing} to \ref{sec:post_processing} detail the different parts of our pipeline (from the raw data through to science ready products) and Section \ref{sec:discussion} summarizes our work.

\section{Overview}
\label{sec:overview}

\subsection{The users of APERO} \label{subsec:users}

\APERO is primarily designed to be used at data centers either at the observatories providing the raw data to be reduced or in collaboration with them. In its current form any user wanting to reduce data with \APERO needs access to all calibration data (to fully calibrate the data: sections \ref{sec:ref_calibrations}, \ref{sec:night_calibrations} and \ref{sec:extraction}) and all hot star data (to fully telluric correct the data: Section \ref{sec:telluric}). Thus in general individual PIs are not expected to use \APERO directly but will, of course, be users of \APERO data (supplied by the observatory or the large collaborations involved with the instrument). For \spirou the Spirou Legacy Survey (SLS, \citealt{donati_spirou_2018, donati_spirou_2020}) collaboration was responsible for producing the data reduction pipeline and thus data centers at the observatory (CFHT) and in Canada and France have been responsible for reducing the data. To allow individual PIs to reduce a single file or single night's worth of data, a release of the full calibration and hot star data sets would be required. However, there is no time frame for such a release of this data.

\subsection{Design} \label{subsec:design}

Although \spirou has been the main driver during the development of \APERO, we took every opportunity possible to keep science algorithms separate from both core-functionality algorithms (such as logging, reading and writing files, database interface and management) and instrument-specific functionality (such as hard-coded value, number of fibers and FITS-header keywords). As an example see Appendix \ref{section:appendix_use_with_nirps} for details on the changes required for use with the Near Infra Red Planet Searcher (\nirps; \citealt{NIRPS2017}).

\APERO is an open-source, publicly available$^{\ref{footnote:apero_git}}$ Python\,3 package. The purpose of \APERO is to take raw data from the telescope, calibrate and correct instrumental and systematic effects where possible, extract and output spectra (2D and 1D) both before and after telluric correction, as well as provide an estimate of the radial velocity and calculate the polarimetry data where required. In this section, we will detail the generic features of \APERO and throughout the rest of the paper concentrate on its use with \spirou.

The package is split into base, core, input-output, language, plotting, science and tool sub-packages (see Figure \ref{fig:apero_design}). The scripts that are run are hereafter referred to as `recipes', as they only contain isolated steps of the specific reduction step, not dissimilar to a cookery recipe (i.e., step 1: do this, step 2: do that). All algorithmic complexity is kept to a minimum in these recipes and stored in one of the above-mentioned sub-packages. Recipes are defined either as reduction recipes or tool recipes. The former usually takes one or a small set of observations in order to reduce some part of the overall data set, and the latter aid the user or developer in a specific task, be it in processing a large amount of data, resetting output directories and databases, maintaining and updating various data structures and databases within \APERO or obtaining logs and statistics.

\begin{figure*}
    \begin{minipage}{16cm}
    \dirtree{.1 apero.\DTcomment{The \APERO package}.
    .2 base.\DTcomment{The most basic functionality}.
    .2 core.\DTcomment{The core functionality}.
    .2 data.\DTcomment{Data files}.
    .2 documentation.\DTcomment{Documentation package}.
    .2 io.\DTcomment{Input/Output functionality}.
    .2 lang.\DTcomment{Language/Text functionality}.
    .2 plotting.\DTcomment{Plotting functionality}.
    .2 recipes.\DTcomment{Top-level recipes}.
    .2 setup.\DTcomment{Installation package}.
    .2 science.\DTcomment{Science algorithms}.
    .2 tools.\DTcomment{\APERO tools package}.
    }
    \end{minipage}
    \caption{The \APERO python package and basic modular design}
    \label{fig:apero_design}
\end{figure*}

The only installation requirement for \APERO is a separate python environment (i.e., with \textsc{conda}\footnote{Conda available at: \url{https://docs.conda.io/en/latest/}} or \textsc{venv}\footnote{Supplied by default with Python \label{footnote:with_python}}). All python modules in the environment are strictly controlled by \APERO via \textsc{pip}$^{\ref{footnote:with_python}}$, thus installing in a general python environment with the user's own modules is not recommended. Installation instructions can be found in the \APERO documentation\footnote{APERO documentation available at \url{http://apero.exoplanets.ca/} \label{footnote:apero_docs}}. Running the installation script provides a full walk-through of all options required to set up \APERO. One specific feature is the use of \APERO `profiles' designed to accommodate multiple setups (i.e., differing constants, configurations, data directories, and database setups, etc). The installation process sets up and checks paths to the data directories, copies all default files and assets, and configures the various databases required for operation. The database installed is either using MySQL, which requires additional installation by the user, or by default, SQLite3 which is supported within python without any additional setup required by the user.

The input, working, and output directories are also designed to work in a flexible way: they can consist of all sub-directories within one primary directory, they can be at unrelated locations on the computer system, and all files can be symbolically linked from elsewhere, for maximum system compatibility. The most basic design for the data directories is shown in Figure \ref{fig:apero_data_dirs}. For a single instrument, it can often be useful to have shared raw data directories for different \APERO profiles (which can have different assets, preprocessed, reduced, directories and database setups, etc).

In addition to the \APERO python package directory (Figure \ref{fig:apero_design}) and the data directories (Figure \ref{fig:apero_data_dirs}) there is also a settings directory (Figure \ref{fig:apero_setting_dir}), where each \APERO profile is stored, containing information on specific instrument and setup parameters). The user has complete control over all installation and database parameters, as well as the ability to manipulate almost all default parameters given for a particular instrument. Explanations of each parameter can be found in the full \APERO documentation$^{\ref{footnote:apero_docs}}$.

\begin{figure*}
    \begin{minipage}{16cm}
    \dirtree{.1 profile directory.\DTcomment{The \APERO profile directory}.
    .2 assets.\DTcomment{A local version of some \APERO data files}.
    .2 raw.\DTcomment{The input files from the telescope}.
    .2 tmp.\DTcomment{The preprocessed files}.
    .2 calibDB.\DTcomment{The calibration database files}.
    .2 reduced.\DTcomment{The \APERO output directory}.
    .2 telluDB.\DTcomment{The telluric database files}.
    .2 runs.\DTcomment{The runs directory}.
    .2 out.\DTcomment{The telescope ready output files}.
    .2 plot.\DTcomment{The plot directory}.
    .2 msg.\DTcomment{The log directory}.
    }
    \end{minipage}
    \caption{An example basic data structure for \APERO reductions}
    \label{fig:apero_data_dirs}
\end{figure*}

\begin{figure*}
    \begin{minipage}{16cm}
    \dirtree{.1 settings directory .\DTcomment{The \APERO settings directory}.
    .2 profile 1.\DTcomment{An example \APERO profile - name determined by user}.
    .3 {user\_constant.ini}.\DTcomment{Constant parameter settings}.
    .3 {user\_config.ini}.\DTcomment{Configuration parameter settings}.
    .3 {profile1.sh.setup}.\DTcomment{Shell Environmental setup - source this if using tcsh or sh}.
    .3 {profile1.bash.setup}.\DTcomment{Bash shell environmental setup - source this if using bash}.
    .3 {install.yaml}.\DTcomment{The most fundamental install parameter settings}.
    .3 {database.yaml}.\DTcomment{The database parameter settings}.
    .2 profile2.\DTcomment{A second \APERO profile similar to profile1}.
    .2 profile3.\DTcomment{A third \APERO profile similar to profile1}.
    }
    \end{minipage}
    \caption{An example of the basic \APERO settings directory}
    \label{fig:apero_setting_dir}
\end{figure*}

\vspace{2cm}

\subsection{Notable features} \label{subsec:features}

Due to \APERO's modular design, it can be run in various ways. The most basic way is to run recipes individually giving all the required arguments (all recipes come with a \lstinline{--help} argument to display all required and optional arguments). All parameters inputted are saved in the FITS outputs as an additional extension.

In addition to running recipes separately, there is the \APERO tool called \lstinline{apero_processing.py} which is designed to automate the reduction based on an input \lstinline{run.ini} file (saved in the \lstinline{runs} directory; see Figure \ref{fig:apero_data_dirs}). The \APERO processing script scans the raw directory and reduces all data requested by the \lstinline{run.ini} file in an automated way, optimizing any steps that can be run in parallel and automatically waiting at the end of steps that require elements of a previous step to be completed. We define sequences of recipes that can be used to reduce specific steps of the full reduction process.

Another notable feature, as mentioned in Section \ref{subsec:design}, is that \APERO can be used for multiple instruments (or the same instrument with different settings) without conflict or duplication of the software - this is managed by having different \APERO profiles (see Section \ref{subsec:design}). \APERO currently works for both \spirou \citep{donati_spirou_2018, donati_spirou_2020} and \nirps (\citealt{NIRPS2017}, see Appendix \ref{section:appendix_use_with_nirps}) data but there are plans to extend compatibility with other instruments, such as SPIP (\citealt{donati_spirou_2018}, Donati et al., in prep.) and ANDES (formerly HIRES, \citealt{HIRES2021}). Any new instrument can use all, some, or none of the scientific algorithms available from other instruments while always benefiting from the \APERO architecture and core functionality.

\begin{figure*}
    \centering
    \includegraphics[width=16cm]{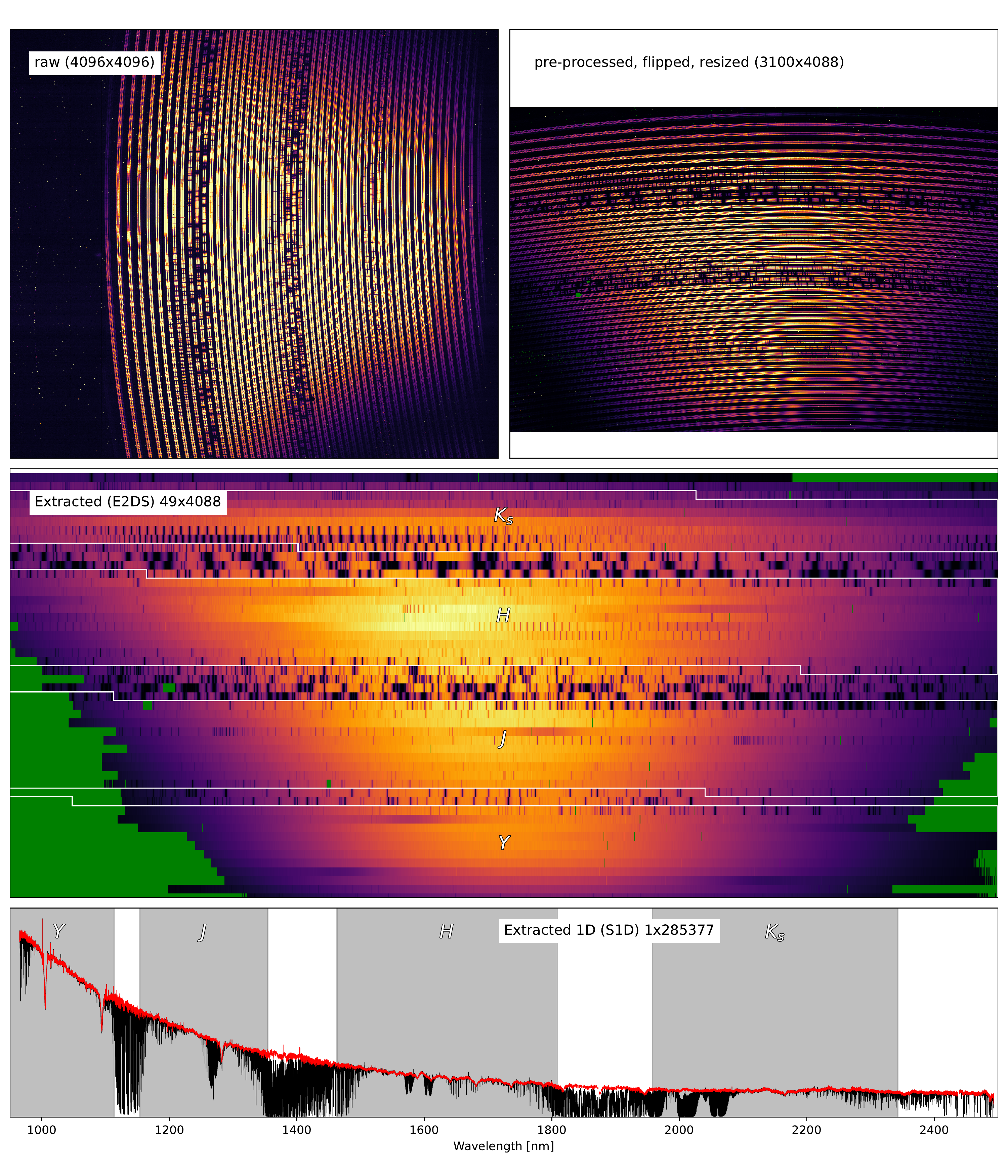}
    \caption{\spirou data at different points in the pipeline. All panels show a hot star observation of data product type (\DPRTYPE) \OBJDARK image (with the OBJ, in this case, being the hot star observation in the science fibers and a DARK in the reference fiber). Top left: the raw \OBJDARK (raw input to \APERO). Top right: the \OBJDARK after pre-processing (flipped and resized, Section \ref{subsec:flip_resize_rescale}). Middle: the extracted \ETDS \OBJDARK with combined flux from the A and B fibers (see Section \ref{sec:extraction}). Bottom: the one-dimensional spectrum of the hot star observation with the telluric correction in red. In the 2D images, \NAN values are shown in green. Note due to the number of pixels in these images we have down-sampled the image leading to apparent aliased structures, which are not present in the real data.}
    \label{fig:size_grid}
    \vspace{1cm}
\end{figure*}

\begin{figure*}
    \centering
    \includegraphics[width=18cm]{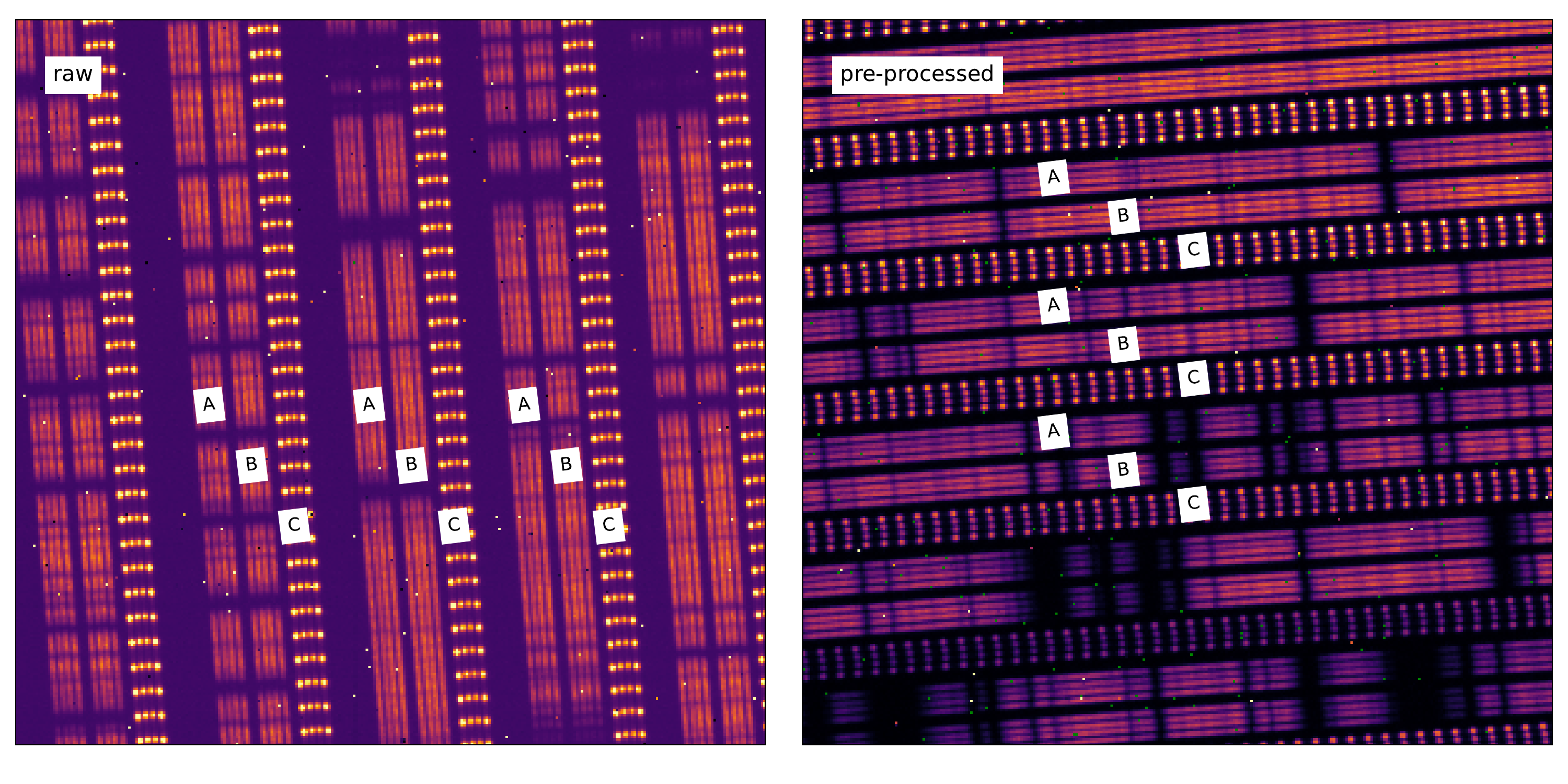}
    \caption{The layout of the three \spirou fibers. Left: An example raw \OBJFP image (an M dwarf star observation in the science fibers and an FP in the reference fiber). Right: The same \OBJFP image but pre-processed. A and B fibers are the science fibers and C is the calibration (or reference) fiber. Pre-processed images are rotated 90-degrees clockwise compared to the raw images (see Section \ref{subsec:image_rot}).}
    \label{fig:fiber_layout}
    \vspace{1cm}
\end{figure*}

\section{APERO - The official CFHT pipeline for \spirou}
\label{sec:spirou}

For the rest of this paper will we only discuss \APERO in terms of \spirou; however, future publications will discuss \APERO's implementation with other instruments. We briefly discuss the \nirps implementation in Appendix \ref{section:appendix_use_with_nirps}. Here we present the aspects of \spirou relevant for this paper. For further details we encourage the reader to consult \cite{donati_spirou_2018, donati_spirou_2020}. \APERO was first adapted from the HARPs DRS \citep{HARPS2003, HARPS2004} with the philosophy of keeping the code accessible and general enough that adding new instruments later would be possible. Note that in addition to \APERO, the \textsc{Libre-Esprit} pipeline from \cite{Donati1997} has also been adapted to handle \spirou data (but is not available publicly). 

\spirou is a near-infrared ($0.98-2.5\mu$m) spectro-polarimeter that saw first light at the Canada France Hawaii telescope (CFHT) in April 2018 \citep{donati_spirou_2018,donati_spirou_2020}. \spirou was designed to have spectral resolving power better than 70\,000 and to achieve precision radial-velocity (\PRV) stability better than 1 \mps. \spirou is composed of three units: the Cassegrain unit (i.e., the polarimeter attached to the Cassegrain focus of the telescope), the calibration unit, and the cryostat, containing the spectrograph. The cryostat is in a vacuum and temperature controlled at the milli Kelvin level, offering state-of-the-art stability. The detector is an \hbox{H4RG-15} HgCdTe\footnote{Manufacturer specifications can be found here: \url{http://www.teledyne-si.com/products-and-services/imaging-sensors/hawaii-4rg}} array \citep{Artigau2018-H4RG, Zandian2016-H4RG, Hodapp2019-H4RG, Hodapp1996-H4RG} with $4096\times4096$ pixels, with 4 of these pixels at the top, bottom, left and right reserved as reference pixels; they are not light-sensitive and are only used for common-mode readout noise rejection. The H4RG is read through 32 amplifiers that are each $128\times4096$ pixels in size. The Cassegrain module has two Fresnel rhombs (an ensemble of prisms used to rotate polarization states) coupled to a Wollaston prism. The Wollaston prism allows the incoming beam (either from the telescope or the calibration unit) to be split into two orthogonally polarised beams. The light of these two beams is conducted to the cryostat by two fluoride fibers (i.e., the science fibers, hereafter fibers A and B, or when combined AB).  In addition to the two science fibers, a third fluoride fiber directly connects the calibration unit to the spectrograph  (hereafter fiber C) providing light from various calibration lamps\footnote{The calibration unit also hosts a laser frequency comb visiting instrument (for use via permission from Jean-Francois Donati), and a thorium argon hollow cathode lamp for arc spectra.}:
\begin{itemize}
    \item a flat field exposure (via a halogen lamp), referred to hereafter as a FLAT.
    \item a uranium neon hollow cathode lamp for arc spectra, referred to hereafter as an HC.
    \item a Fabry-P\'erot etalon with tens of thousands of lines, referred to hereafter as an FP.
\end{itemize}

\noindent as well as providing an option for an unilluminated dark signal, hereafter referred to as a DARK, where the fiber sees a cold source inside the calibration unit. More details about the calibration unit can be found in \cite{Boisse2016}. The light from all three fibers is passed through a slicer to increase the spectral resolution for a given fiber size; \cite{Micheau2018}, leading to four closely packed slices per fiber.

The spectrograph itself is cross-dispersed in the perpendicular direction using an R2 echelle grating; this allows the \hbox{H4RG} detector to capture the entire spectral range of \spirou on the detector with no wavelength gaps but does lead to curved echelle orders with some overlap in wavelength between consecutive orders. We extract 49 orders\footnote{It is possible to extract 50 orders; however, \APERO does not extract the bluest order (diffraction order \#80) due to low SNR. \label{footnote:fifty_orders}} with each order spread along the 4088 pixels (grating diffraction orders \#79 to \#31). The \APERO input spectrum through various stages of the pipeline can be seen in Figure~\ref{fig:size_grid} and the layout of the three fibers (two science and one calibration) can be seen in Figure~\ref{fig:fiber_layout} in the raw and preprocessed rotations (see Section \ref{sec:preprocessing} for details).

The \spirou detector control software reads the detector continuously every 5.57\,s and produces a 2D image (\hbox{$4096\times4096$}) constructed from the linear fit of the pixel value versus time (henceforth the slope, as well as an intercept, error, and number of frames used for quality checks -- see Section \ref{subsec:cosmic_ray_reject}). This is the raw 2D `ramp' image used by \APERO as an initial input. An overview of this can be found in Appendix \ref{section:appendix_fitstoramp} but this software is not provided as part of \APERO (but the raw cubes are stored by CFHT). The `ramp' images are supplied by CFHT (via CADC\footnote{CADC is the Canadian Astronomy Data Center and is accessible via \url{https://www.cadc-ccda.hia-iha.nrc-cnrc.gc.ca/en/}}) and are thus referred to as the raw images for input into \APERO.

\subsection{Definitions of \texorpdfstring{\APERO}\' input files} \label{subsec:input_definitions}

One of the first actions of \APERO is to identify data product types (\DPRTYPE) for each possible input file taken from the telescope. For this purpose, we use the notation\footnote{In special cases the notation is $\{AB\}\_\{C\}\_\{KEY\}$ where the `KEY' is added when further information is required in the \DPRTYPE to distinguish two modes of the same type.} $\{AB\}\_\{C\}$ where fibers A and B are the science fibers and C is the reference fiber. The values of each correspond to what light was present in the fiber; a science signal ($OBJ$ or $POLAR$, see below for details) or a calibration signal (i.e., $DARK$, $FLAT$, $HC$ or $FP$). Note fibers A and B can have a calibration signal but fiber C cannot have a science signal.

\vspace{2cm}

Some of the most frequently used \DPRTYPE are thus:
\begin{itemize}
    \item \DARKDARKINT, where the INT added indicates that the DARK signal in the science AB channel comes from the calibration unit (at parking position) and is fed through the Cassegrain unit (i.e., the thermal contribution of the calibration unit, the polarimeter, and the fiber feedthrough into the cryostat; this is the dark used to calibrate calibration exposures).
    \item \DARKDARKTEL, where the TEL added indicates that the DARK signal in the science AB channel sees the sky (in fact the mirror covers) which includes the thermal contributions from the Cassegrain module and the fiber feedthrough into the cryostat. This is the dark used for on-sky exposure calibration.
    \item \FLATDARK and \DARKFLAT, where we have either a FLAT (halogen lamp) in the A and B fibers and a DARK (INT) in the C fiber, or a DARK (INT) in the A and B fibers and a FLAT in the C fiber respectively.
    \item \FLATFLAT, where all three fibers have a FLAT signal.
    \item $\xspace$\FPFP, where all three fibers have the FP (Fabry Perot etalon) signal.
    \item \DARKFP, where fibers A and B are DARK and fiber C has the FP.
    \item $\xspace$\HCHC, where all three fibers have the HC (UNe hollow cathode) signal. The HCONE here distinguishes the UNe hollow cathode from the ThAr hollow cathode lamp (HCTWO). 
    \item \OBJDARK and \POLDARK, where fibers A and B have a science signal (either in spectroscopic or polarimetric configuration) and the C fiber has a DARK (INT) signal.
    \item \OBJFP and \POLFP, where fibers A and B have a science signal (either in spectroscopic or polarimetric configuration) and the C fiber has an FP signal.
\end{itemize}

We list all usable combinations of fibers A, B, and C and how they are identified from SPIRou header values in Appendix \ref{section:appendix_inputs}.

On a standard night of operations, daily calibrations are taken both before  and after observations. The standard set of observations for each calibration set is as follows:

\begin{tasks}[style=enumerate](2)
    \task \DARKDARKINT ($\times2$)
    \task \DARKDARKTEL ($\times2$)
    \task \DARKFLAT ($\times5$)
    \task \FLATDARK ($\times5$)
    \task \FLATFLAT ($\times5$)
    \task \DARKFP ($\times2$)
    \task \FPFP ($\times5$)
    \task \HCHC ($\times2$)
\end{tasks}

\noindent These calibrations are tuned to be optimal for the extraction of all objects, avoiding saturation whilst taking a minimal amount of time to obtain.

SPIRou has two science signal modes: spectroscopy mode, and polarimetry mode. In \APERO this is distinguished via the definition of the reduction mode \DRSMODE (\textsc{spectroscopy} or \textsc{polar}) and is seen in \DPRTYPE leading to the distinction between \OBJDARK and \POLDARK, and, \OBJFP and \POLFP. For \DRSMODE set to \textsc{spectroscopy}, the rhomb position must be $P16$ for fiber A and $P16$ for fiber B (which means no polarization in either fiber). For \DRSMODE set to \textsc{polar} any other combination of rhomb positions ($P2$, $P4$, $P14$, $P16$) is deemed a polarimetric setup, however, only certain combinations of rhomb positions are used and are valid for calculating the polarimetric products (these are dealt with in the polarimetry code, see Section \ref{sec:polar}).

\subsection{Overview of the reduction process} \label{subsec:reduction_process}

\begin{figure*}
    \centering
    \includegraphics[width=18cm]{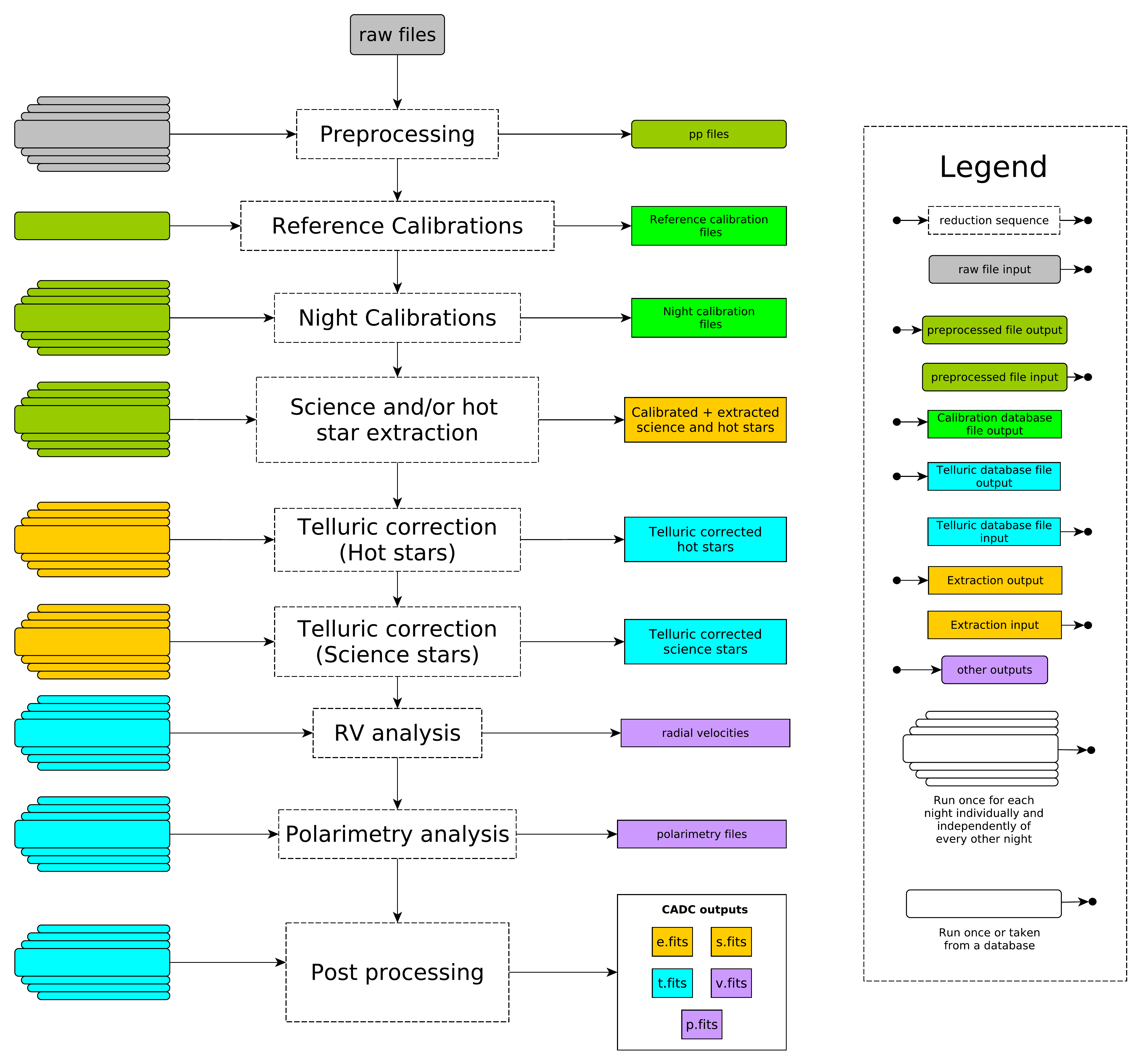}
    \caption{Overall flow chart for the \APERO reduction process for \spirou. Raw files are shown in gray, preprocessed (pp) files are shown in green, extracted products are in yellow, telluric products are in blue, and radial velocity and polarimetry products are shown in purple. The stacked inputs show that these inputs are used from every night of observation, but run on a per-file or per-night basis. Note that other inputs are required (e.g., calibration files) and detailed in the individual sections.}
    \label{fig:apero_overview}
\end{figure*}

The reduction process for \spirou is separated into eight main steps: the pre-processing (Section \ref{sec:preprocessing}), the reference calibrations (Section \ref{sec:ref_calibrations}), the nightly calibrations (Section \ref{sec:night_calibrations}), the extraction (Section \ref{sec:extraction}) for science observations and hot stars, the telluric absorption correction (Section \ref{sec:telluric}), the RV analysis (Section \ref{sec:rv}), the polarimetry calculations (Section \ref{sec:polar}), and the post-processing (Section \ref{sec:post_processing}). These steps are summarized in Figure \ref{fig:apero_overview}.

Although each of these eight steps can be run individually, and vary from a single recipe to a set of recipes for each step, the primary reduction method is automated to provide efficiency and to be reproducible wherever the data are being reduced. Note that \APERO is also compatible with the CFHT automation scripts. The CFHT automation scripts were developed to handle the constant flow of incoming raw data and handle sending the outputs after \APERO has the final products. As such we define `sequences' which can contain multiple or all steps. Example sequences are a `full sequence' having all steps, a `limited sequence' having all steps for a particular subset of astrophysical objects, a `calibration sequence' designed just to reduce calibration observations, a `telluric sequence' to create all files required for telluric correction, a `science sequence' designed to be run once all calibration and telluric steps have been run (either on one, multiple or all astrophysical objects), or an `engineering sequence' to do special steps required only for engineering and test purposes. This is all controlled via a user's run file (a $run.ini$ file). The \Aprocessing recipe reads the run file containing information such as:
\begin{itemize}
    \item which observation directories (e.g., which nights) to reduce or skip.
    \item which science targets to reduce.
    \item which sequence or sequences to use.
    \item which individual recipes of a sequence should be run.
    \item which recipes should be skipped or repeated.
    \item the number of physical cores to use.
\end{itemize}

Based on this run file and the raw data on disk, \APERO works out all recipes that should be run and the order in which these should be done, and most importantly which of these recipes can be run at the same time in parallel and which cannot - ultimately making the most efficient use of any machine resources that \APERO is being run on.

For a standard run (where no data has been previously processed) the \Aprocessing recipe will process the sequences in the order mentioned above without any human intervention (assuming nominal input data). It can work for both a single night of data or a complete re-reduction of all data since the first light and has been shown to produce reproducible results at multiple data centers.

\section{Pre-processing}
\label{sec:preprocessing}

The raw images (those retrieved from the telescope after the ramp fitting algorithm has been run) require some preliminary processing to remove detector artifacts. These artifacts are documented in this section. All frames independent of \DPRTYPE are preprocessed in the same manner before any other step of \APERO is run.

\subsection{Header fixes and object resolution} \label{subsec:obj_res}

The \spirou header provides the required information to process files. However, to facilitate data reduction a few header keys are added or updated.

The first header key we add is the \APERO object name (\OBJECTNAME), this header key is the object name used throughout \APERO. In general, it is the object name taken from the raw input file but all punctuation and white spaces are removed and replaced with underscores ($\_$) and all characters are capitalized, while `+' and `-' are  replaced with `P' and `M' respectively. This avoids object names with slightly different formats being considered as different objects (e.g., \textit{TRAPPIST-1} vs \textit{Trappist 1}) and allows for use in filenames. Next, the target type ($TRG\_TYPE$) with a value of either $TARGET$, $SKY$ or a blank string is added. This key exists in the raw file header of newer files (2020 and later) but has been found to be incorrect or missing for older files, especially when dealing with some sky frames (older sky frames can usually be identified by a suffix or prefix `sky` in the object name if not already identified as a sky by the target type header key). As well as this a mid-exposure time (\MIDEXPOSURE) is added which is equivalent to the time recorded at the end of exposure minus half the exposure time (\MJDEND $- $ \EXPTIME $/2$). \MIDEXPOSURE time is used throughout \APERO and is the recommended time to use, as opposed to other header keys such as \MJSTART, which isn't strictly the start of observation time but the time the observation request is sent. The last two keys added, as mentioned in Section \ref{subsec:reduction_process}, are the \DRSMODE and \DPRTYPE.

Once the headers are fixed with the above additions and corrections, if the raw files are of \DPRTYPE \OBJFP, \OBJDARK, \POLFP, or \POLDARK we cross-match the \OBJECTNAME with an object database of object names, positions, motions, parallax, known radial velocity estimates, temperatures and aliases. These are mostly sourced directly from SIMBAD \citep{wenger2000}, and cross-matched with the most up-to-date proper motion and parallax catalogs (based on an id cross-match from SIMBAD with Gaia EDR3; \citealt{Gaia_edr3_2021}, DR2; \citealt{gaia_dr2}, DR1; \citealt{gaia_dr1_2016}, UCAC4; \citealt{ucac4_2013} or Hipparcos; \citealt{hipp_1997}). This ensures the object name given is not already known by another object name, and all astrometric parameters are consistent even for observations from differing PIs. This is important for steps in the telluric process where we combine all objects of the same \OBJECTNAME where possible (see Section \ref{sec:telluric}). This local database of object names can be updated and is maintained in such a way as to keep consistency and inform users when updates have been made. All reductions of a single \OBJECTNAME should always be done with a single set of astrometric parameters.

\subsection{File corruption check} \label{subsec:pp_qc}

Not every raw file contains usable data. For example, a rare occurrence where the detector acquisition system has a synchronization issue in retrieving the pixel stream leads to a 1-pixel offset of the readout. Therefore as part of the pre-processing, we check for corrupt files. We do this by comparing images to a list of known hot pixels. We verify that hot pixels are at the expected position. If they are not at the expected position, this is corrected by registering the pixel grid to the nominal pixel position. Missed lines or columns at the edge of the array are replaced by \NAN values. This does not lead to a loss in science pixels as the 4-pixel edge of the array consists of non-light-sensitive reference pixels.

\subsection{Top and bottom pixel correction} \label{subsec:top_bottom}

\begin{figure*}
    \centering
    \includegraphics[width=18cm]{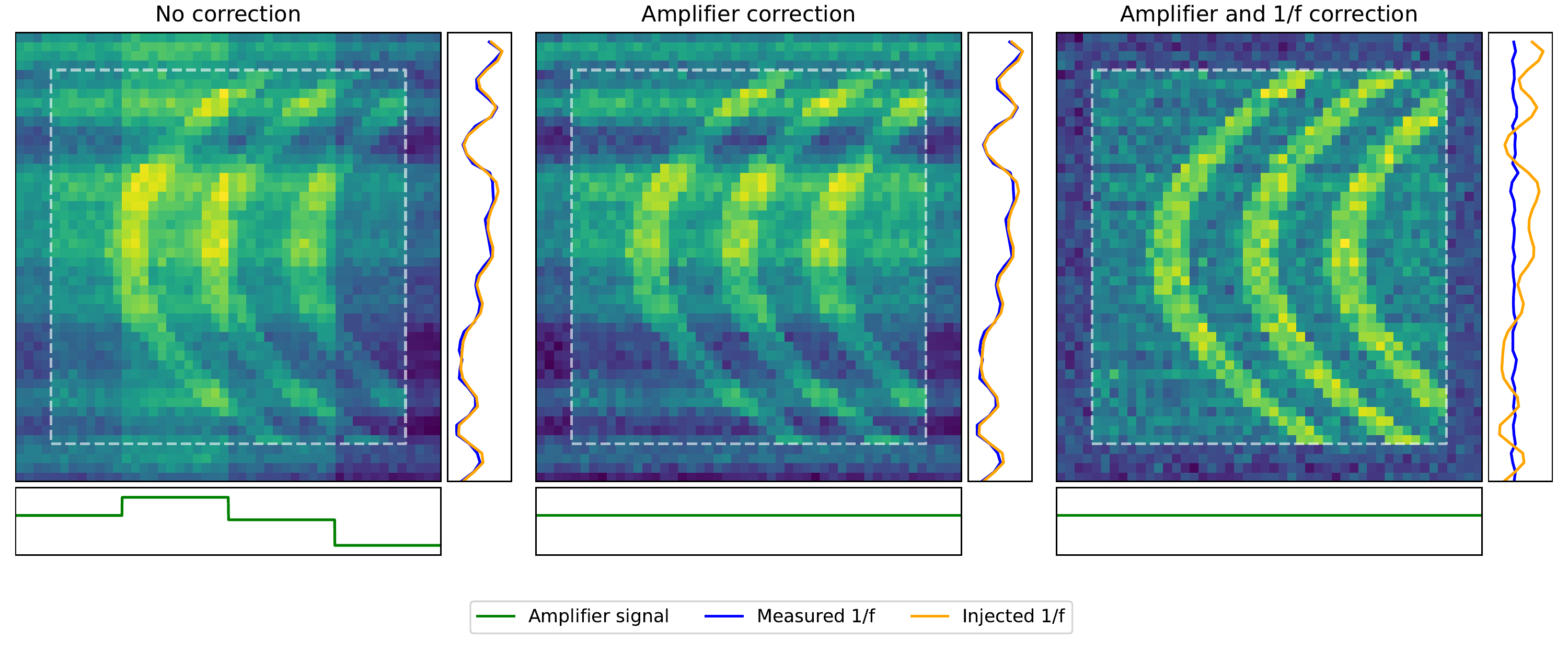}
    \caption{Schematic image to illustrate median filter dark and \oneoverf noise correction. The left image is equivalent to a raw data image with common-mode noise between amplifiers and \oneoverf noise present. The middle image shows the common-mode noise between the amplifiers fixed. The right image shows the image corrected for the common-mode noise and the \oneoverf noise. The reference pixels are those outside the white-dashed line. This image has been exaggerated as the full image has 4088 non-reference pixels and is thus impossible to view the 4-wide reference pixels. Illustrative spectral traces have been added to guide the eye.}
    \label{fig:dark_amp_one_over_f_corr}
\end{figure*}

The first part of the correlated noise filtering accounts for gradients along the long axis of the amplifiers readout by removing the slope between the first and last read reference pixels within each amplifier. We take a median of the amplifier `bottom' and `top' reference pixels and subtract for each amplifier the slope between these regions. This accounts for fluctuations in the detector electronics on timescales comparable to or longer than the readout time. Higher-frequency noises are handled as a common-mode between amplifiers in the following step (Section \ref{subsec:one_over_f_noise}). High-frequency readout noise that is not correlated between amplifiers cannot be corrected as it overlaps with science data and cannot be measured independently; it represents the limiting factor for the fainter targets observed with \spirou. This correction is represented by the correction of the `amplifier signal` in Figure \ref{fig:dark_amp_one_over_f_corr}.

\subsection{\oneoverf noise correction} \label{subsec:one_over_f_noise}

The \oneoverf noise component arises from the detector readout electronics that induce structures that are common to all amplifiers and sampled by the reference pixels. The \oneoverf noise manifests itself as  stripes perpendicular to the amplifiers.

\begin{figure*}
    \centering
    \includegraphics[width=16cm]{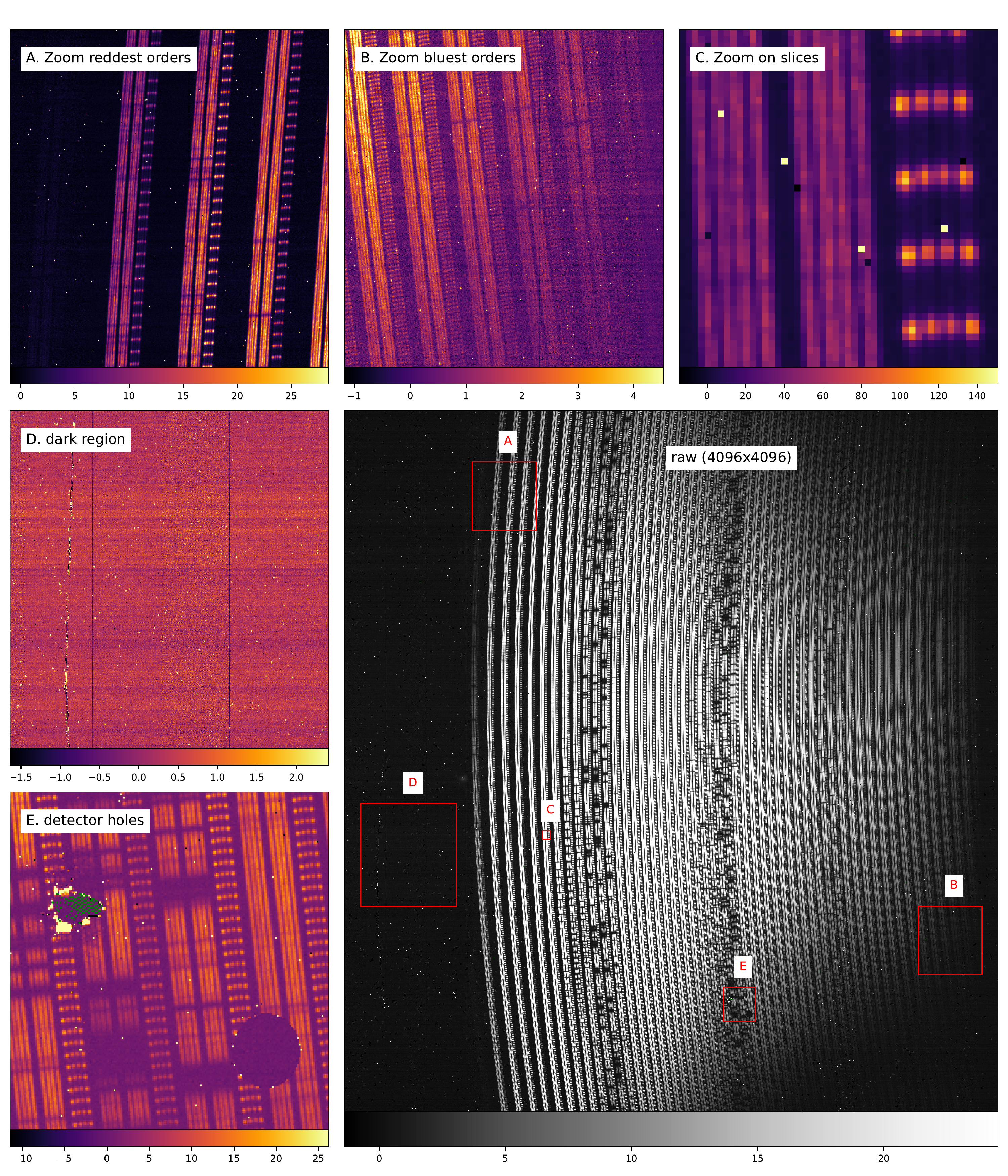}
    \caption{Features and characteristics of the \spirou raw images (\OBJFP). Panel A: A zoom-in of some of the reddest orders (showing the striping in the across-order direction). Panel B: A zoom-in of the bluest orders (also showing the striping in the across-order direction as well as a vertical band in the along-order direction). Panel C: A zoom-in on the individual slices for fibers A, B, and C (showing the shape of the individual slices within a fiber). Panel D: The unilluminated region (showing examples of the detector effects to be removed during pre-processing, as well as one of the large-scale defects present on the detector). Panel E: two of the large detector holes. The detector artifacts highlighted here are corrected in the preprocessing step (Section \ref{sec:preprocessing}) and this figure is reproduced after preprocessing in Figure \ref{fig:pp_features}. Color bars have units of ADU/s.
    \label{fig:raw_features}}
\end{figure*}

This noise has power at all frequencies and they affect the entire array (i.e., both light-sensitive and reference pixels). Ideally, we would fully correct with the 8 reference pixels, but the high-frequency components have an SNR that is too low to measure them robustly within these pixels. We correct the low-frequency components of the \oneoverf noise ($>$32\,pixels) with the reference pixels and then measure the high-frequency component with the unilluminated part the array (the large ($\sim$800-pixel wide beyond $K$-band orders; `dark' region in Figure \ref{fig:raw_features}). Once measured, this common-mode \oneoverf noise is subtracted from all columns of the science array. A cartoon of the \oneoverf signal is shown in Figure \ref{fig:dark_amp_one_over_f_corr}.

\begin{figure*}
    \centering
    \includegraphics[width=16cm]{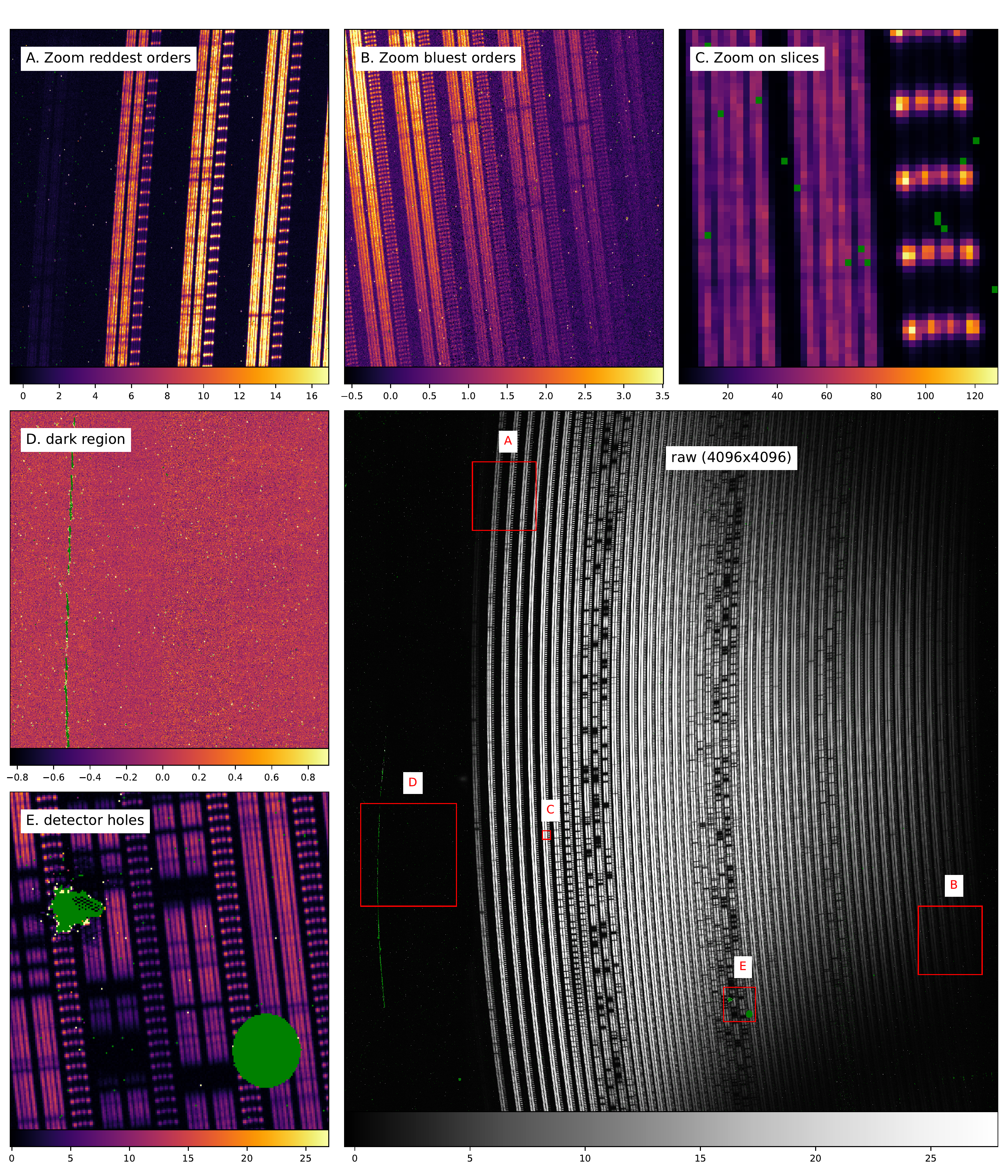}
    \caption{Same as Figure \ref{fig:raw_features} but after pre-processing (without the \APERO rotation applied to match the orientation of the same raw \OBJFP image). Panel A: A zoom-in of some of the reddest orders (showing that the striping in the across-order direction has been removed). Panel B: A zoom-in of the bluest orders (also showing the striping in the across-order direction has been removed as well as a vertical band in the along-order direction having been removed). Panel C: A zoom-in on the individual slices for fibers A, B, and C showing some flagged hot pixels - given a \NAN value, in green). Panel D: The unilluminated region (showing how well the pre-processing has cleaned the unilluminated region, as well as one of the scratches present on the detector). Panel E: shows two of the large-scale defects also with many of the pixels flagged as unusable (given a \NAN value, in green). Color bars have units of ADU/s.
    \label{fig:pp_features}}
\end{figure*}

\subsection{Cosmic ray rejection} \label{subsec:cosmic_ray_reject}

 Cosmic rays hits are easier to flag with infrared arrays than they are with CCD data sets due to the acquisition through multiple readouts. Pixels without a cosmic ray hit are expected to see an accumulation of electrons in their well that is linear with time while a cosmic ray hit would induce a glitch in that accumulation that can easily be flagged. One could attempt to reconstruct a ramp while including a discontinuity at the moment of the hit (e.g. \citealt{Giardino2019}); considering that cosmic rays are rare and that this would add a significant burden in terms of data processing, we opt to simply flag pixels hit by a cosmic ray as invalid (\NAN values). The flagging of cosmic rays is done in two steps.
 
 First, we check for the consistency between the total number of photons received over the entire ramp and the formal ramp error statistics from the linear fit. Discrepant points, even if they remain within the unsaturated regime of the pixel dynamic range, are flagged as invalid. Second, the ramp fitting of the pixel value provides both a slope and an intercept. The slope is the signal used for scientific analysis, and the intercept is discarded. This intercept value corresponds to the state of the detector prior to the first readout, which, for HxRG arrays, is a structured signal. The intercept values have a typical dispersion of $\sim1000$\, ADUs, and discrepant values indicate that photons within a given pixel do not follow a linear accumulation with time. The consistency of the intercept value with expected statistics is used to further flag invalid pixels within a ramp. The flagged cosmic ray pixels can be seen when comparing Figure \ref{fig:raw_features} to Figure \ref{fig:pp_features}.

\subsection{Rotation of image} \label{subsec:image_rot}

The pre-processed images are then rotated to match the HARPS orientation. This is a legacy change left over from when some algorithms shared a common ancestry with the HARPS DRS pipeline  (\citealt{HARPS2003, HARPS2004}). For \spirou data this is equivalent to a 90-degree clockwise rotation (shown in the top left and right of Figure \ref{fig:size_grid}).

\section{Reference calibrations}
\label{sec:ref_calibrations}

While many calibrations are taken on a nightly basis, as one expects slight drifts in the instrument that need to be calibrated, a number of instrument parameters are expected to be fixed on a timescale of years in the absence of a major upgrade (e.g., changing optical elements or science array). Furthermore, a number of calibrations, in particular those linked to the wavelength solution, require an approximate solution to derive an accurate one more reliably. In addition, having a reference calibration is very useful when performing quality assessments of nightly calibrations. We therefore define a night that has been vetted for spurious calibrations as the `reference calibration night'. In the future, we may have reason to split the data between reference nights (i.e., before and after a thermal cycle). Currently, the reference night uses calibration data from 2020-08-31 but we have experimented with different reference nights. We postulate that using two or more reference nights would split the data and that any and all data from different reference nights should not be combined until the last possible moment (i.e., when comparing radial velocities, for \PRV work).

\begin{figure*}
    \centering
    \includegraphics[width=18cm]{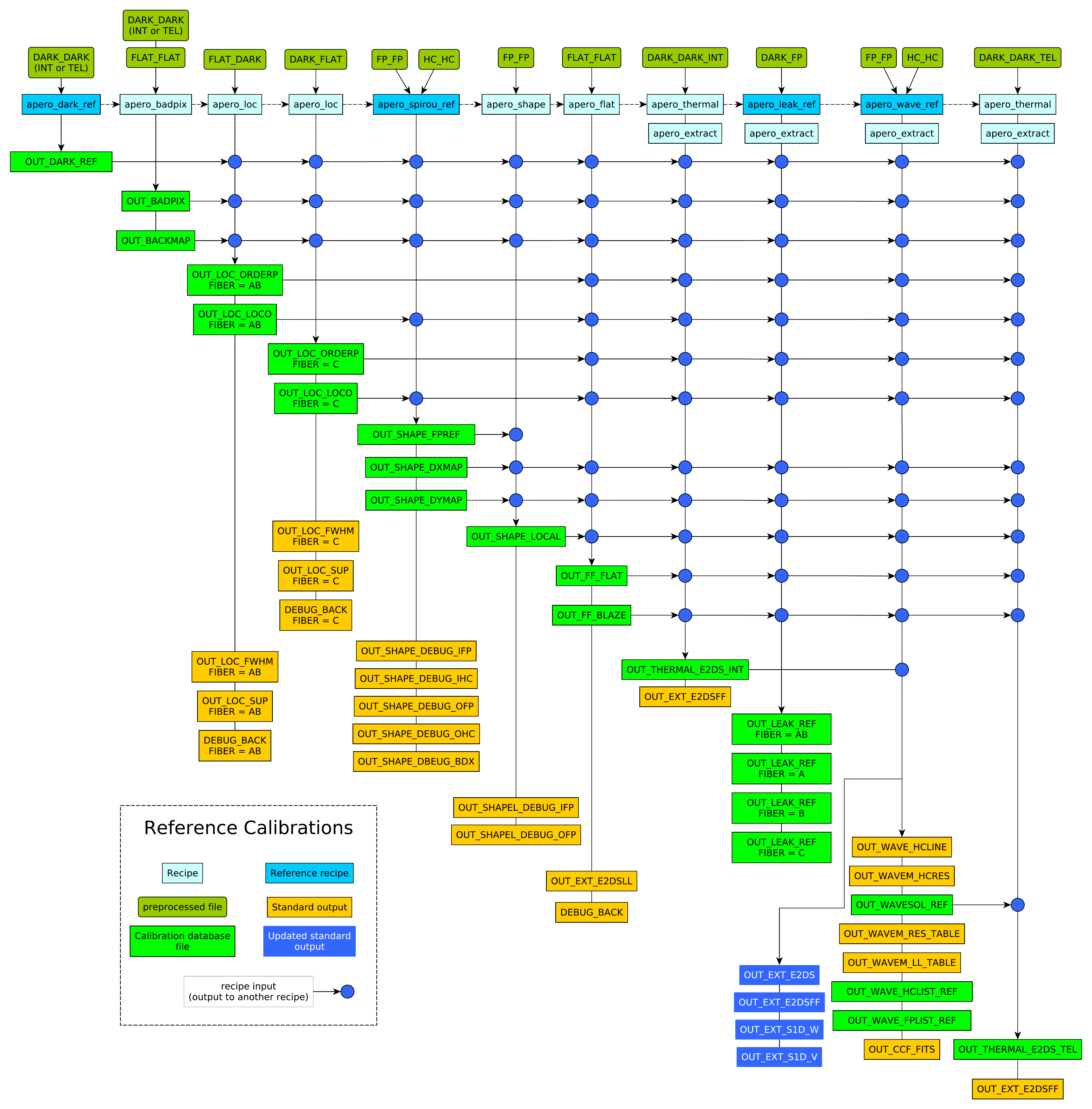}
    \caption{Reference calibration sequence: The input of each step is a pre-processed file. All outputs of reference calibration steps are checked using quality control and then added to the calibration database for use later in \APERO. Only one night for the full \spirou data set is used as a reference observation night. To run some steps of the reference calibration sequence, nightly calibrations must be generated for the reference observation night,
    thus nightly calibrations are shown in the diagram in the position they are run, between reference recipes. Reference calibrations include the reference dark, shape reference, leak correction reference, and reference wavelength solution. The nightly steps required are bad pixel correction, localization, flat and blaze correction, and thermal correction recipes.}
    \label{fig:overview_ref_calib}
\end{figure*}

The reference recipes rely on certain nightly recipes to be run for the reference night; we thus include some nightly calibrations from Section \ref{sec:night_calibrations} in the reference sequence of recipes. Note that the nightly calibration recipes are used as part of the reference sequence and are not equivalent to running a nightly calibration sequence before running the reference recipes (this also means a nightly calibration sequence on the reference night must be run again as part of the nightly calibrations). The order of the reference sequence is as follows:
\begin{itemize}
    \item \Adarkref: a high-pass reference dark from a large subset of preprocessed \hbox{\DARKDARK} files on disk (Section \ref{subsec:ref_dark}).
    \item \Abadpixel: a nightly bad pixel map on the reference night (Section \ref{subsec:bad_pixel}).
    \item \Alocalisation: a nightly measurement of the order position on the reference night (Section \ref{subsec:localization}).
    \item \Ashaperef: a reference map of each order's spectral and spatial shape using a large subset of preprocessed \hbox{\FPFP} files on disk and the reference night \hbox{\HCHC} preprocessed files (Section \ref{subsec:ref_shape}).
    \item \Ashapenight: a nightly snapshot of each order's shape on the reference night (Section \ref{subsec:night_shape}).
    \item \Aflatblaze: a nightly measurement of the blaze and flat profile on the reference night (Section \ref{subsec:flat_blaze}).
    \item \Athermal: a nightly extracted internal dark for determining thermal correction on the reference night (Section \ref{subsec:thermal_calibration}).
    \item \Aleakref: a reference map of FP leakage from the calibration to science fibers using all preprocessed \hbox{\DARKDARK} files on the reference night (Section \ref{subsec:ref_leak}).
    \item \Awaveref: a reference wavelength solution using the reference night \hbox{\FPFP} and \hbox{\HCHC} preprocessed files (Section \ref{subsec:ref_wave}).
    \item \Athermal: a nightly extracted telescope dark for determining thermal correction on the reference night (Section \ref{subsec:thermal_calibration}).
\end{itemize}

The overall reference calibration sequence flow can be seen in Figure \ref{fig:overview_ref_calib}.

\subsection{Generating the reference dark calibration file} \label{subsec:ref_dark}

As \spirou has no moving internal parts for increased stability, one cannot move the fiber out of view and independently measure the detector's dark current. Thus dark frames are non-trivial to construct, as there are two independent contributions to the `dark' image, one arising from the dark current of the science arrays and the other from thermal emission. This problem is mainly seen in the $K$ band and is shared with any \PRV spectrograph for which the fiber thermal emission is commensurate with the per-pixel dark current.

The thermal background manifests itself as a very low-level contribution (typically 0.015 e-/s/pixel), well below the typical target flux, but has a `high flux' tail of much brighter pixels. As the \spirou science array has an extremely stable temperature (sub-milli Kelvin), one expects the pixel dark current to be very stable. From all preprocessed \hbox{\DARKDARK} files, across all nights, we select a subset of 100 \hbox{\DARKDARK} files, uniformly distributed in time as much as possible using a sorting function. If there are less than 100 \hbox{\DARKDARK} files across all available nights we use all files; this becomes our reference dark.

One could use this as the single step for dark correction, but a significant challenge arises. The fiber train is always connected and the science array always sees the thermal emission from the fibers and the hermetic feedthrough connecting the fibers to the cryostat. This thermal emission changes with the temperature of the fiber train and moves, at the pixel level, on timescales of months to years following thermal cycles and maintenance of the instrument. Applying a simple scaling of the dark current, including the thermal background from the fiber, would lead to erroneous subtraction in science data, with sometimes an over subtraction of $\sim2.4\,\mu$m flux, leading to negative flux. We opt for a decoupling of the two contributions in the data calibration. We construct a high-frequency median dark current, which contains pixel-to-pixel detector contributions and low-frequency components from the thermal background of the fiber train. The high-frequency component can be scaled with integration time while the low-frequency one needs to be adjusted (see Section \ref{subsec:thermal_calibration}). This high-pass reference dark image is then saved to the calibration database for use throughout \APERO.

\subsection{Generating the reference shape calibration files} \label{subsec:ref_shape}

In \PRV measurements, constraining the exact position of orders on the science array, both in the spectral and spatial dimensions, is key as the position of our spectra on this science array encodes the sought-after velocity of the star. The diffraction orders of \spirou, and nearly all \PRV spectrographs, follow curved lines, and the image slicer has a 4-point structure (See figures \ref{fig:raw_features} and \ref{fig:pp_features}, panel C) that is not parallel to the pixel grid. 

Within the \APERO framework, we decided to split the problem into two parts: a reference shape calibration (this section) and a nightly shape calibration (Section \ref{subsec:night_shape}). For the reference step, we constrain the bulk motion, as defined through an affine transformation and register all frames to a common pixel grid to well below the equivalent of 1\mps. We perform the order localization and subsequent steps on a nightly basis as it has the significant advantage that registered frames have all orders at the same position to a very small fraction of a pixel. Furthermore, having registered frames allows for better error handling within \APERO; one does not expect pixel-level motions between calibrations after this step.

The reference shape recipe takes preprocessed \hbox{\FPFP} and \hbox{\HCHC} files (as many as given by the user or as many as occur on the nights being used via \Aprocessing). The reference shape recipe combines the \hbox{\FPFP} files into a single \hbox{\FPFP} file and the \hbox{\HCHC} files into a single \hbox{\HCHC} file (via a median combination of the images). After combining, the \FPFP and \hbox{\HCHC} images are calibrated using our standard image calibration technique (see Appendix \ref{section:appendix_standard_image_calibration}). In addition to the combined \FPFP and \HCHC, we create a reference FP image. This reference FP image is created by selecting a subset of 100 \hbox{\FPFP} files (uniformly distributed across nights) and combining these with a median. This reference FP image is then saved to the calibration database for use throughout \APERO.

The registration through affine transformations is done using the \hbox{\FPFP} calibrations. We take the combined \hbox{\FPFP} files and localize in the 2D frame the position of each FP peak and measure the position of the peak maxima. Considering the 3 \spirou fibers and 4 slices (i.e., 12 2D peaks per FP line), this means there are $>100\,000$ peaks on the science array. These are taken as `reference' positions. For each calibration sequence, we then find the affine transformation that minimizes the RMS between the position of the FP and the FP reference image calibration. The resulting affine transformation consists of a bulk shift in $dx$, $dy$, and a $2\times2$ matrix that encodes rotation, scale, and shear. These values are kept and can be useful to identify shifts in the optics (e.g., after earthquakes or thermal cycles) as well as very slight changes in plate scale and angular position of the array which can be of interest in understanding the impact of engineering work onto the science data products. For example, we can readily measure a 10$^{-5}$ fractional change in the \spirou plate scale following a maintenance thermal cycle of the instrument; the ratio of the point-to-point RMS to the median of the plate scale value is at the $1.7\times10^{-7}$ level. The interpolations between pixel grids are done with a 3$^{\rm rd}$ order spline.  We note that changes in the FP cavity length arise from a number of reasons such as gas leakage and temperature and will lead to a motion of FP peaks on the array that is not due to a physical motion of the array or optical elements within the cryostat. Considering that typical drifts are at the $\sim0.3$\,m/s/day level, to first order this leads to a typical $10^{-9}$/day fractional increase in the plate scale along the dispersion direction. This effectively leads to a minute change in the effective dispersion of the extracted file wavelength solution. As this change is common to both the FP, the HC, and the science data, it is accounted for when computing the wavelength solution and cavity length change (see sections \ref{subsec:ref_wave} and \ref{subsec:wave_correction}).

Once the affine transformation has been applied, images are registered to a common grid (the reference FP image). We then construct a transform that makes the orders straight and corrects for slicer structure in the dispersion direction. This leads to the construction of two maps corresponding to $x$ and $y$ offsets that need to be applied to an image to transform it into a rectified image from which a trace extraction can be performed directly through a 1-D collapse in the direction perpendicular to the dispersion of a rectangular box around the order. The $y$ direction (see Figure~\ref{fig:backest} for orientation) map is computed from the order-localization polynomials (Section \ref{subsec:localization}). The $x$ direction map is determined by first collapsing the straightened orders of a \FPFP calibration and cross-correlating each of the spectral direction pixel rows to find its offset relative to the collapsed-extracted spectrum. The $x$ and $y$ offsets are then saved to the calibration database for use throughout \APERO.

\subsection{Generating the reference leak calibration file} \label{subsec:ref_leak}

\begin{figure*}
    \centering
    \includegraphics[width=18cm]{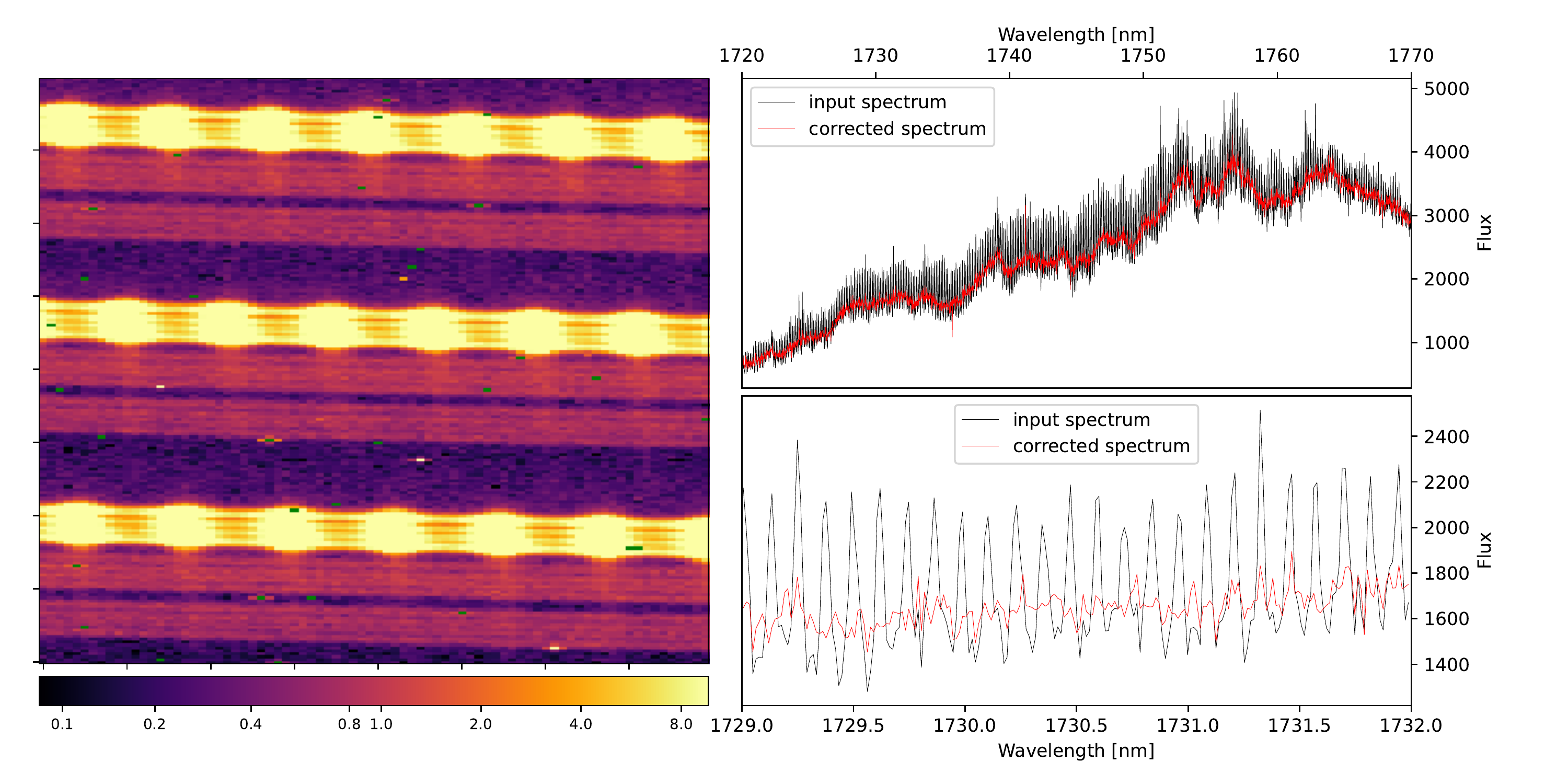}
    \caption{Example of the leakage from the calibration fiber to the science fibers, when the science fiber is unilluminated (DARK). Left: The pre-processed \DARKFP image with a Logarithm stretch in flux (flux measured in electrons). Pixels with a \NAN value are shown in green. Right top: The extracted \DARKFP for order \#44 before correction (black, input spectrum) and after (red, corrected spectrum). Right bottom: A zoom-in on the extracted \DARKFP for order \#44.}
    \label{fig:leak_plot}
\end{figure*}

For \PRV observations, the observational setup is most often one with a science object in the A and B fibers and an FP illumination in the C fiber (i.e., \hbox{\OBJFP} or \hbox{\POLFP}). Considering that the \spirou slicer has sharp edges in its pupil, there is a diffraction pattern that leads to a spike in the cross-fiber direction and a modest cross-fiber component in the leakage. The leakage of the FP spectrum onto the science spectrum is constant through time as it is solely due to pupil geometry, and can therefore be calibrated and subtracted. The reference leak recipe finds all \hbox{\DARKFP} files in the raw directory (from the reference night). Each \hbox{\DARKFP} file is then extracted (see the extraction process in Section \ref{sec:extraction}). Once all \hbox{\DARKFP} files are extracted they are combined for each fiber: AB, A, B, and C (via a median across all extracted \ETDS files) creating one image ($49\times4088$) per fiber. Conceptually, the leak correction is straightforward: take the combined \DARKFP, normalize each C fiber FP to unity (using the $5^{\rm th}$ percentile of FP flux within the order) and measure the recovered spectrum in the A and B fibers. For any given \hbox{\OBJFP} or \hbox{\POLFP} observation, one simply measures the C fiber FP flux and scales the leakage in A and B accordingly. An example of this is shown in Figure \ref{fig:leak_plot} for a DARK in the science fibers and an FP in the calibration fiber. The method has been tested over the lifetime of \spirou and subtracts the high-frequency component of the leakage at a level better than 1 in 100 in the most contaminated orders. The reference leak calibration file (\REFLEAK) is then saved to the calibration database for use throughout \APERO.

\subsection{Generating the reference wavelength calibration files} \label{subsec:ref_wave}

The wavelength solution generation follows the general idea of \cite{Hobson2021} however since publication there has been an overall reshuffling of the logic. As such we present an overview of the process here but refer the reader to \cite{Hobson2021} for further specific details.

The reference wavelength solution recipe takes preprocessed \hbox{\FPFP} and \hbox{\HCHC} files (as many as given by the user or as many as occur on the nights being used via \Aprocessing) from the reference night. It combines the \hbox{\FPFP} and \hbox{\HCHC} files into a single \hbox{\FPFP} and a single \hbox{\HCHC} file (via a median combination of the images). These combined \hbox{\FPFP} and \hbox{\HCHC} files are then extracted (see the extraction process in Section \ref{sec:extraction}).

We first consider the combined flux in fibers A and B (the AB fiber). We locate the \hbox{\HCHC} lines, starting with a line list generated as in \cite{Hobson2021}, fitting each peak with a Gaussian and measuring the position of the peak, and inferring peak wavelength from an initial guess at the wavelength solution from physical models. The first time this HC finding is performed we allow for a global offset between the current \hbox{\HCHC} file and the initial guess at the wavelength solution (this is important when our reference night is far in time from when our initial wavelength solution data was taken).

For the \FPFP AB fiber, a similar process is followed. However, instead of a single Gaussian, an Airy function is used (to account for the previous and following FP peak in the fitting process):

\begin{equation}
    F_{airy} = A\left( 0.5 \left(1 + \frac{2\pi(x-x_0)}{w} \right) \right)^{\beta} + DC
    \label{equation:airy_function}
\end{equation}

\noindent where $F$ is the modeled flux of the FP, $A$ is the amplitude of the FP peak, $x_0$ is the central position of the FP peak, $w$ is the period of the FP in pixel space, $\beta$ is the shape factor of the FP peak and $DC$ is a constant offset. Once we have found all HC and FP lines in the AB fiber we calculate the wavelength solution.

The accurate wavelength solution for reference night is then found through the following steps:
\begin{itemize}
    \item From FP peak spacing within each order, derive an effective cavity length per order.
    \item Fit the chromatic dependency of the cavity with a 5$^{\rm th}$ order polynomial and keep that cavity in a reference file; through the life of the instrument, we will assume that cavity changes are achromatic relative to this polynomial.
    \item From the chromatic cavity solution, we find the FP order value of each peak, typically numbering from $\sim9 600$ to $\sim24\,500$ respectively at long and short wavelength ends of the \spirou domain.
    \item From the peak numbering, which is known to be an integer, we can refine the wavelength solution within each order. This solution is kept as a `reference' wavelength solution.
\end{itemize}

The finding of the fiber AB HC and FP lines and the calculation of the wavelength solution is repeated multiple times (in an iterative process). We essentially `forget' the locations of the HC and FP lines and re-find them as if we hadn't found them before, only this time instead of the initial guess wavelength solution we use the previous iteration's calculated solution and the previous iterations calculated cavity width fit as a starting point.

Finally, after three iterations, which is sufficient to converge to floating point accuracy, we re-find the HC and FP lines for the AB fiber one last time using the final reference wavelength solution and final cavity width fit. We also make an estimate of the resolution, splitting the detector into a grid of 3$\times$3 and using all HC lines in each sector to estimate the line profile and thus the resolution of each sector. We then process each fiber (A, B, and C) in a similar manner to the AB fiber (finding HC and FP lines from the extracted images and calculating the wavelength solution) the only difference being we do not fit the cavity width nor do we fit the chromatic term; we force the coefficients to be the ones found with the AB fiber.

For quality control purposes we calculate an FP binary mask using the cavity width fit and use this to perform a cross-correlation function between the mask and the extracted FP for all fibers (AB, A, B, and C). We use the cross-correlation function to measure the shift of the wavelength solutions measured in fiber AB compared to fibers A, B, and C and confirm that this is less than 2 \mps. As a second quality control, we match FP lines (found previously) between the fibers and directly calculate the difference in velocity between these lines as a second metric on the radial velocity shift between the fibers' wavelength solutions. Note that typically for the reference night the value of these quality control metrics is around 10-20 \cmps between fibers (i.e. $AB-A$, $AB-B$, $AB-C$).

The reference wavelength solution file (\REFWAVE) for each fiber, a cavity fit file, and a table of all HC and FP lines found are then saved to the calibration database for use throughout \APERO. A resolution map is also saved. The \hbox{\HCHC} and \hbox{\FPFP} extracted files have their headers updated with the reference wavelength solution.

\begin{figure*}
    \centering
    \includegraphics[width=17cm]{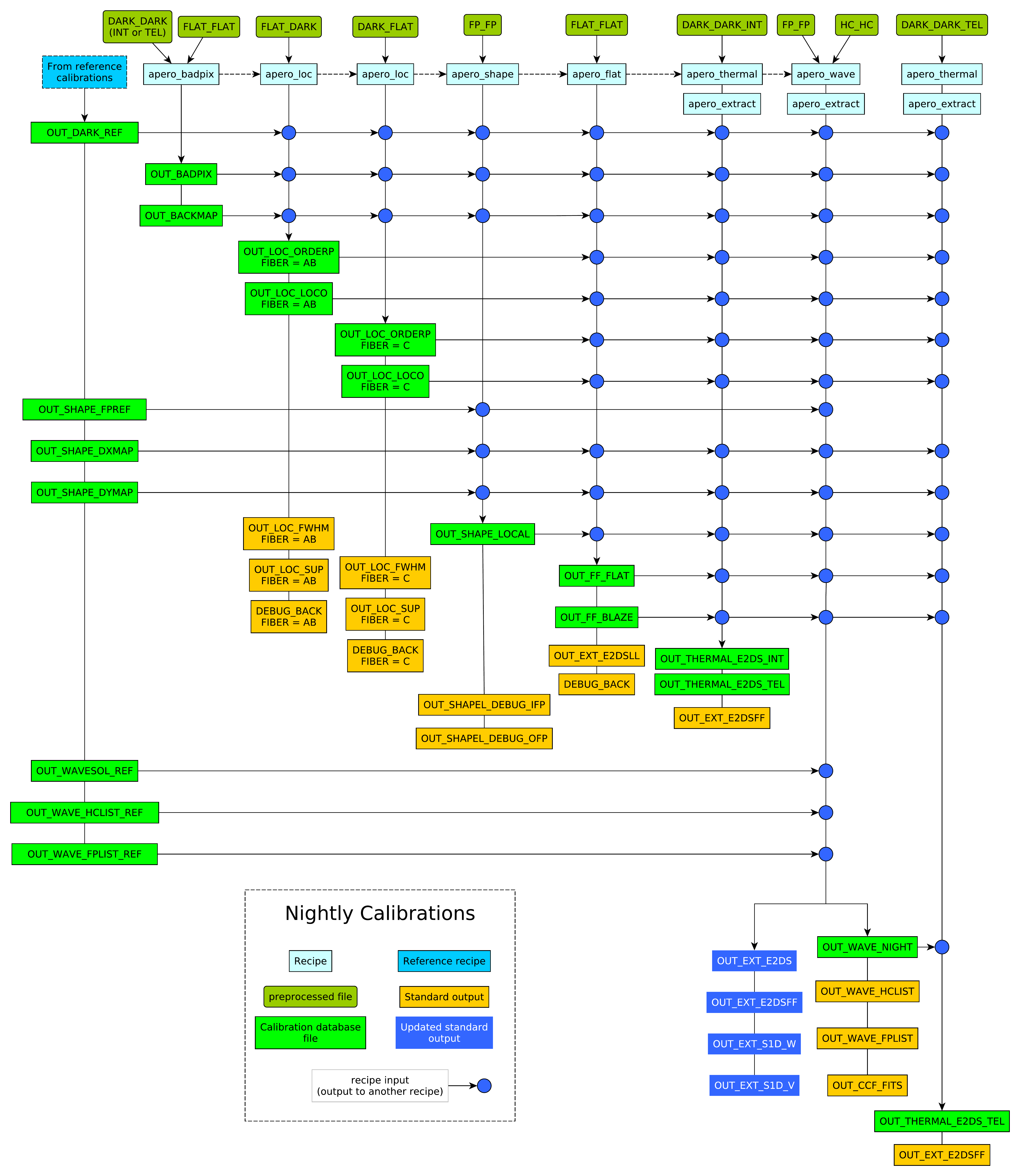}
    \caption{Nightly Calibration sequence: The input of each step is a pre-processed file. All outputs of nightly calibration steps are checked using quality control and then added to the calibration database for use later in \APERO. Steps include bad pixel correction, localization, shape correction, flat and blaze correction, thermal correction, and wavelength solution calibration files.}
    \label{fig:overview_night_calib}
\end{figure*}

\section{Nightly calibrations}
\label{sec:night_calibrations}

\spirou calibrations are taken twice a day: once before the start of the science observations and once after all science observations are completed (i.e., once in the evening and once in the morning). When processing, each step (i.e., each recipe type) is run on all applicable data for all nights before moving on to the next recipe unless a user is running one night at a time (e.g., processing new data from one night or one run only). For an individual night calibrations from before and after science observations are combined.

The order of the night calibration sequence is as follows:
\begin{itemize}
    \item \Abadpixel~- a bad pixel map for each night (Section \ref{subsec:bad_pixel}).
    \item \Alocalisation~- a measurement of the order position on each night (Section \ref{subsec:localization}).
    \item \Ashapenight~- a snapshot of each order's shape on each night (Section \ref{subsec:night_shape}).
    \item \Aflatblaze~- a measurement of the blaze and flat profile on each night (Section \ref{subsec:flat_blaze}).
    \item \Athermal~- an extracted internal dark for determining thermal correction of calibrations on each night (Section \ref{subsec:thermal_calibration}).
    \item \Awavenight~- a wavelength solution measurement on each night (Section \ref{subsec:wave_correction}).
    \item \Athermal~- an  extracted telescope dark for determining thermal correction of on-sky observations on each night (Section \ref{subsec:thermal_calibration}).
\end{itemize}

The overall nightly calibration sequence flow can be seen in Figure \ref{fig:overview_night_calib}.

\subsection{Generating bad pixel calibration files} \label{subsec:bad_pixel}

\begin{figure*}
    \centering
    \includegraphics[width=16cm]{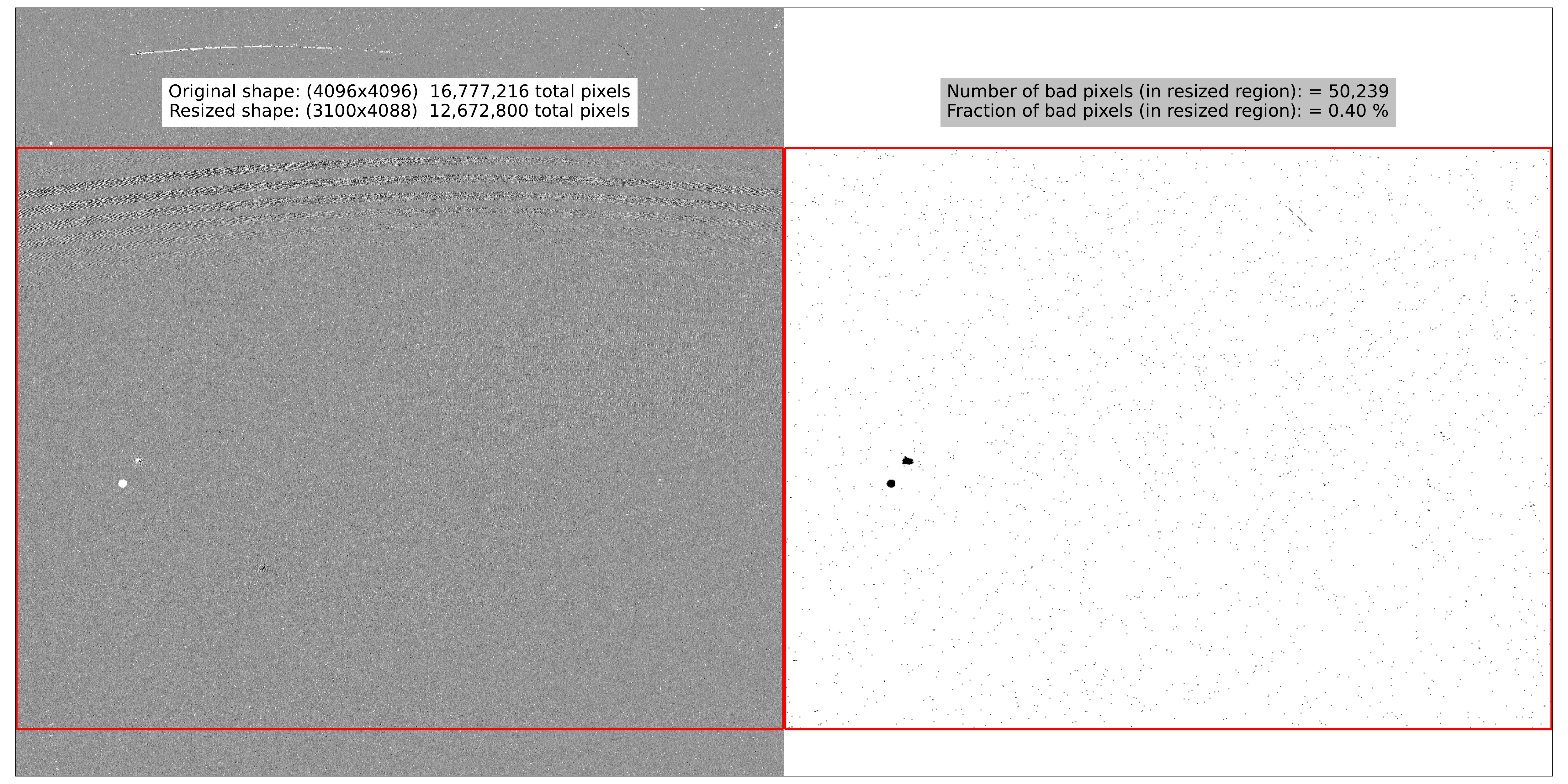}
    \caption{Left: a typical \DARKDARK image, right: an example combined bad pixel map (the final bad pixel map used for a night calibration). For this bad pixel map example, we identify $\sim50\,000$ bad pixels (0.4\%, the majority being located in two large clusters). In both the red square represents the resized area (see Section \ref{subsec:flip_resize_rescale}), bad pixels are only kept in this region.}
    \label{fig:badpix_map}
\end{figure*}

The bad pixel recipe takes preprocessed \DARKDARK and \FLATFLAT files (as many as given by the user or as many as occur on the nights being used via \Aprocessing). It combines all \DARKDARK files and all \FLATFLAT files into a single \DARKDARK and a single \FLATFLAT (via a median combination of the images). Bad pixels are then identified in the \FLATFLAT by using Equation \ref{equ:badpix_flat}.

\begin{equation}
   \label{equ:badpix_flat}
    M_{\text{flat } i,j} = \left\{ \begin{array}{cl}
      1 : & FLAT_{i,j} \text{ is not finite} \\
      1 : & \mid (FLAT_{i,j} / FLAT_{\text{med } i,j}) - 1 \mid > \text{cut\_ratio} \\
      1 : & FLAT_{\text{med } i,j} < \text{illum\_cut} \\
      0 : & \text{otherwise} \\
    \end{array} \right.
\end{equation}

\vspace{2cm}
\noindent where $FLAT_{i,j}$ is the flux in \ARRIJ of the \FLATFLAT image; $FLAT_{\text{med }}$ is the median filtered flat image (using a filtering width of 7 pixels) and $M_{\text{flat } i,j}$ is 1 to flag a bad pixel or 0 otherwise, cut\_ratio is 0.5 (flagging pixels with a response less than 50 percent of their neighbors or unphysically brighter than neighbors) and illum\_cut is 0.05 (flagging pixels at the edge of the blaze response). $FLAT$ and $FLAT_{\text{med }}$ have first been normalized by the $90^{\rm th}$ percentile of flux in the median filtered flat image. Thus $ M_{\text{flat}}$ is a Boolean flag map of bad pixels on the flat image. For the \DARKDARK image, bad pixels are identified using Equation \ref{equ:badpix_dark}.

\begin{equation}
    \label{equ:badpix_dark}
    M_{\text{dark } i,j} = \left\{ \begin{array}{cl}
      1 : & DARK_{i,j} \text{ is not finite} \\
      1 : & DARK_{i,j} > 5.0 \text{ADU/s}  \\
      0 : & \text{otherwise} \\
    \end{array} \right.
\end{equation}

\noindent where $DARK_{i,j}$ is the flux in \ARRIJ of the dark image. Thus $M_{\text{dark}}$ is a Boolean flag map of bad pixels on the dark image. We choose a value of 5.0 ADU/s as it is representative of the pixel flux of a typical science target. Including pixels with a brighter level of dark current than this leads to a loss in SNR rather than a gain. We note that this threshold could be target-dependent but for simplicity we use a single value.

In addition to this bad pixels in a full detector engineering flat ($FULLFLAT$ taken during commissioning) are also identified using Equation \ref{equ:badpix_fullflat}.

\begin{equation}
    \label{equ:badpix_fullflat}
    M_{\text{full-flat } i,j} = \left\{ \begin{array}{cl}
      1 : & \mid FULLFLAT_{i,j} - 1 \mid > 0.3  \\
      0 : & \text{otherwise} \\
    \end{array} \right.
\end{equation}

\vspace{0.5cm}
\noindent where $FULLFLAT_{i,j}$ is the flux in \ARRIJ of the full detector engineering flat. Thus $ M_{\text{full-flat}}$ is a Boolean flag map of bad pixels on the full detector engineering flat image. We chose 0.3 as this flagged the defective regions identified manually on the detector. The 1$\sigma$ dispersion of the full detector engineering flat image is 2 percent.

These three bad pixel maps are then combined into a single bad pixel map (Equation \ref{equ:badpix_comb}) that is shown in Figure \ref{fig:badpix_map}.

\begin{equation}
    \label{equ:badpix_comb}
    M_{i, j} = M_{\text{flat } i,j} \text{ or } M_{\text{dark } i,j} \text{ or } M_{\text{full-flat } i,j}
\end{equation}

\vspace{0.5cm}
\noindent where $M_{i,j}$ is the Boolean flag value for \ARRIJ used as the final bad pixel map. A standard bad pixel map can be seen in Figure \ref{fig:badpix_map}.

The final step with the bad pixel maps is to dilate clumps of large bad pixels, identifying additional pixels from the edges of these clumps. This is done using the \lstinline{binary_erosion} and \lstinline{binary_dilation} functions in \modscipy \citep{Virtanen_2020} using circular apertures of 5 and 8 pixels respectively for the erosion and dilation. The erosion is used to remove small bad pixel clumps and isolate pixels from a copy of the bad pixel map. This then allows the remaining larger bad pixel clumps to be identified. The dilation then increases the size of these large clumps flagging pixels around their edges as bad. This copy of the bad pixel map is then merged back into the original bad pixel map.

\begin{figure*}
    \centering
    \includegraphics[width=\textwidth]{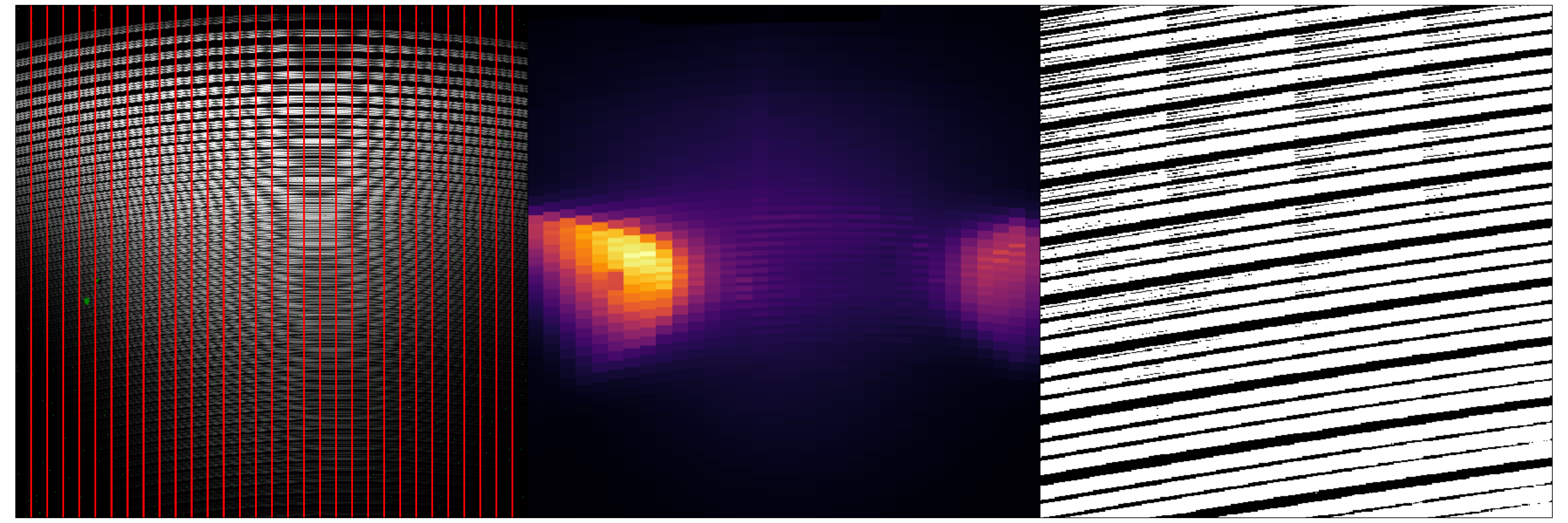}
    \caption{Left: The \FLATFLAT combined and preprocessed input image to the bad pixel calibration recipe. Over plotted in red are the ribbons defined at 128\, pixel steps in the across order direction, Middle: the low spatial frequency background estimate ($BACK\_EST$) based on the $5^{\rm th}$ percentile of $128\times128$ chunks, Right: a zoom in on the out of order region mask ($M_{\text{back }}$) calculated from Equation \ref{equ:badpix_backmap}, white area is identified as in-order pixels, black as out-of-order pixels.}
    \label{fig:backest}
\end{figure*}

In addition to this, we use the bad pixel map along with the \FLATFLAT image to define a mask of the out-of-order regions. It is defined by slicing the image into ribbons of width 128 pixels (in the along-order direction); this width is chosen such that it is small enough for the orders not to show a significant curvature within each ribbon (essentially splitting the image into 32 $4088\times128$ rectangles). We take a median of each ribbon in the along-order direction (creating 32 vectors of length 4088) creating a profile of the orders for every 128 pixels across the order. We split each of these 32 order profiles into 128-pixel regions and estimate the background of these as the 5$^{th}$ percentile of the flux in this sub-ribbon. We then set this background estimate for all pixels in that $128\times128$ box and thus produce $BACK\_EST$, a $4096\times4096$ image where each $128\times128$ sub-region is set to the $5^{th}$ percentile value. The out-of-order region mask is then set by Equation \ref{equ:badpix_backmap}. The ribbon regions, background estimate ($BACK\_EST$), and out-of-order region mask are shown in Figure \ref{fig:backest}, where bad pixels (with \NAN value) are set to green.

\begin{equation}
    \label{equ:badpix_backmap}
    M_{\text{back } i,j} = \left\{ \begin{array}{cl}
      1 : & FLAT_{i,j} < BACK\_EST_{i,j}  \\
      0 : & \text{otherwise} \\
    \end{array} \right.
\end{equation}

\vspace{0.5cm}
\noindent where $M_{\text{back } i,j}$ is the out of order region mask, $FLAT_{i,j}$ is the flux in \ARRIJ of the \FLATFLAT image and $BACKEST_{i,j}$ is the crude $5^{\rm th}$ percentile background estimate for \ARRIJ (from the $128\times128$ sub-regions).

Both the bad pixel map (\BADPIX) and out-of-order region mask (\BACKMAP) are saved to the calibration database for use throughout \APERO. The bad pixel map and background map are used in our standard image calibration (see Appendix \ref{section:appendix_standard_image_calibration}).

\subsection{Generating localization calibration files} \label{subsec:localization}

The localization recipe takes preprocessed \DARKFLAT or \FLATDARK files (as many as given by the user or as many as occur on the nights being used via \Aprocessing). It is run twice, once for the C fiber localization (with a set of \DARKFLAT) and once for the AB fiber localization (with a set of \FLATDARK). It combines the \DARKFLAT files or the \FLATDARK files into a single \DARKFLAT or \FLATDARK (via a median combination of the images). After combining, the images are calibrated using our standard image calibration technique (see Appendix \ref{section:appendix_standard_image_calibration}).

The first step in the localization code is to take the combined and calibrated \DARKFLAT or \FLATDARK and apply a weighted box median (shown in Equation \ref{equ:loc_orderp}).

\begin{equation}
    \label{equ:loc_orderp}
    IM_{\text{orderp } j} = \left\{ \begin{array}{ll}
      \text{MED}(IM_{j=0:j=k+1}):     & k < 5            \\
      \text{MED}(IM_{j=k-5:j=4088}):  & k > 4088 - 5     \\
      \text{MED}(IM_{j=k-5:j=k+5+1}): & \text{otherwise} \\
    \end{array} \right.
\end{equation}

\vspace{0.5cm}
\noindent where $IM_{\text{orderp } j}$ is the order profile flux for all rows in the $j^{th}$ column, $IM_{j=x:j=y}$ is the combined, calibrated \DARKFLAT or \FLATDARK, that spans all columns from $j=x$ to $j=y$, and $k$ is the column index number and ranges from $j=0$ to $j=4088$. 

This produces the order profile image of the \DARKFLAT or \FLATDARK which is used for the optimal extraction (see Section \ref{subsec:optimal_extraction}) and to locate the orders.

To locate the orders we use the \modscikit \lstinline{measure.label} algorithm \citep{scikit-measure-label1, scikit-measure-label2} which labels connected regions. Two pixels are defined as connected when both themselves and their neighbors have the same value. We use a connectivity value of 2 meaning that any of the 8 surrounding pixels can be neighbors if they share the same value. 

In order to facilitate the labeling we first perform a $95^{th}$ percentile of a box (of size $25\times25$\, pixels) across the whole image, as true illuminated pixels' flux is location-dependent. We set a threshold at half that value and label all pixels above this threshold as one and all pixels below this to a value of zero. We then perform the \lstinline{measure.label} on this Boolean map (referred to from this point on as $Mask_{orders}$). This is just a first guess of the order positions and usually returns many labeled regions that are not true orders. 

To remove bad labels we first remove any labeled region with less than 500\,pixels. We then remove any pixel within a labeled region that has a flux value less than 0.05 times the $95^{th}$ percentile of all pixels in that given labeled region and remove this pixel from $Mask_{orders}$. We then median filter each row of $Mask_{orders}$ to clean up the labeled edges and apply a binary dilation (\modscipy \lstinline{ndimage.binary_dilation}) algorithm. This binary dilation essentially merges labeled regions that are close to each other together by expanding regions marked with ones around the edges of these regions. After $Mask_{orders}$ has been updated we re-run the labeling algorithm. As a final filtering step, we remove any region center that does not overlap with the central part of the image in the along-order direction (i.e., the center $\pm$ half the width of the detector, 2044$\pm$1022 pixels).

Once we have the final set of labeled regions we use $Mask_{orders}$ on each order to fit a polynomial fit (of degree 3) to the pixel positions in that labeled region forcing continuity between orders by fitting each coefficient across the orders. We also use the $Mask_{orders}$ pixel positions to linearly fit the width of each order.

\begin{figure*}
    \begin{minipage}{1.0\textwidth}
        \centering
        \includegraphics[width=15.5cm]{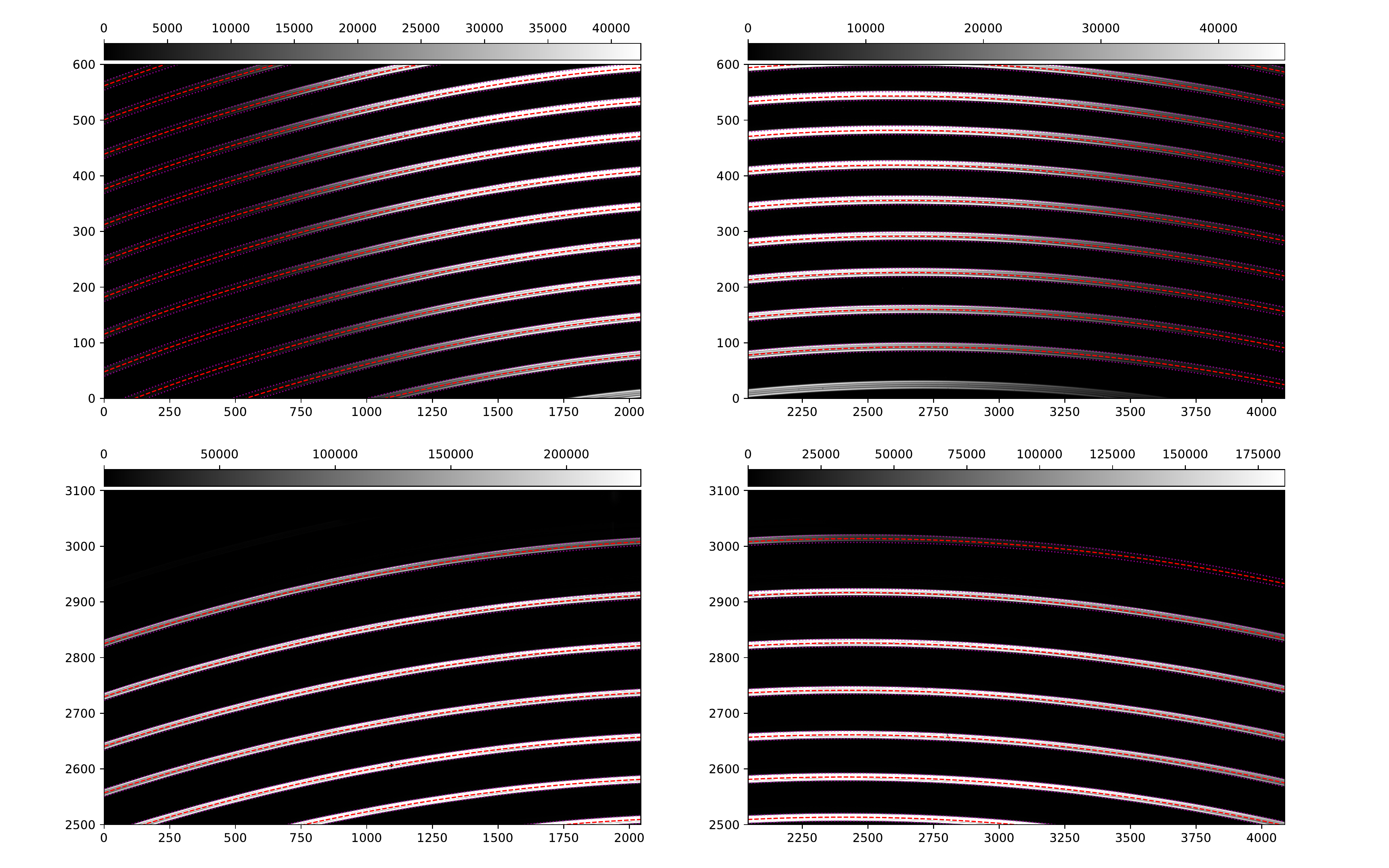}
        \caption{The localization polynomial fits for position and width in the four corners of the \DARKFLAT image (for fiber C). In total we fit 49 orders$^{\ref{footnote:fifty_orders}}$ for fiber C. The y-axis shows the across-order direction and the x-axis shows the along-order direction. Color bar is flux in electrons.}
        \label{fig:loc_corner}
    \end{minipage}
    
    \begin{minipage}{1.0\textwidth}
        \centering
        \includegraphics[width=15.5cm,trim={0 0 0 0.825cm}, clip]{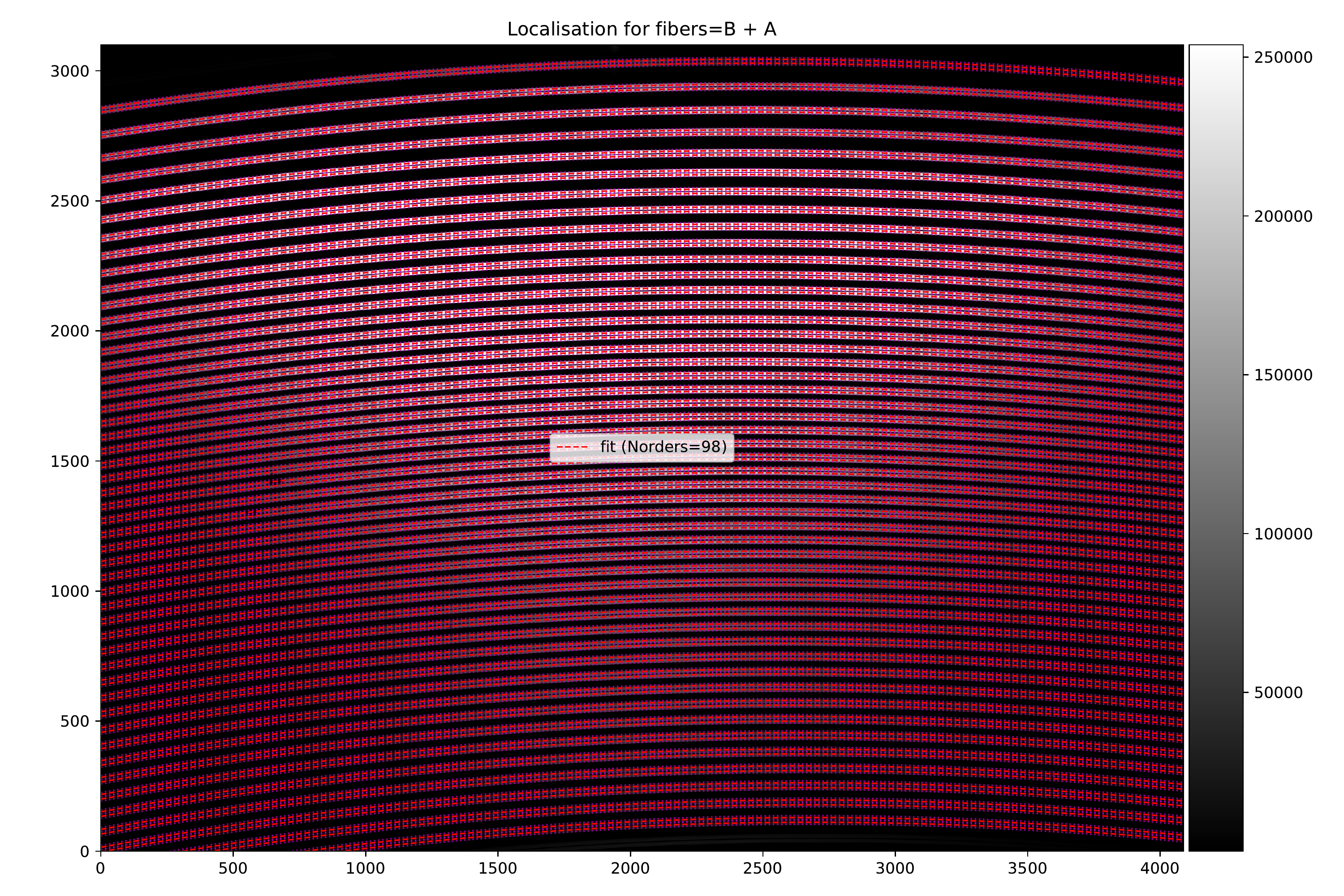}
        \caption{The localization polynomial fits for position and width across all orders of the \FLATDARK image (for fibers A and B). In total we fit 49 orders$^{\ref{footnote:fifty_orders}}$ for fiber A and 49 orders$^{\ref{footnote:fifty_orders}}$ for fiber B. The y-axis shows the across-order direction and the x-axis shows the along-order direction. The color bar is flux measured in electrons.}
        \label{fig:loc_full}

    \end{minipage}
\end{figure*}

For a \DARKFLAT, this produces polynomial fits and coefficients for 49 orders for the C fiber. For a \FLATDARK input, this produces polynomial fits and coefficients for 98 orders (49 orders for A and 49 orders for B). These polynomial coefficients for the positions of the orders and the widths of the orders are then converted into values as a function of position across each order. These fits can be seen in figures \ref{fig:loc_corner} and \ref{fig:loc_full}.

As part of quality control we check that:
\begin{itemize}
    \item the number of orders is consistent with the required number of orders (49 for fiber C, 98 for fibers A+B). 
    \item the across-order value at the center of the detector is always larger than the value of the previous order
\end{itemize}

The order profile (\ORDERP), locations of the orders (\LOCO), and widths of the orders are saved to the calibration database (if both quality control criteria are met) for use throughout \APERO.

\subsection{Generating nightly shape calibration files} \label{subsec:night_shape}

Before extracting the spectrum, we need to transform the image into a format that is amenable to a simple 1-dimensional collapse. Given our reference FP grid and the  $x$ and $y$ displacements maps from Section \ref{subsec:ref_shape}, on a given night, we only need to find the affine transform that registers FP peaks onto the reference FP image and updates the $x$ and $y$ transform maps within the affine contribution. This assumes that the order curvature is constant through the life of the instrument and that the slicer shape is stable. We note that as the order profiles are determined in each nightly calibration, a slight (sub-pixel) modification of the position of orders would have no impact on the extracted spectra which are extracted with the profile measured for the corresponding night.

The nightly shape recipe takes preprocessed \FPFP files (as many as given by the user or as many as occur on each of the nights being used via \Aprocessing). It combines the \FPFP files into a single \FPFP per night (via a median combination of the images). After combining, the \FPFP images are calibrated using our standard image calibration technique (see Appendix \ref{section:appendix_standard_image_calibration}). We take the \REFFP, \SHAPEX and \SHAPEY calibrations from the calibration database (created in Section \ref{subsec:ref_shape}). If multiple exist we use the closest in time (using the header key \MIDEXPOSURE from the header). To find the linear transform parameters (dx, dy, A, B, C, and D) between the reference \FPFP and this night's \FPFP we find all the FP peaks in the reference \FPFP image and in the nightly \FPFP image. Once we have the linear transform parameters we shift and transform the combined and calibrated nightly \FPFP via our shape transform algorithm (see Appendix \ref{section:affine_transformation}, using the linear transform parameters, \SHAPEX and \SHAPEY) and save the transformed image and un-transformed image to disk (for manual comparison to the input \FPFP image). 

As part of quality control, we check that the RMS of the residuals in both directions (across order and along the order) are less than 0.1\, pixel, which has been found to be optimal to flag pathological cases. The transformation parameters (dx, dy, A, B, C, and D, henceforth \SHAPELOCAL) are then saved to the calibration database (if both quality control criteria are met) for use throughout \APERO.

\subsection{Generating flat and blaze calibration files} \label{subsec:flat_blaze}

An essential part of the extraction process is calibrating the flat field response (removing the effect of the pixel-to-pixel sensitivity variations) and calculating the blaze function. The blaze can be seen visually in the raw and preprocessed images (e.g., region B in figures \ref{fig:raw_features} and \ref{fig:pp_features}) as a darkening of the orders, especially at the blue end, towards the sides of the detector (in the along-order direction).

The nightly flat recipe takes preprocessed \FLATFLAT files (as many as given by the user or as many as occur on each night being used via \Aprocessing). It combines the \FLATFLAT files into a single \FLATFLAT per night (via a median combination of the images). After combining, the \FLATFLAT images are calibrated using our standard image calibration technique (see Appendix \ref{section:appendix_standard_image_calibration}). The combined, calibrated \FLATFLAT file is then extracted (using the same extraction algorithms presented in Section \ref{sec:extraction}). The rest of the flat and blaze recipe is handled per order. Once extracted, the \ETDS ($49\times4088$) is median filtered (with a width of 25 pixels) and all pixels with flux less than 0.05 the $95^{th}$ percentile flux value or greater than 2 times the $95^{th}$ percentile flux value are removed. Each \FLATFLAT \ETDS order is then fit with a $sinc$ function (Equation \ref{equ:blaze_sinc}).

\begin{equation}
    \label{equ:blaze_sinc}
     \begin{array}{cc}
        \text{B}_i = AS(sin(\theta)/\theta)^2 \\
        \\
        S = 1 + s(x_i - L) \\
        \theta = \pi \bar{x_i} / P \\
        
        \bar{x_i} = (x_i - L) + Q(x_i - L)^2 + C(x_i - L)^3
    \end{array}
\end{equation}

\vspace{0.5cm}
\noindent where $\text{B}_i$ is the blaze model for the $i^{th}$ \ETDS order, $A$ is the amplitude of the $sinc$ function, $P$ is the period of the $sinc$ function, $s$ is the slope of the $sinc$ function, $x_i$ is the flux vector of the \ETDS order, $L$ is the linear center of the $sinc$ function, $Q$ is a quadratic scale term, and $C$ is a cubic scale term. The terms fit in the $sinc$ function are $A$, $P$, $L$, $Q$, $C$ and $s$ as a function of $x_i$.

\begin{figure*}
    \centering
    \includegraphics[width=18cm]{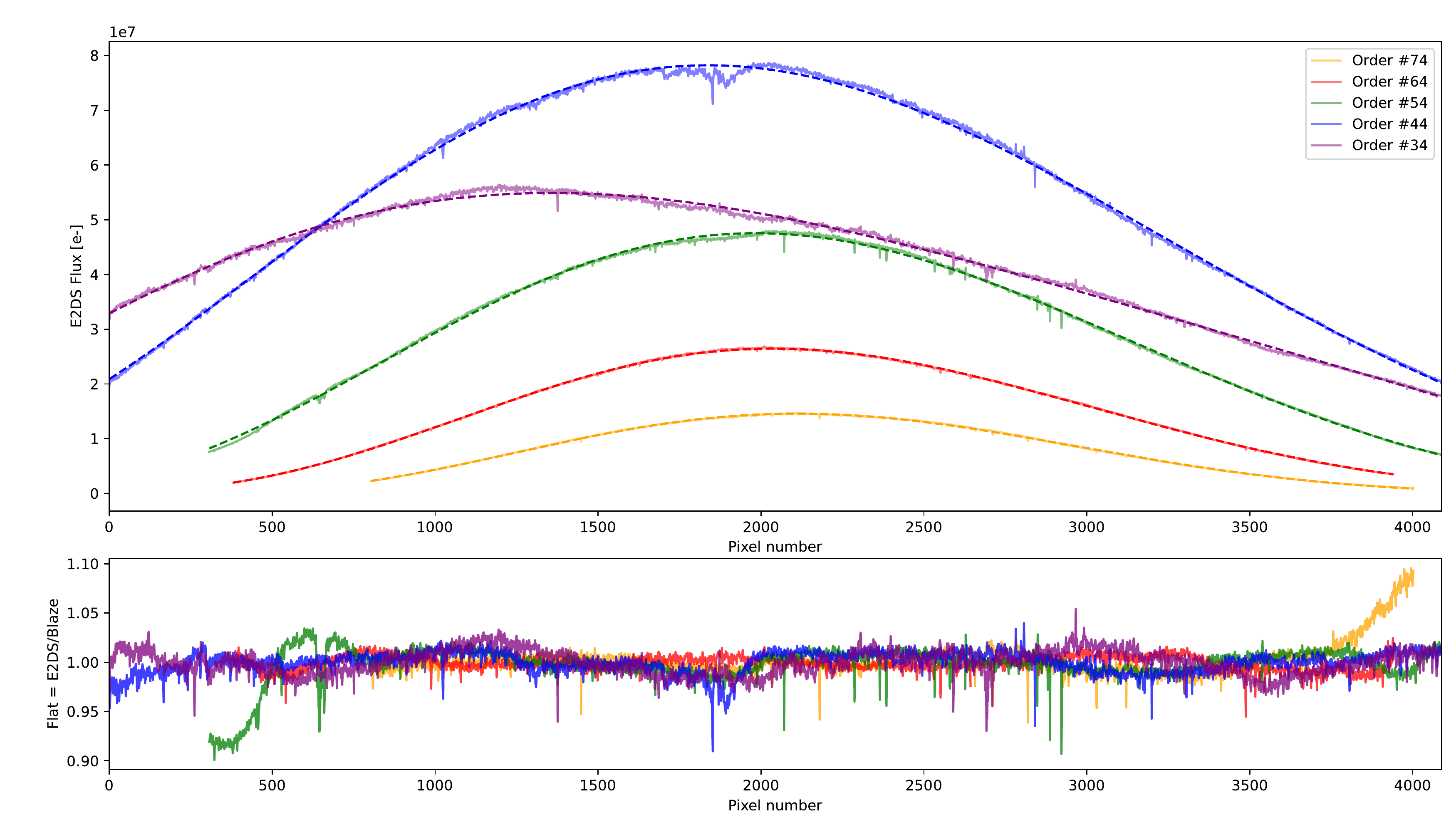}
    \caption{Top: The blaze fit to several of the \spirou \FLATFLAT \' echelle orders (fiber AB). Bottom: the ratio of the blaze and input spectrum are used as the flat calibration file, here a perfect detector would have a value of one.}
    \label{fig:flat_and_blaze}
\end{figure*}

Once we have a set of parameters the blaze function for this order is $\text{B}_i$ for all values of the flux for this order. The original \ETDS order is then divided by the blaze function and this is used as the flat profile. A standard deviation of the flat is also calculated for quality control purposes. This process is repeated for each order producing a full blaze and flat profile (49$\times$4088) for the input \FLATFLAT files. An example blaze fit and the resulting flat can be seen in Figure \ref{fig:flat_and_blaze}. To avoid erroneous contributions to the flat any outlier pixels (outside 10$\sigma$ or within $\pm$0.2 of unity) are set to \NAN. Note that the multiplication of the blaze and the flat is equivalent to the full response function of the detector. For some orders (\#34 and \#74), there is a large residual at one edge of the blaze falloff. This is due to the mismatch between the analytical function used and the actual profile; the flat-field correction (Figure \ref{fig:flat_and_blaze}, bottom panel) accounts for this mismatch.

For quality control, we check that the standard deviation of the flat for each order is less than 0.05. The flat (\FLAT) and blaze (\BLAZE) profiles are then saved to the calibration database (if the quality control criteria are met) for use throughout \APERO.

\subsection{Generating thermal calibration files} \label{subsec:thermal_calibration}

The nightly thermal recipe takes preprocessed \DARKDARKINT files or \DARKDARKTEL files (as many as given by the user or as many as occur on each of the nights being used via \Aprocessing). It combines the \DARKDARKINT or \DARKDARKTEL files into a single \DARKDARKINT or \DARKDARKTEL respectively (via a median combination of the images). These combined \DARKDARKINT or \DARKDARKTEL files are then extracted (see the extraction process in Section \ref{sec:extraction}).

The thermal background seen by \spirou in a science exposure is the sum of the black body contribution of the sky, the Cassegrain unit (at the temperature of the telescope), the calibration unit (for the reference channel), and the thermal emission of the hermetic feedthroughs that connect the fibers into the cryostat. A small contribution also arises from the Earth's atmosphere itself. This emissivity is proportional to one minus the telluric transmission at the corresponding wavelength and if left unaccounted for in the thermal model would lead to emission-like features in the thermal-corrected spectrum in the strongest absorption lines. From a series of sky-dark frames, we measured that the median additional emissivity from the saturated absorption line is at the $4\%$ level of the black body envelope. We account for the additional contribution by using a median sky absorption spectrum and adding a small contribution proportional to the excess emissivity due to the Earth's atmosphere in strong absorption lines. Note this contribution is only added for the \DARKDARKTEL files (as the \DARKDARKINT images do not see the sky). For this reason, we split generating the thermal calibration files into two steps: we generate the \DARKDARKINT thermal calibration files, then after a wavelength solution has been generated we generate the \DARKDARKTEL thermal calibration files (which require a nightly wavelength solution to add the contribution due to the emission-like features).

Considering that the telescope and front-end temperature change through the night, one needs to apply a thermal correction that is adjusted per frame (this is done as part of the extraction recipe in Section \ref{sec:extraction}). While the slope of the black body contribution changes very little over the $2.1-2.5$\,$\mu$m domain, within which the thermal background is significant, the amplitude of the contribution varies by a factor of $>2$ between nights (typically a factor 2 for every 8$^\circ$C) and needs to be adjusted for individual observations. While we have no external measurement of the thermal background, there are a number of completely saturated telluric water absorption features $2.4-2.5$\,$\mu$m that provide a measure of the total thermal emission seen by \spirou. These regions are used to scale the thermal background model such that they have a median flux of zero.

The thermal calibration files (\THERMALI and \THERMALT) are then saved to the calibration database for use throughout \APERO. The \THERMALI calibrations are used for correcting internal lamp spectra (i.e., other calibrations) and \THERMALT calibrations are used to correct all science spectra (Section \ref{sec:extraction}).

\subsection{Generating the nightly wavelength solution files} \label{subsec:wave_correction}

Considering that the wavelength solution is central in the anchoring of \PRV measurement and that the instrument will drift through time, one needs to obtain a wavelength solution as close as possible in time to the science exposures, ideally on a nightly basis. The nightly wavelength solution captures sub-$\mu$m level motions within the optical train and high-order changes in the focal plane that are not captured by the affine transform used to register frames as described in sections \ref{subsec:ref_shape} and \ref{subsec:night_shape}. The nightly wavelength solution recipe takes preprocessed \FPFP files and \HCHC ~files (as many as given by the user or as many as occur on each of the nights being used via \Aprocessing). It combines the \FPFP and \HCHC ~files into a single \FPFP and a single \HCHC ~file (via a median combination of the images). These combined \FPFP and \HCHC files are then extracted (see the extraction process in Section \ref{sec:extraction}). 

The rest of the process is similar to the reference wavelength solution (Section \ref{subsec:ref_wave}). The wavelength solution is determined as follows:
\begin{itemize}
\item Under the assumption that the reference wavelength solution is correct at the pixel level, identify HC lines (catalog wavelength) and FP peaks (FP order).
\item By combining the reference chromatic FP cavity length and position of FP peaks of known FP order,  fit a per-order wavelength solution.
\item Using that wavelength solution, measure the velocity offset in the position of HC lines ($\Delta v_{\rm HC}$) and derive an achromatic increment to be applied to the FP cavity
\item Scale the $0^{\rm th}$ order term of the $N^{\rm th}$ order cavity polynomial by $1-\frac{\Delta v_{\rm HC}}{c}$, where $c$ is the speed of light in the units of $\Delta v_{\rm HC}$.
\item Iterate the last two steps until $\Delta v_{\rm HC}$ is consistent with zero.
\end{itemize}

The main difference with the reference wavelength solution for fiber AB is that while we start the calculation of the wavelength solution with the cavity fit and wavelength solution from the reference wavelength solution calibration, we only allow for changes in the achromatic term. This is because the chromatic dependence of the cavity width is related to the coating of the FP etalon, and is therefore not expected to change rapidly. An achromatic shift, on the other hand, corresponds to a change in the cavity length of the FP, due in part to pressure or temperature variations, which may happen between nights. Meanwhile, for fibers A, B, and C we fit nothing and use the fiber AB wavelength cavity coefficients. The FP mask for quality control is also not re-generated. Therefore all cross-correlations between fibers AB and A, B, and C are done relative to the reference night wavelength solution (however we only check quality control on $AB-A$, $AB-B$, and $AB-C$). As with the reference wavelength solution recipe, a wavelength solution for each fiber, and the FP and HC lines founds during the process, are then saved to the calibration database for use throughout \APERO.

\begin{figure*}
    \centering
    \includegraphics[width=14cm]{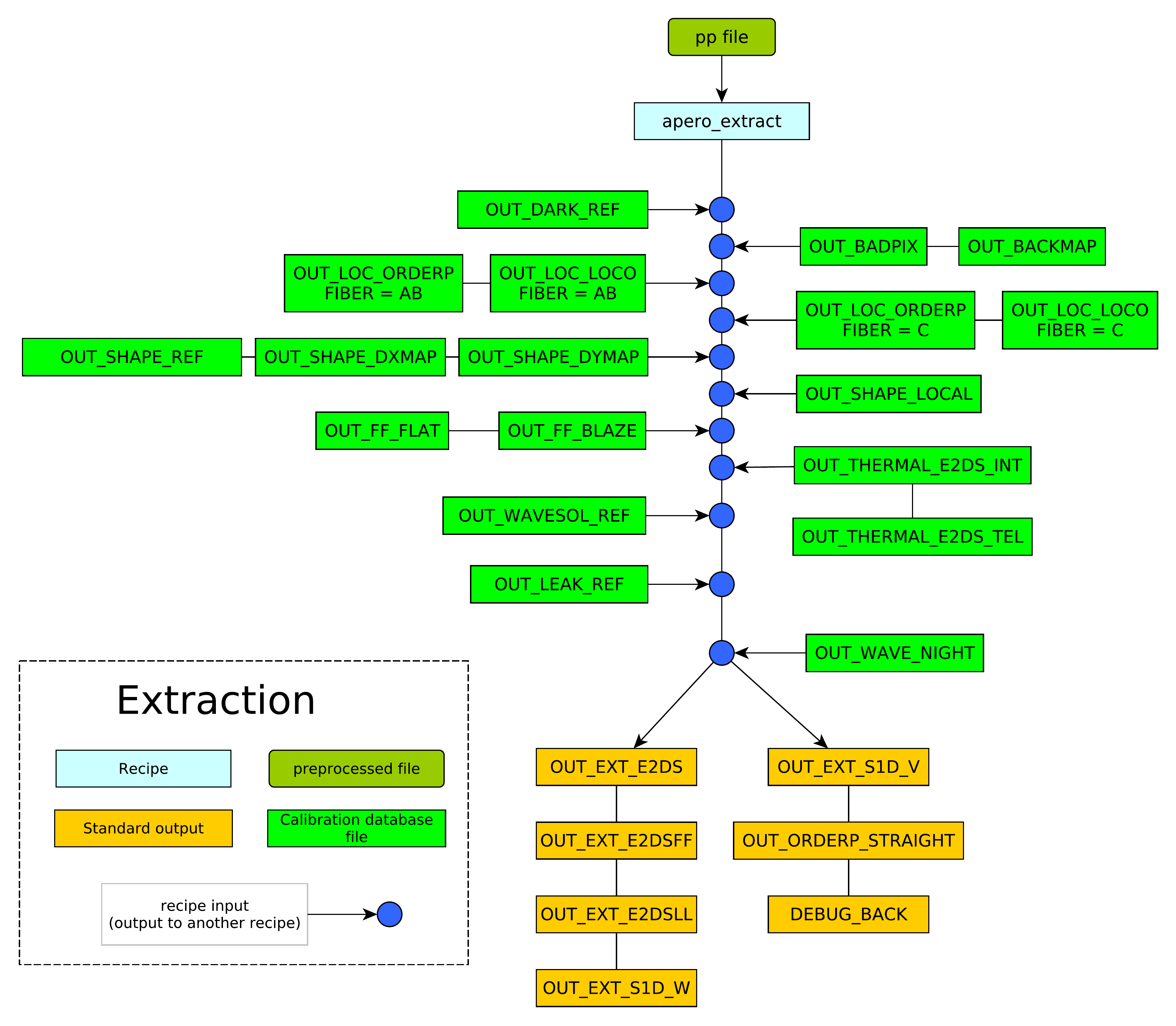}
    \caption{Extraction sequence: The input is a pre-processed file, that is then calibrated and extracted. The outputs are the 2D (per order) and 1D spectra for fibers AB, A, B, and C as well as some debugging outputs.}
    \label{fig:overview_extraction}
\end{figure*}

\section{Extraction}
\label{sec:extraction}

\begin{figure*}
    \centering
    \includegraphics[width=18cm]{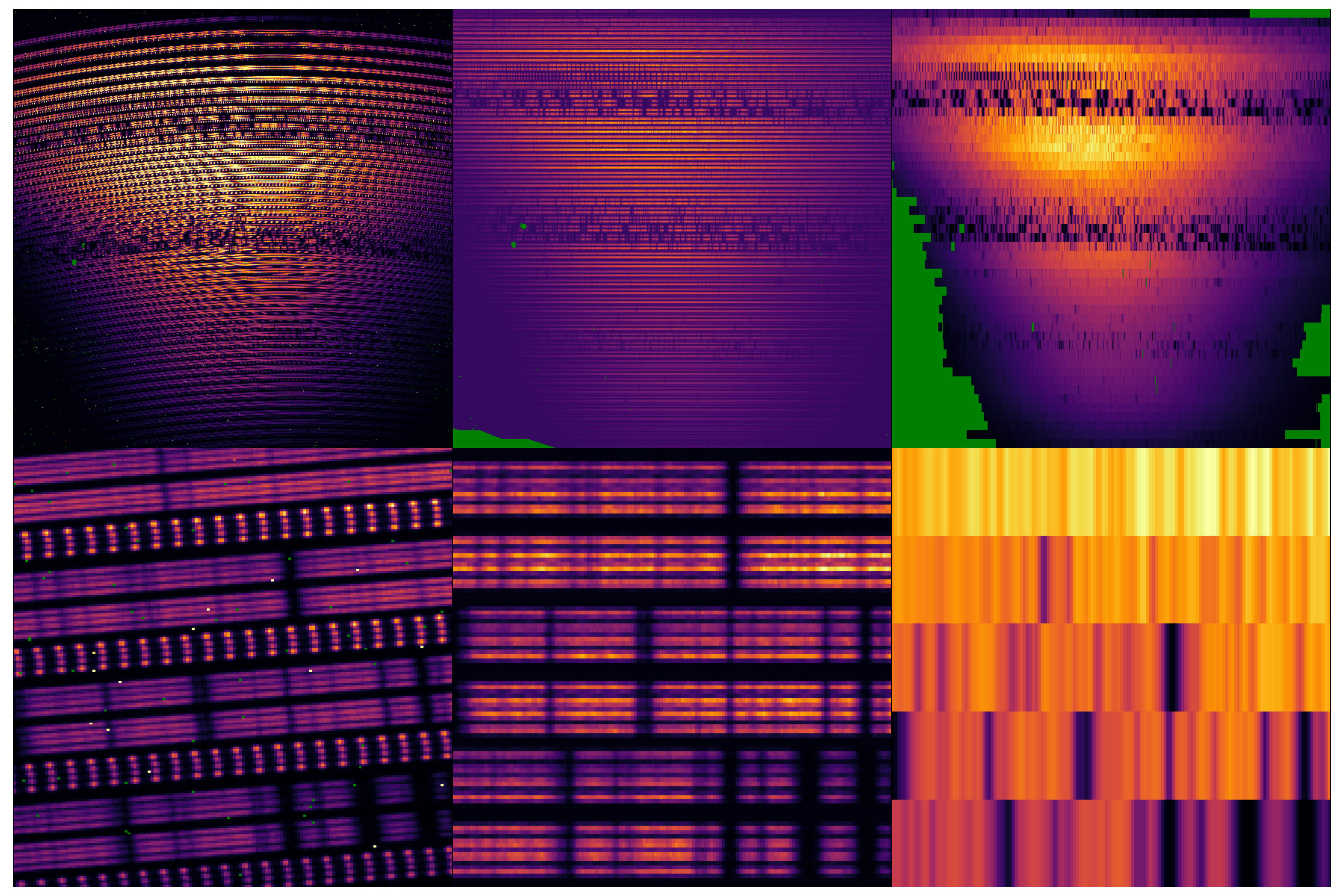}
    \caption{Top left: A full 2D image of a star after \Apreprocessing with a science target in the A and B fibers and an FP in the C fiber. Top middle: The full image is corrected for the slicer shape and tilt and straightened to remove the curvature due to the echelle orders - this is directly prior to summing in the extraction process (i.e. the optimal weights have been applied and is known as the \ETDSLL), here we see the AB extraction i.e before summing we see both the A and B fiber. Top right: The \ETDS -- the extracted flux summed along pixel columns across the A and B fiber (for an AB extraction) covering the 4 slices in A and 4 slices in B. Bottom left, middle and right, the corresponding zoom-ins of the same point in the extraction recipe as at the top (bottom left shows 4 orders for fibers A, B and C uncorrected, the bottom middle shows 3 orders for fibers A and B corrected, and bottom right shows 5 orders of the combined A and B flux). The tilt and curvature due to the echelle orders can be clearly seen on the left and are clearly corrected in the middle. Hot pixels from the full 2D image of the star after \Apreprocessing to the middle panels were removed during the calibration process (see Appendix \ref{section:appendix_standard_image_calibration}). In all panels, \NAN values are shown in green.}
    \label{fig:e2ds_grid}
\end{figure*}

The extraction recipe takes any preprocessed file (as many as given by the user but in general just one single file). The files are combined (if requested) and are calibrated using our standard image calibration technique (see Appendix \ref{section:appendix_standard_image_calibration}). Once calibrated, the correct (closest in time) order profile (\ORDERP), positions of the orders (\LOCO), \SHAPELOCAL, shape reference ($x$ and $y$ maps), and wavelength solution are loaded for each fiber (AB, A, B, and C). The order profiles and input image are transformed to the reference FP grid using the affine transformation mentioned in Section \ref{subsec:ref_shape}, Section \ref{subsec:night_shape} and Appendix \ref{section:affine_transformation}, and using the shape $x$ and $y$ maps the image is corrected for the slicer geometry, the tilt and the bending due to the echelle orders.

The extraction recipe then extracts the flux (using optimal extraction, Section \ref{subsec:optimal_extraction}), calculates the barycentric correction (Section \ref{subsec:berv_corr}), corrects contamination from the reference fiber (if an FP is present in the reference fiber, Section \ref{subsec:leak_corr}), corrects for the flat (Section \ref{subsec:flat_corr}), corrects for the thermal contribution (Section \ref{subsec:thermal_corr}) and generates the 1D spectrum (Section \ref{subsec:s1d}). An overview of the extraction procedure can be seen in Figure \ref{fig:overview_extraction}.

\vspace{1cm}

\subsection{Optimal extraction} \label{subsec:optimal_extraction}

Once the image and the order profile (from localization) have been corrected for the slicer geometry and curvature of the echelle orders we extract out the combined flux in the science channels (fibers A and B) to create a fiber AB, as well as extracting out the flux in A and B (for polarization work) and C separately (for the reference fiber calibrations). As the orders are already straightened we use just the localization coefficient value at the center of the image to extract vertically along each order. We then divide the image by the order profile to provide a weighting across the order (i.e., an optimal extraction, \citealt{Horne_optimal_1986}). This produces the image in Figure \ref{fig:e2ds_grid} (the \ETDSLL, top middle and bottom middle panels) where we show the image for fiber AB (A+B). The final step of the optimal extraction is to sum vertically across the columns accounting for cosmic rays by using a sigma clip $|flux|>10\sigma$ away from the median value for that column. This creates our \ETDS (extracted 2D spectrum) and for \spirou, this leads to images with 49 orders and 4088 pixels along the orders.

\vspace{1cm}

\subsection{BERV correction} \label{subsec:berv_corr}

Ideally, any stellar spectrum observed would be measured from a point stationary with respect to the barycenter of the Solar System \citep{wright2014}. However, ground-based observations are subject to: the orbit of the Earth, the rotation of the Earth, precession and other Earth motions, and to a lesser extent gravitation time dilation, leap-second offsets, and factors affecting the star itself (i.e., parallax, proper motions, etc). We use the term  BERV (Barycentric Earth Radial Velocity) hereinafter to collect all these terms into a single measurement which can be used to correct a specific spectrum at a specific point in time. We calculate the BERV using the \textsc{barycorrpy} package \citep{barycorrpy2020_1, barycorrpy2020_2}, which uses the astrometric parameters fed in at the preprocessing level (Section \ref{subsec:obj_res}). The calculation from \textsc{barycorrpy} includes the estimate for the BERV itself and the corrected or barycentric Julian Date (BJD) at the mid-exposure time. \textsc{barycorrpy} has a precision better than the \cmps level. We also estimate the maximum BERV value for this object across the year. If for any reason the BERV calculation fails\footnote{The \textsc{barycorrpy} algorithm requires an internet connection to update leap seconds and requires downloading ephemeris data. There are circumstances where either at the user end or the server end these services are unavailable and thus \textsc{barycorrpy} fails.} with \textsc{barycorrpy} we calculate an estimate of the BERV (precise to $\sim$10 \mps, modified from \lstinline{PyAstronomy.pyasl.baryvel}; a python implementation of \textsc{helcorr}; \citealt{pyasl_helcorr_2002}) and flag that an estimated BERV correction was calculated. This estimated BERV is not precise enough for \PRV work but is sufficient to allow for acceptable telluric correction.

\subsection{Leak Correction} \label{subsec:leak_corr}

For scientific observations, the reference fiber either has a DARK or an FP illuminating the pixels in this fiber. For \PRV an FP allows a simultaneous RV measurement of an FP alongside the measurement of the stellar RV; this allows precise tracking of the instrumental drift when the simultaneous FP is compared to the \FPFP from the nightly wavelength solution calibration, see Section \ref{sec:rv}. However, as mentioned in Section \ref{subsec:ref_leak}, light from the FP has been shown to slightly contaminate the science fibers and thus we provide a correction for such calibration.

During the reference sequence (Section \ref{sec:ref_calibrations}) many \DARKFP are combined (and extracted) to form a model of the light seen in the science fibers when no light (other than the contribution from the DARK) was present as well as an extracted reference fiber measurement of the FP flux that caused said contamination in the science fibers. Using these models, the contamination measured in the science channels of the reference leak recipe is then scaled to the flux of the simultaneous FP of the observation (using the extracted flux from this scientific observation we are trying to correct). Then, this model is subtracted from the original science observation for each of the science fibers (AB or A or B), order-by-order, as in Equation \ref{equ:leak_corr}.

\begin{equation}
     \begin{array}{cc}
        ratio_{i} = \frac{\Sigma(L[C]_{i}S[C]_{i})}{\Sigma(S[C]_{i}^2)} \\
        \\ 
        scale_{i} = \frac{L[AB,A,B]_{i}}{ratio_{i}} \\
        \\
        S[AB,A,B]_{i,corr} = S[AB,A,B]_{i} - scale_{i} \\

    \end{array}
    \label{equ:leak_corr}
\end{equation}

\noindent where $L[C]$ is the model of the FP from the leak reference recipe, $S[C]$ is the 2D extracted spectrum in the reference fiber (fiber C), $L[AB,A,B]$ is the model of the contamination from the FP from the leak reference recipe in the science fibers (either AB or A or B), $S[AB,A,B]$ is the 2D extracted flux in the science fibers (either AB or A or B), $S[AB,A,B]_{corr}$ denotes the leak-corrected 2D extracted spectrum in the science fibers (either AB or A or B) and $i$ denotes that this is done order-by-order.

\subsection{Flat Correction} \label{subsec:flat_corr}

Whether or not the reference fiber was corrected for contamination the next step is to correct for the flat. From the nightly calibrations (Section \ref{sec:night_calibrations}) we expect a \FLAT file for each night of observations (and one for each fiber: AB, A, B and C) to be present in the calibration database. To correct for the flat we simply divide the extracted spectrum (one for each fiber) by the corresponding flat spectrum as in Equation \ref{equ:flat_corr}.

\begin{equation}
    \begin{array}{cc}
        S[AB,A,B,C]_{corr} = S[AB,A,B,C] / FLAT[AB,A,B,C] \\
    \end{array}
    \label{equ:flat_corr}
\end{equation}

\noindent where $S[AB,A,B]_{corr}$ denotes the flat corrected 2D extracted spectrum,  $S[AB,A,B,C]$ denotes the 2D extracted spectrum prior to correction and $FLAT[AB,A,B,C]$ denotes the flat applied from the calibration database created using the method described in Section \ref{subsec:flat_blaze}. 

As both the extracted spectrum and the \FLAT are 2D and equivalent in shape there is no need to do this per order. Note that it is highly recommended to use a flat from the same night as the observation, and where possible this is the default option, however, nights near in time to the observation are used when no closer calibration is available. The flat fielded extracted spectrum is denoted by \ETDSFF (to distinguish it from the non-flat fielded extracted spectrum \ETDS). Note also that no 2D spectrum is ever saved with the blaze correction applied, however, we provide blaze spectra that can be used to correct any spectrum for the blaze. As with the flat, using the blaze for that night is highly recommended, and where possible is the default blaze used by \APERO when a blaze correction is required.

\subsection{Thermal Correction} \label{subsec:thermal_corr}

The reference dark, applied during the standard image calibration phase (see Appendix \ref{section:appendix_standard_image_calibration}), removes the high-frequency components of the dark; however, the thermal contribution still remains (and varies on a night-by-night basis). For this reason, we use nightly extracted \DARKDARK files to model the thermal contribution present in an observation during the night. The thermal correction model comes in two flavors, one for science observations where we assume there is some sort of continuum to the spectrum and telluric contamination as well as a small contribution arising from the Earth's atmosphere itself, described in Section \ref{subsec:thermal_corr_a}, and one for HC or FP extractions where these assumptions are not true, as detailed in Section \ref{subsec:thermal_corr_b}.

\subsubsection{Thermal Correction of a science observation} \label{subsec:thermal_corr_a}

\begin{figure*}
    \centering
    \includegraphics[width=16cm]{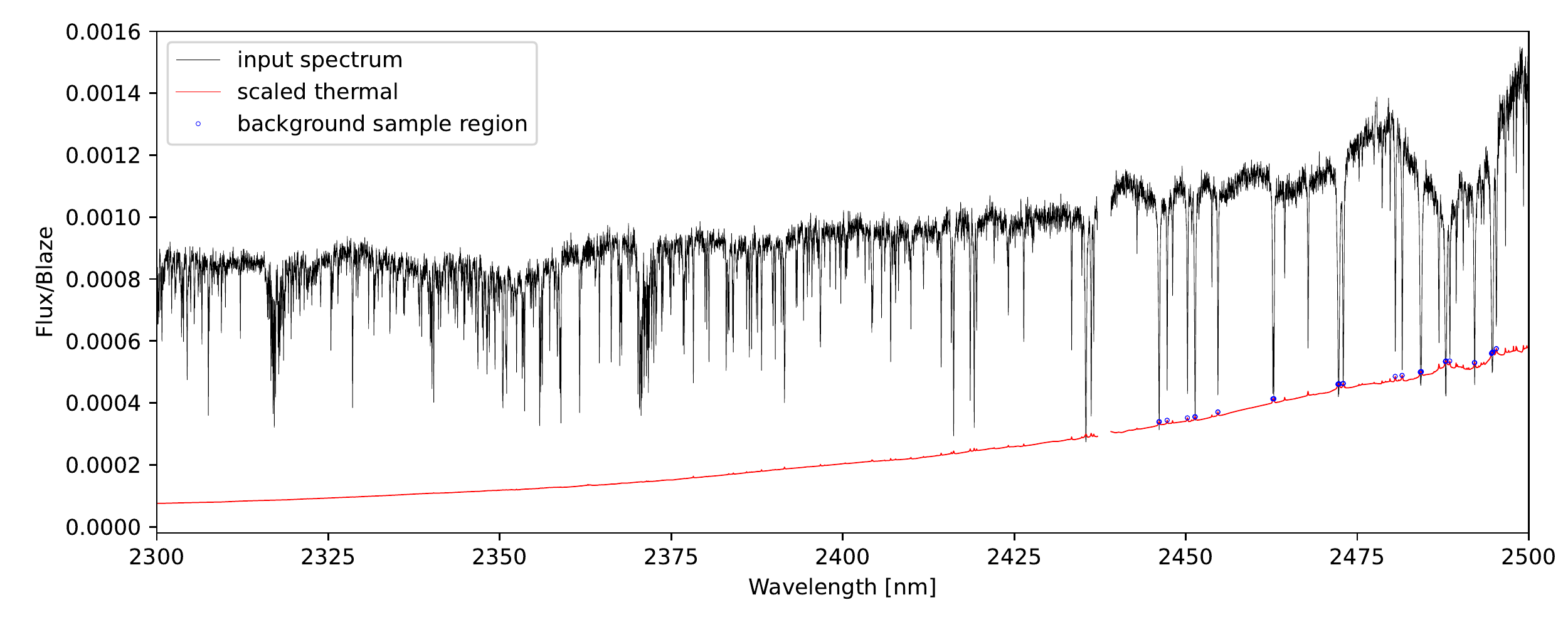}
    \caption{Thermal correction of a science observation. Thermal correction is scaled to a science spectrum via mapping flux lower than 0.01 in an inputted transmission spectrum from TAPAS (background sample region, blue dots on the plot). In black is the extracted last order of Gl699 and in red is the scaled thermal background taken from the \DARKDARKTEL thermal calibration file. The gaps in the spectrum are due to the blaze cut off at the edge of orders and no overlap in this domain. Some deep lines between 2.48 and 2.50\micron ~will be over-corrected (they lie below the scaled thermal contribution). This arises from the finite resolution of the spectrographs that tends to make narrow absorption features shallower than wider features, such as doublets. This problem will be addressed in the next version of APERO.}
    \label{fig:thermal_corr}
\end{figure*}

In the case where we have a scientific observation, a \DARKDARKTEL (where the calibration fiber sees the cold source and the science fibers see the mirror covers) is used. The extracted \DARKDARKTEL (from \ref{subsec:thermal_calibration}) is then median filtered with a width of 101 pixels (on a per-order basis). This width was chosen to be big enough to capture large-scale structures in the dark and not be significantly affected by readout noise. A fit is then made to the red most orders ($>2450$\,nm) using only flux lower than 0.01 from a transmission spectrum from the Transmissions of the AtmosPhere for AStromomical data tool (TAPAS; \citealt{TAPAS2014}) -- i.e., a domain where transmission is basically zero. We assume that we can safely use any flux with a transmission of order zero to scale the thermal background to this zero transmission value (see Equation \ref{equ:thermal_corr_a} and Figure \ref{fig:thermal_corr}).

\begin{equation}
    \begin{array}{cc}
    
        mask = \left\{ \begin{array}{cl}
        1: & TAPAS < 0.01  \\
        0: & \text{otherwise} \\
        \end{array} \right. \\
        \\
        ratio = median\left( \frac{TT[AB,A,B,C]\times mask}{S[AB,A,B,C] \times mask} \right) \\
        \\
        S[AB,A,B,C]_{corr} = S[AB,A,B,C] - \frac{TT[AB,A,B,C]}{ratio} \\
    \end{array}
    \label{equ:thermal_corr_a}
\end{equation}
\vspace{0.5cm}

\noindent where $TAPAS$ is the TAPAS spectrum, $TT[AB,A,B,C]$ is a nightly extracted \DARKDARKTEL spectrum, $S[AB,A,B,C]$ denotes the 2D extracted spectrum prior to correction and $S[AB,A,B]_{corr}$ denotes the thermally corrected 2D extracted spectrum.

\subsubsection{Thermal Correction of a calibration} \label{subsec:thermal_corr_b}

In the case where we have an HC or an FP observation, a \DARKDARKINT (where all three fibers see only the cold source, not the sky nor the mirror covers) is used. The extracted \DARKDARKINT is then median filtered (again with a width of 101 pixels on a per-order basis) and a fit is made using an envelope to measure the thermal background in the reddest orders ($>2450$\, nm). The envelope is constructed by using the flux below the 10$^{\rm th}$ percentile (i.e., not in the HC or FP peaks). This is then converted into a ratio and scaled to the observation we are correcting (see Equation \ref{equ:thermal_corr_b}).

\begin{equation}
    \begin{array}{cc}
        ratio = median\left( \frac{TI[AB,A,B,C]}{P_{10}(TI[AB,A,B,C])} \right) \\
        \\
        S[AB,A,B,C]_{corr} = S[AB,A,B,C] - \frac{TI[AB,A,B,C]}{ratio} \\
    \end{array}
    \label{equ:thermal_corr_b}
\end{equation}
\vspace{0.5cm}

\noindent where $P_{10}$ is the 10$^{th}$ percentile value, $TI[AB,A,B,C]$ is a nightly extracted \DARKDARKINT spectrum (median filtered with a width of 101 pixels), $S[AB,A,B,C]$ denotes the 2D extracted spectrum prior to correction and $S[AB,A,B]_{corr}$ denotes the thermally corrected 2D extracted spectrum.

\begin{figure*}
    \centering
    \includegraphics[width=16cm]{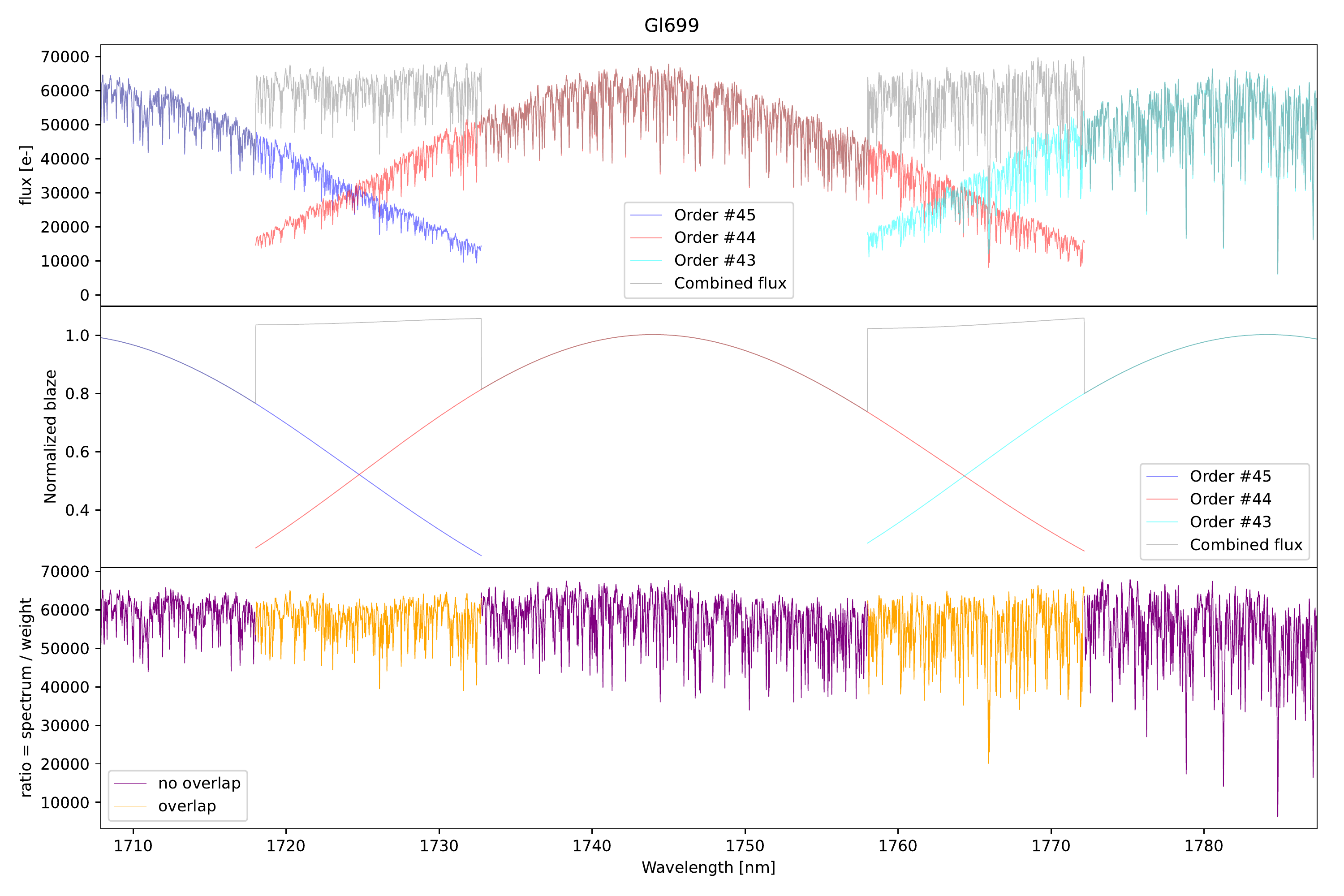}
    \caption{Construction of the \SONED data product. In constructing the \SONED from the \ETDSFF frame, one consecutively splines individual orders from the science and files onto the destination grid. The top panel shows a sample region within the $H$ band with 3 consecutive orders and the sum of all fluxes. The middle panel shows the same operation performed for the blaze. The bottom panel shows the ratio of the sum of the science flux and blaze, providing seamless stitching of orders.}
    \label{fig:s1d_demo}
\end{figure*}

\subsection{\SONED Generation} \label{subsec:s1d}

The \ETDS and \ETDSFF formats are not necessarily the most convenient for science analysis, having duplicated wavelength coverage at order overlap and slightly varying velocity sampling with each order and between orders. We therefore transform the \ETDSFF file into the \SONED format. The \SONED is sampled on a constant grid for all objects. We have two differing \SONED formats, one with a uniform step in wavelength (0.05\,nm/pixel) and one with a constant step in velocity (1 \kmps/pixel), both being sampled between 965\,nm and 2500\,nm. Numerically, to construct the \SONED, we use as an input the \ETDSFF file prior to blaze correction and the blaze file (Section \ref{subsec:flat_blaze}) as inputs. We create two \SONED vectors, one corresponding to the total flux and one corresponding to the total blaze on the destination wavelength grid. We use a 5$^{th}$ order polynomial spline to project the flux of a given order onto the flux grid and perform the same operation with the blaze onto the weight vector. We do not consider the blaze below 20\% of the peak blaze value and values on the destination wavelength grids that are out of the order's range are set to zero. We loop through orders and sum the contribution of each order onto the respective destination grids for the \ETDSFF science flux and blaze. The \SONED generation is summarized in Figure \ref{fig:s1d_demo}. Note that the \SONED generation only depends on the blaze calibration. As such any spectrum (regardless of emission lines, low flux, or strong bands) can be converted to \SONED format and we generate \SONED for \HCHC ~and \FPFP as well as science targets.

\vspace{3cm}

\section{Telluric correction}
\label{sec:telluric}

The detailed performances of \spirou's telluric correction will be presented in a forthcoming publication (Artigau in prep); we summarize the main steps here. Most telluric-correction schemes are either model-based (e.g., Molecfit, \citealt{Smette2015} or \citealt{Allart2022}) or purely empirical \citep{Artigau2014, Wobble2019}. Here we adopt a hybrid method with both a simplistic model-based correction and a further empirical correction of residuals. 

The first part of the telluric correction uses the average TAPAS absorption spectrum  \citep{TAPAS2014} for Maunakea for 6 chemical species, H$_2$O, O$_2$, CO$_2$, N$_2$, CH$_4$ and NH$_3$; we combine these into two absorption spectra, one for water and one for all other components, dubbed the `dry' spectrum, at full resolution. We correct the absorption of water and dry components by adjusting the optical depth of each telluric spectra by nulling a cross-correlation function (CCF) of unsaturated absorption lines of the corresponding absorber. To account for the overlapping stellar and telluric features, prior to the CCF, the science spectrum is divided by a stellar template of the object of interest. The `dry' optical depth has been found to be remarkably consistent with the geometric airmass, at the 0.03 airmass level RMS. The `water' optical depth cannot be compared to a measurement of precipitable water vapor comparable in accuracy to the geometric airmass but is expected to be similar in accuracy as it is measured through the same method and includes a larger number of lines in its CCF.

\begin{figure*}
    \centering
    \includegraphics[width=18cm]{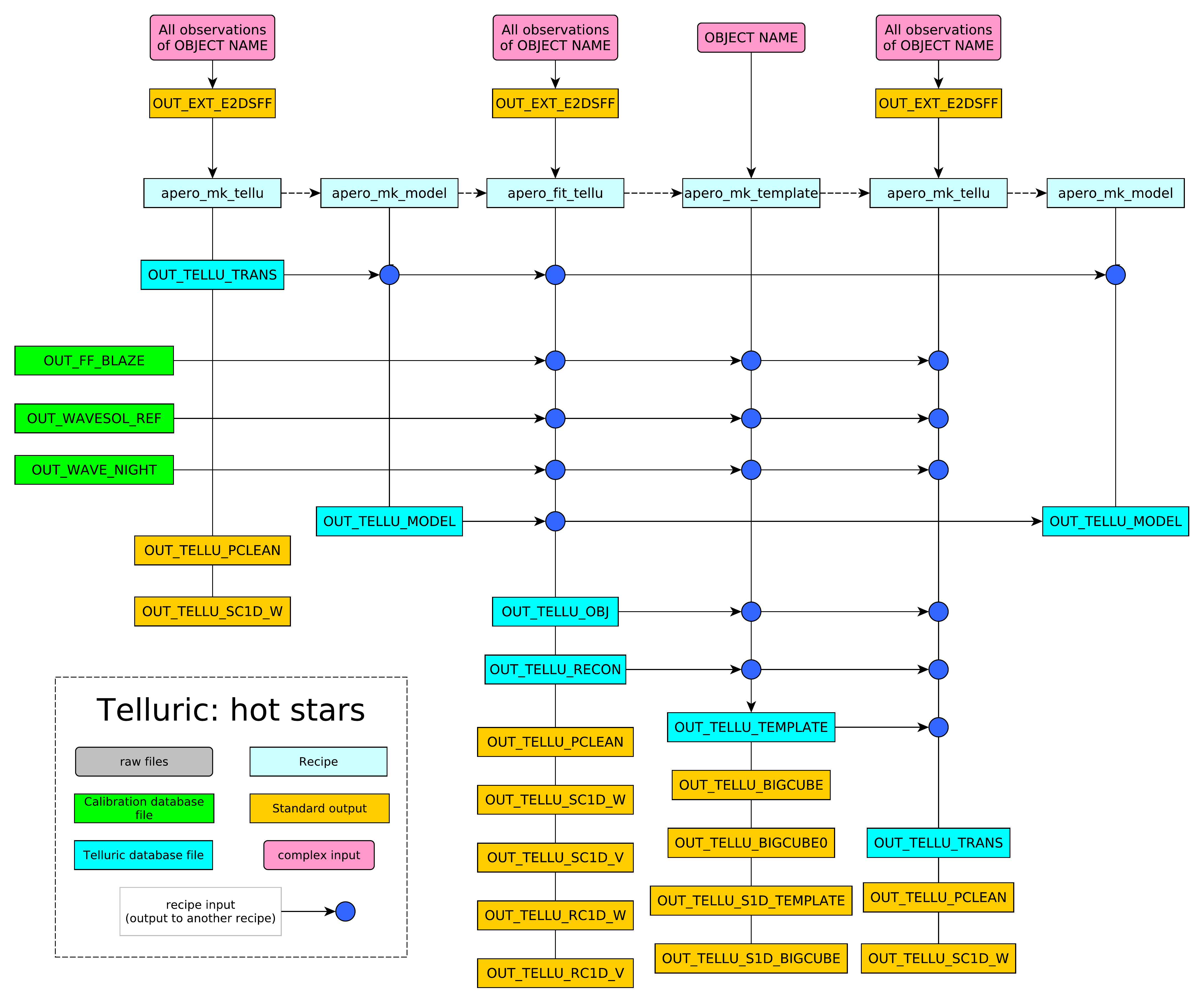}
    \caption{Telluric hot star sequence}
    \label{fig:overview_tellu_hotstar}
\end{figure*}

\begin{figure*}
    \centering
    \includegraphics[width=16cm]{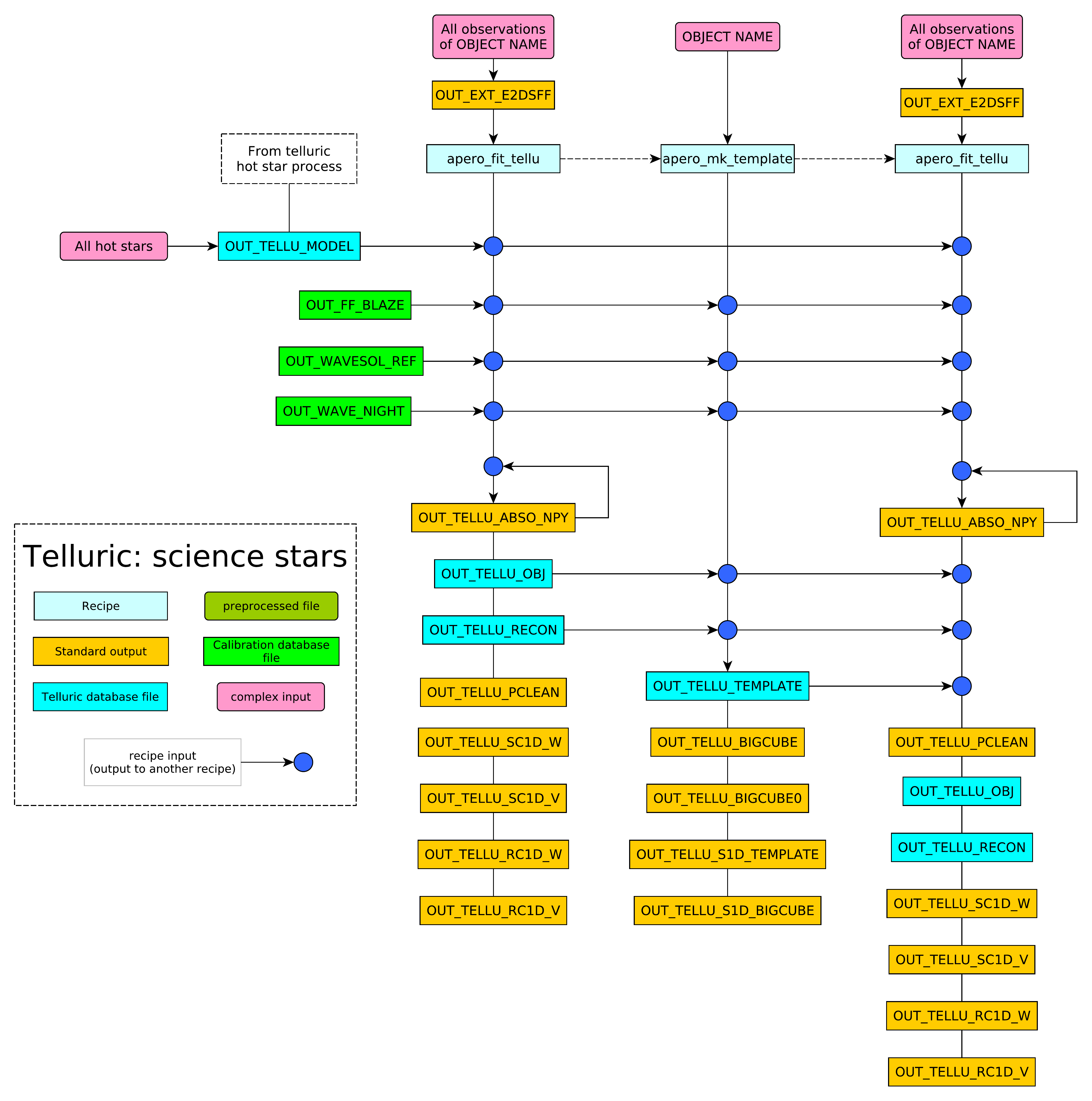}
    \caption{Telluric science sequence}
    \label{fig:overview_tellu_science}
\end{figure*}

In \APERO the telluric correction is an automated iterative process and an overview schematic of the process can be seen in figures \ref{fig:overview_tellu_hotstar} and \ref{fig:overview_tellu_science}. 

We need only a few specific hot stars (currently around 30) chosen with the following criteria:

\begin{itemize}
    \item fast rotators with a published $v\sin i>200$ \kmps (to easily identify telluric lines).
    \item B or A spectral type, avoiding Be stars that have Balmer-series emission lines.
    \item bright to minimize overheads, with an $H$-band brightness of $3-6$\, mag.
    \item observable from Maunakea.
    \item for stars close in sky position ($<10^{\circ}$), the brighter one is preferred.
\end{itemize}

The database of hot stars is constructed as follows:
\begin{enumerate}
    \item model residual transmission of hot stars (\Amktellu, Section \ref{subsec:make_tellu}).
    \item model the absorption contributions from water and from `dry` components (essentially all contributions that aren't water, \Amkmodel, Section \ref{subsec:tellu_model}).
    \item correct the telluric absorption in these hot stars (\Aftellu, Section \ref{subsec:fit_tellu}) using the models of residual transmission.
    \item construct a template of these hot stars (using all available observations that pass quality control, \Amtemp, Section \ref{subsec:tellu_templates}).
    \item re-model the residual transmission (again using \Amktellu, Section \ref{subsec:make_tellu}, but this time with the addition of a template).
    \item re-make the model of the absorption contributions from water and from dry components (using \Amkmodel, \ref{subsec:tellu_model}).
\end{enumerate}

Once this database of hot stars has been constructed science observations can be corrected. This is also an automated iterative process, and is done in the following order:

\begin{enumerate}
    \item correct the telluric absorption for each science observation (\Aftellu, Section \ref{subsec:fit_tellu}) using the models of residual transmission from the hot stars.
    \item construct a template for each science object (using all available observations that pass quality control, \Amtemp, Section \ref{subsec:tellu_templates}).
    \item re-correct the telluric absorption for each science observation (again using \Aftellu, Section \ref{subsec:fit_tellu}, but this time with the addition of a template for that object).
\end{enumerate}

\begin{figure*}
    \centering
    \includegraphics[width=18cm]{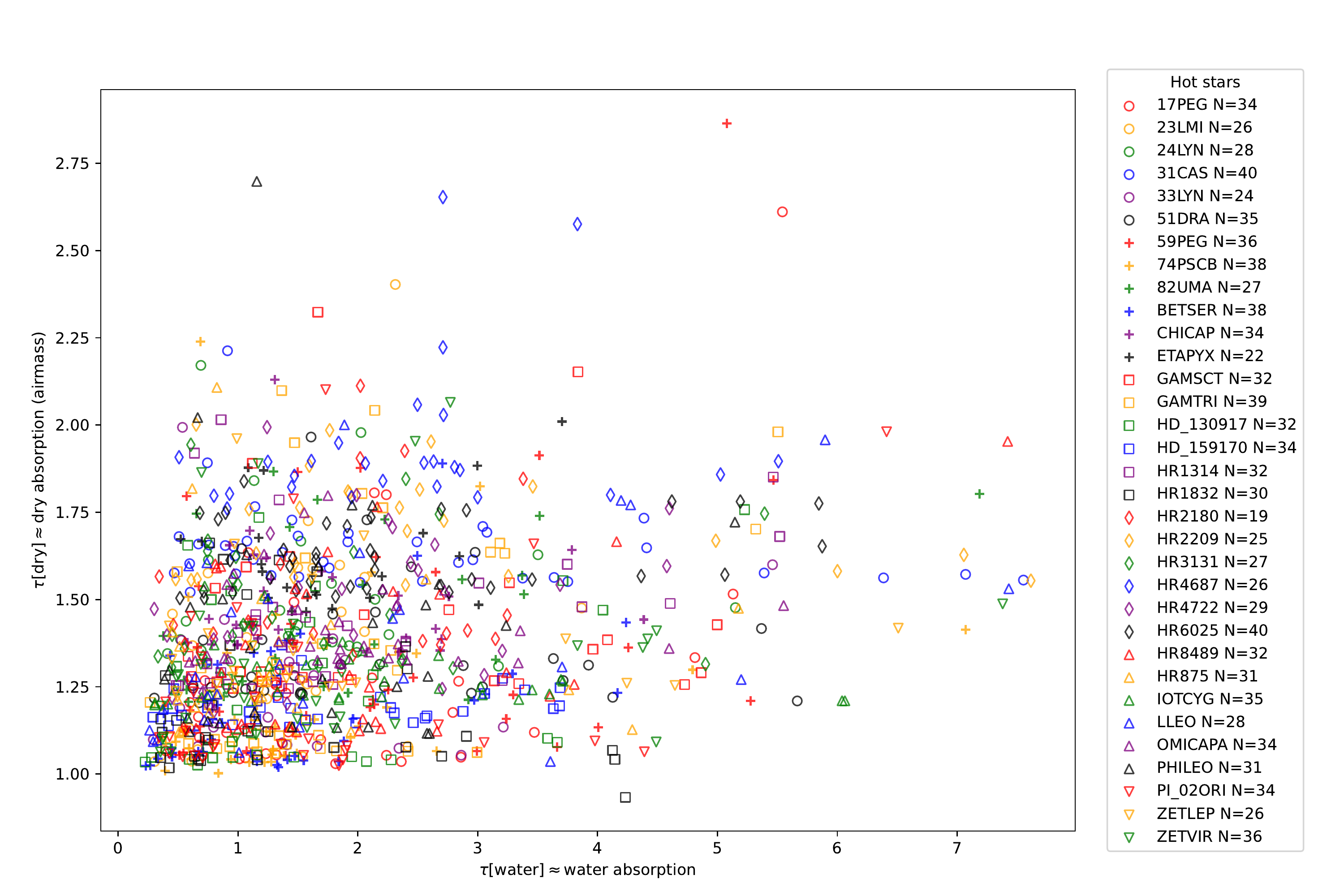}
    \caption{Hot stars (used for telluric correction) coverage in the water and dry components of absorption (fractional throughput) used to construct a library of telluric residuals after telluric pre-cleaning.}
    \label{fig:tellu_hotstar_coverage}
\end{figure*}

This process is beneficial as it does not require hot star observations every night but relies on a well organized observing strategy such that hot star observations cover a wide range of airmass and water vapor conditions. The \spirou observation coverage in airmass and water vapor can be seen in Figure \ref{fig:tellu_hotstar_coverage}. The requirement that these hot stars are bright also means that no more than a few minutes are required to get a sufficient SNR, without taking away too much valuable telescope time for science observations.

One complication of this method is more data is always available to add to these models and templates of objects. In ideal circumstances, one would regenerate all hot star and science target templates with all data for those objects and subsequently. Then one would redo the later steps of the hot star and the science observation corrections (steps 4 to 6 and 2 to 3 respectively). However, once a sufficient number of observations of a specific object (either a hot star or science target) is reached, we find that it is not necessary to re-make templates for normal processing. However, whenever it is possible we do re-reduce all data including these telluric steps to provide the most optimal template possible. The exact number of telluric stars needed for `sufficient' correction has been the subject of much debate without a clear-cut answer. One way to solve this is to set a requirement that the telluric correction is a minor contribution to the SNR of any given observation. Considering that observations of the brightest science targets have an SNR of 300, if we want telluric stars to have a $<5$\% contribution to this value, then the effective photon noise of the telluric stars should be $>1000$. Considering that typical hot star observations have an SNR of $\sim$200, this would imply that one needs at least 25 observations. The full calculation is somewhat more complex as these uncertainties propagate in a linear model and the effective noise will depend on where any given observation falls within the $\tau[{\rm dry}]$ versus  $\tau[{\rm water}]$ diagram. As an order of magnitude, we recommend having $>100$ hot star observations spanning the range of expected observing conditions. There are also criteria on BERV coverage as to whether a template is worth constructing in the first place (see Section \ref{subsec:tellu_templates}).

\subsection{Residual transmission of hot stars (\Amktellu)} \label{subsec:make_tellu}

The residual transmission recipe (\Amktellu) takes a single hot star observation (an extracted, flat-fielded 2D spectrum). The first step is a pre-cleaning correction which essentially removes the bulk of the telluric absorption, producing a corrected 2D spectrum as well as an absorption spectrum, sky model, and an estimate of the water and dry components of the absorption (Artigau in prep). The pre-cleaning uses a stellar template, if available, to better measure the water and dry components. The corrected 2D spectrum is then normalized by the 95$^{th}$ percentile of the blaze per order and the residual transmission map is created by using a low-pass filter (per order) on the hot star (and dividing by a template if present).

We make sure the pre-cleaning was successful (i.e., the water component exponent is between 0.1 and 15 and the dry component exponent is between 0.8 and 3.0) and check that the SNR for each order is above a $100$; subsequently, the hot star residual transmission maps are added to the telluric database.

\subsection{Water and dry component models (\Amkmodel)} \label{subsec:tellu_model}

During the pre-cleaning process (Artigau in prep.) for the hot stars (done as part of \Amktellu, Section \ref{subsec:make_tellu}) we calculate the water and dry exponents of absorption. Once we have observed a sufficiently large library of telluric hot stars (see Figure \ref{fig:tellu_hotstar_coverage}), typically a few tens under varying airmass and water column conditions, we take all of the residual transmission maps that passed quality control and calculate a linear minimization of the parameters. The linear minimization is done per pixel per order, across all transmission maps (removing outliers with a sigma clipping approach) against a three-vector sample (the bias level of the residual, the water absorption exponent, and the dry absorption exponent). The output is three vectors each the same size as the input 2D spectrum (49$\times4088$), one for each of the three vector samples. These are used in every \Aftellu recipe run (Section \ref{subsec:fit_tellu}) to correct the telluric residuals after telluric cleaning. The three vectors are saved and added to the telluric database.

\subsection{Correcting telluric absorption (\Aftellu)} \label{subsec:fit_tellu}

\begin{figure*}
    \centering
    \includegraphics[width=18cm]{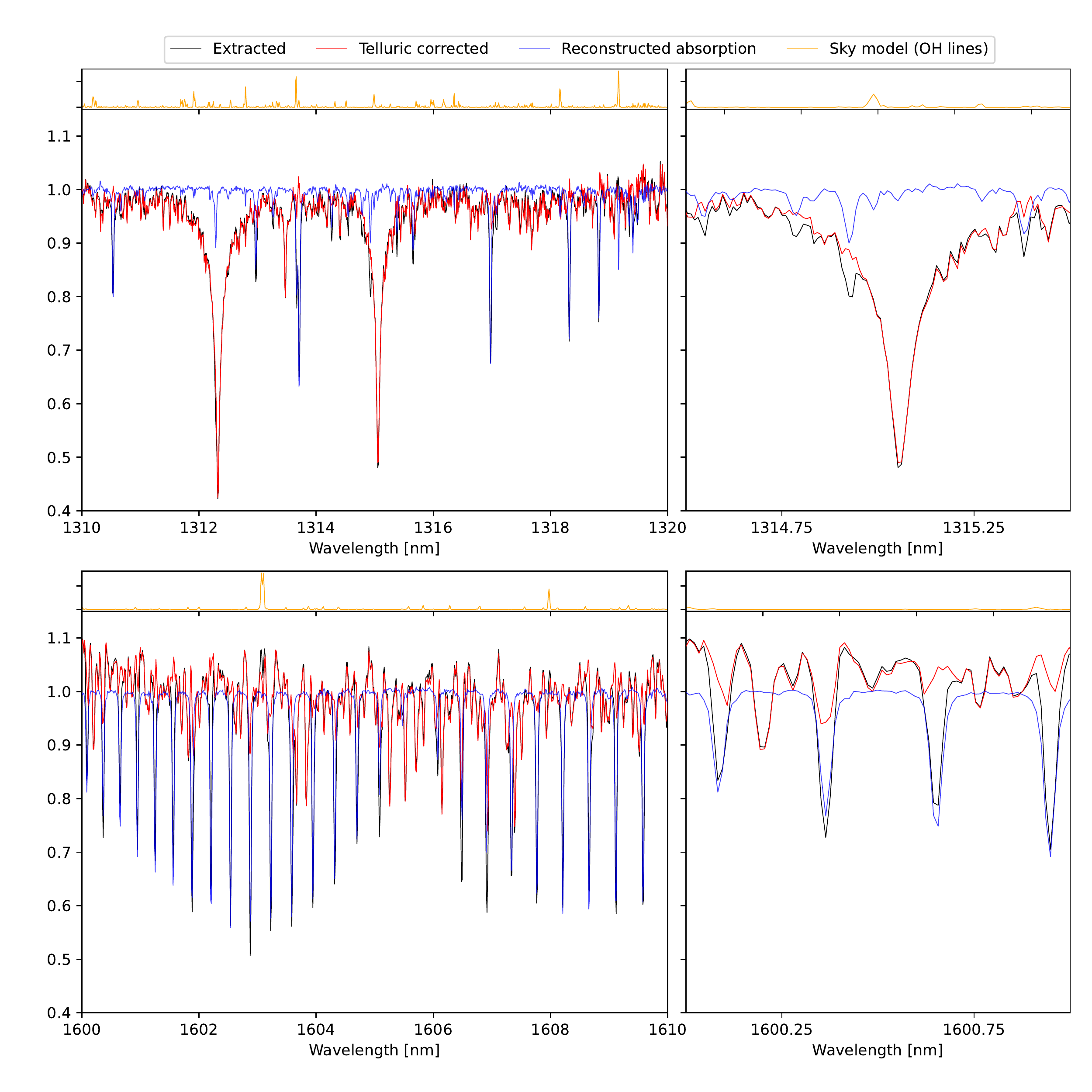}
    \caption{Example telluric correction. The input data (\ETDSFF) is in black, the calculated reconstructed absorption is in blue, and the final telluric corrected spectrum is in red. Each wavelength domain shown here has a subplot showing the modeled sky emission from OH lines (also corrected for) with arbitrary units.}
    \label{fig:example_tellu_correction}
\end{figure*}

All hot stars and science targets are corrected for telluric absorption. The first step, as with \Amktellu, is the pre-cleaning correction. Then, we correct the residuals of the pre-cleaning at any given wavelength by fitting a linear combination of water and dry components. We assume that any given absorption line in the TAPAS absorption spectrum has a strength that is over or underestimated relative to reality, the residuals after correction will scale, as a first order, with the absorption of the chemical species. The same is true with line profiles; if the wings of a line are over or underestimated, the residuals will scale with absorption. An example corrected telluric spectrum can be seen in Figure \ref{fig:example_tellu_correction}. We correct the telluric absorption on the combined AB extracted spectrum and subsequently use the same reconstructed absorption (for fiber AB) to correct the extracted spectra for fibers A and B individually.

\vspace{2cm}
\subsection{Template generation (\Amtemp)} \label{subsec:tellu_templates}

Templates for each astrophysical object are created simply by shifting all observations (in BERV) from their nightly wavelength solution to the reference wavelength solution. This effectively creates a cube\footnote{In practice some astrophysical objects have thousands of observations so a median is done in parts (splitting into bins in time), combining the median cubes together to produce one final cube, to reduce computational requirements.} of observations for specific astrophysical objects which are then normalized (per observation) by the median for each order. We pass a low-pass filter over this cube and then the cube is reduced to a single 2D (extracted and telluric-corrected) spectrum by taking a median in the time dimension (across observations). The same process is done for the 1D spectrum. The 2D templates are copied to the telluric database for use in the rest of the telluric cleaning process (the second iterations of \Amktellu and \Aftellu), except if the BERV change throughout all epochs is below 8 \kmps. The 1D spectrum is saved as a useful output of \APERO.

\section{RV Analysis}
\label{sec:rv}

One major requirement for \spirou is precision radial velocity measurements. \APERO provides a per-observation estimate of radial velocity using a standard CCF method, cross-correlating an observation with an appropriate set of lines (a mask). This radial velocity estimate is precise enough for quality checks (i.e., at the several tens of \mps level) but not at the precision required for pRV. For pRV in the near infrared, access to multiple observations of a single object is required. As of the publishing of this paper, the newly developed line-by-line (LBL) recipes \citep{LBL2022} are self contained and publicly available\footnote{LBL is available on GitHub and documented here: \url{https://lbl.exoplanets.ca/} \label{footnote:lbl_docs}}. However, currently, the LBL recipes cannot be called inside \APERO. Thus LBL RVs are not automatically produced as part of the fully automated \APERO framework. Adding the call to the LBL recipes in the \APERO automated function is one of the main goals for the next version of \APERO (version 0.8).

\begin{figure*}
    \centering
    \includegraphics[width=15cm]{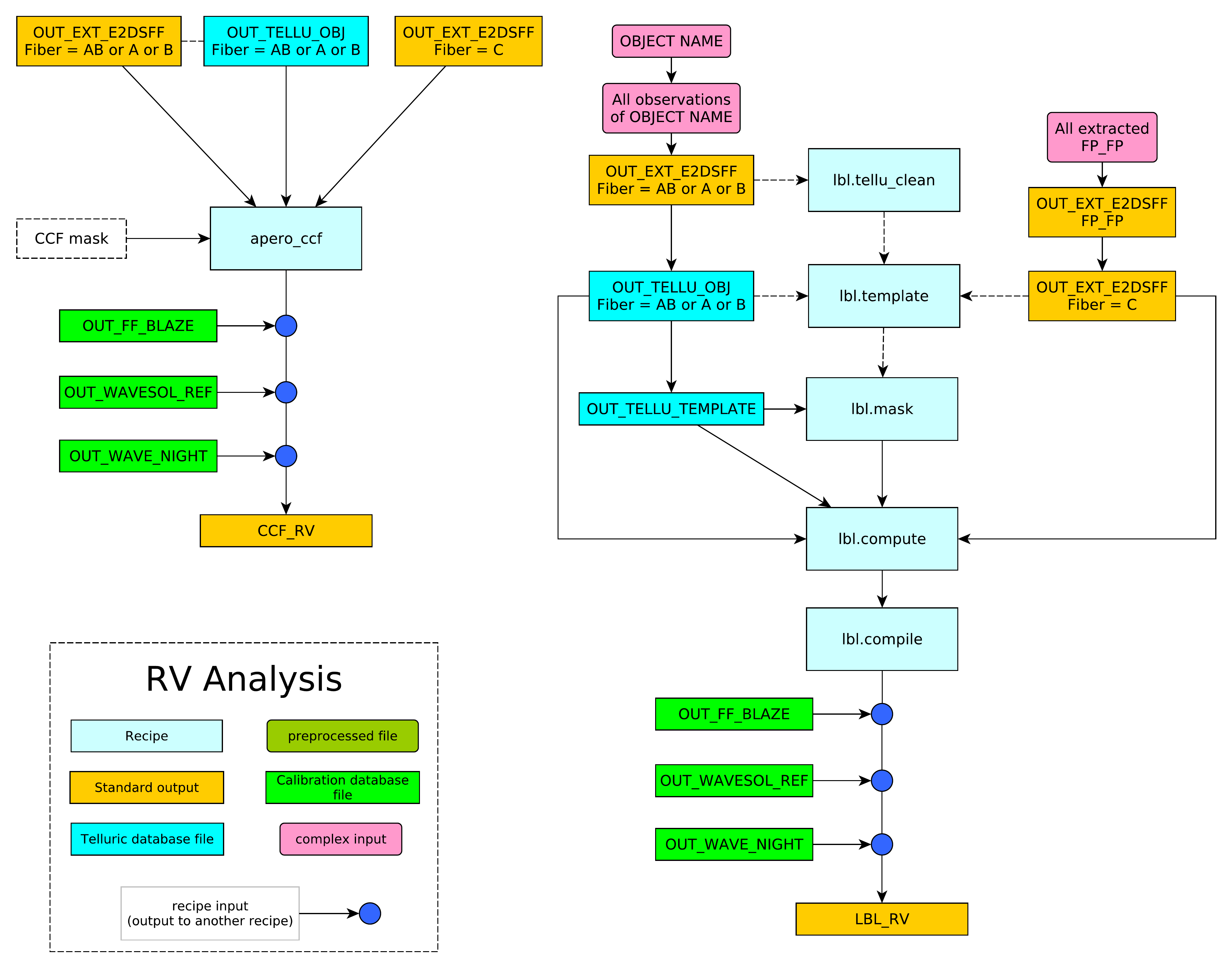}
    \caption{RV sequence}
    \label{fig:overview_rv}
\end{figure*}

Below we describe the current CCF radial velocity estimate provided by \APERO and briefly describe two alternate methods (developed alongside \APERO), the LBL method \citep{LBL2022} and an external CCF method \citep{Martioli2022}, both giving precision far closer to that required for science. The internal CCF method and LBL are shown in Figure \ref{fig:overview_rv}.

\subsection{The \APERO CCF recipe} \label{subsec:ccf}

The CCF method is very often used for \PRV work, particularly in the optical domain. In the early \APERO effort, it was the main tool to derive precise RV values. When implementing a near-infrared version of the CCF, a number of challenges appeared. The near-infrared domain is plagued with telluric absorption, and even after telluric correction, some wavelength domains are expected to have significant excess noise levels. Deep or saturated telluric lines cannot be corrected and are better left as gaps (represented as \NAN) in the spectrum that are fixed for the entire time series considered. When computing a CCF, how does one account for gaps in the data? The star's yearly line of sight variations will cause this gap to shift against the stellar spectrum by up to $\pm$32\,\kmps depending on ecliptic latitude. In the optical, one can simply reject the entire domain affected by the gap ($64$\,\kmps plus the gap width); however, at optical wavelengths, deep absorption lines are sufficiently sparse that the overall loss in wavelength domain due to telluric absorption is small, which is not the case in the near-infrared.

To further obfuscate the issue, telluric absorption varies between nights, so if one went down this path of masking, it would end with the masking of a large window affected by any line that gets deeper than a given threshold, even if only once in a time-series that may include hundreds of visits. The combination of varying conditions and yearly BERV excursions leads to a loss of domain that is simply unacceptable, especially considering that the parts of the near-infrared that are richest in sharp spectroscopic features (See Figure 4 in \citet{LBL2022}) are at the blue and red edges of the $H$ band, which are affected by telluric water absorption.

We opted for a CCF that correlates weighted delta functions against the spectrum but set the weight to zero when reaching a point below 0.5 telluric transmission (where unity is no telluric absorption). This is done on a spectrum-to-spectrum basis, to minimize the effective throughput losses. This CCF measurement is performed per spectrum using one of the 3 standard masks available in \APERO depending on the star's temperature (GL\,846, Gl\,699, GL\,905 respectively for \teff $>3500$\,K, $3000-3500$\,K, $<3000$\,K). We derive per-order as well as global CCFs. These data products are useful to confirm the systemic velocity of the star, avoiding eventual target misidentifying, as well as for flagging spectroscopic binaries. For time-series analysis, it can be significantly improved upon by using all observations to perform a spectral cleaning (see Section \ref{subsec:rv_ext}) to obtain a much cleaner CCF or through completely different methods, such as the line-by-line algorithm (Section \ref{subsec:lbl}).

\subsection{LBL Analysis} \label{subsec:lbl}

The most accurate RV measurements with \spirou have been obtained within the line-by-line (LBL$^{\ref{footnote:lbl_docs}}$; \citet{LBL2022}) framework. The LBL recipes exist externally to \APERO and the LBL outputs are currently not produced as part of \APERO. However, the \APERO outputs are directly used with the LBL scripts (\APERO data was the first data to work with LBL and \spirou \APERO data the main driver for the development of the LBL algorithms). In short, the LBL algorithm builds on the \citet{Bouchy2001} work that defines an RV measurement as the projection of a spectrum-to-template residual onto the derivative of the template. Instead of performing a spectrum-wide RV measurement, the LBL algorithm subdivides the spectrum into thousands of `lines' and derives a per-line velocity. It then uses a mixture model to derive an average velocity, allowing for the presence of velocity outliers. This allows for the filtering of outlying values in the spectra that arise, among other things, from imperfect telluric correction, OH line residuals, and detector defects. The framework is strongly inspired by the work of \citet{Dumusque2018} that used a per-line analysis to mitigate the impact of activity in \PRV time series of Sun-like stars. The LBL algorithm also provides a measurement of higher-order derivatives that can be used as tracers of activity. The second derivative of the spectrum notably provides an accurate measurement of changes in the effective FWHM of lines, even in the presence of dense molecular bands.

As detailed in Section 5 of \citet{LBL2022}, the analysis of a long time series of Barnard's star (from \spirou data with \APERO) produces an RV dispersion at the 2.59\,m/s that is consistent with the 2.57\,m/s uncertainty, as propagated from noise estimates. This shows that any systematic error from the \APERO framework, unaccounted for by zero-point corrections (Vandal et al., in prep), is at the sub-\mps level.

\APERO data analyzed with the LBL algorithm has been presented in \citet{Gan2022_2136}, \citet{Martioli2022}, and \citet{Cadieux2022}, setting constraints at, or below, the \mps-level on phase-folded RV curves of sub-Neptune TESS discoveries. A major goal of the next version of \APERO (version 0.8) is to call LBL recipes inside \APERO to allow the LBL products to be automatically produced and integrated into the post-processed results (see Section \ref{subsec:v_fits}).

\subsection{Comparison with an external CCF routine} \label{subsec:rv_ext}

In addition to the \APERO CCF recipe and the LBL analysis, we also have an external CCF recipe (\texttt{spirou-ccf}). The external CCF package \texttt{spirou-ccf}\footnote{\url{https://github.com/edermartioli/spirou-ccf}} has been developed to run an independent CCF analysis on the \spirou data, where one can obtain radial velocity measurements with meter-per-second precision, which is comparable to the results obtained with the LBL method. As mentioned in Section \ref{subsec:ccf}, a direct application of the method introduced by \cite{weightedCCF2002} to calculate the CCFs for the \spirou spectra will not provide very precise velocity measurements due to the fact that the line depths do not reflect the true statistical weight of each line as they are polluted by residual noise left by telluric absorption. In addition, the telluric features move throughout the time series and the signal-to-noise ratio is variable with time. 

The methods implemented in this package are described in \cite{Martioli2022}. In summary, all spectra of the time series are first loaded and a low-order flux correction is applied so that all spectra match a median template. The time dispersion of each spectral bin is calculated to estimate an empirical statistical weight for each region of the spectrum, which is then used to update the weights in the CCF calculation. An iterative sigma-clip filter is also applied and the spectral regions with large gaps are removed. The gaps are caused by some masking done in earlier stages of the pipeline, for example, by detector defects or by regions with deep telluric absorption where the telluric correction has failed. A global continuum is also calculated and used to normalize all spectra uniformly. Finally, a CCF is calculated for each spectrum and a CCF template-matching algorithm is applied to the entire time series to obtain the final relative velocity shifts for all spectra. 

\cite{LBL2022} also compares this external CCF approach to the LBL, thus we encourage the reader to consult that article for a more in-depth comparison.

\section{Polarimetry}
\label{sec:polar}

The polarimetry module for \APERO was adapted from the \lstinline{spirou-polarimetry} module\footnote{A prototype version of the \spirou polarimetry code is available on \url{https://github.com/edermartioli/spirou-polarimetry}}. Figure \ref{fig:overview_polar} shows a schematic flowchart of the polarimetry processing in \APERO. Table \ref{tab:aperopolarparams} shows the APERO input parameters required by the polarimetry module.

\spirou as a polarimeter can measure either circular (Stokes V) or linear (Stokes Q or U) polarization in the line profiles. Each polarimetric measurement is performed by 4 exposures obtained with the Fresnel rhombs set at different orientations (see Section 3.1 of \citealt{donati_spirou_2020}). In Table \ref{tab:rhombpositions} we provide the index position of each Fresnel rhomb, as they appear in the FITS header, for each exposure in the corresponding polarimetric mode. These indices are used by \APERO to recognize exposures within a polarimetric sequence, and then correctly apply the method introduced by \cite{Donati1997} to calculate polarimetric spectra.

\begin{table*}
\centering
\caption{APERO pipeline polarimetric parameters.}
\label{tab:aperopolarparams}
\begin{tabular}{lcc}
    \hline
    \hline
    Description of parameter & Default value & Accepted values \\
    \hline
    Use telluric corrected flux & \texttt{True} & \texttt{True,False} \\
    Use BERV corrected wavelength & \texttt{True} & \texttt{True,False} \\
    Use source RV corrected wavelength & \texttt{False} & \texttt{True,False} \\
    Interpolate flux values & \texttt{True} & \texttt{True,False} \\
    Polarimetry method & \texttt{'Ratio'} & \texttt{'Ratio','Difference'} \\
    Apply polarimetric sigma-clip & \texttt{True} & \texttt{True,False} \\
    Clipping threshold in units of sigma  & 4 & positive real \\
    Polarimetric continuum detection method &  \texttt{'IRAF'} &  \texttt{'MOVING\_MEDIAN','IRAF'} \\
    Stokes I continuum detection method &  \texttt{'IRAF'} &  \texttt{'MOVING\_MEDIAN','IRAF'} \\
    Continuum bin size [number of points] &  900 & positive integer \\
    Continuum fit function & \texttt{'spline3'} &  \texttt{'polynomial','spline3'} \\
    Order of continuum polynomial & 5 & positive integer \\
    Stokes I continuum fit number of knots & 50 & positive integer \\
    Remove continuum polarization & \texttt{True} & \texttt{True,False} \\
    Normalize Stokes I by the continuum & \texttt{False} & \texttt{True,False} \\
    \hline
    Initial velocity of LSD profile [\kmps] & -150 & negative real \\
    Final velocity of LSD profile [\kmps] & 150 & positive real \\
    Number of points in LSD profile & 151  & positive integer \\
    LSD mask lines are in air-wavelength  & \texttt{False} & \texttt{True,False} \\
    Minimum line depth for LSD analysis & 0.03  & $(0.0,1.0)$ \\
    Min. Landé for LSD analysis & 0 & $(0.0,\infty)$ \\
    Max. Landé for LSD analysis & 10 & $(0.0,\infty)$\\
    \hline
    \end{tabular}
\end{table*}

\begin{table*}
\centering
\caption{Index positions of the Fresnel rhombs (RHB1 and RHB2) for exposures taken in each observing mode of SPIRou.}
\label{tab:rhombpositions}
\begin{tabular}{c|cc|cc|cc|cc}
\hline
\hline
Observing  & \multicolumn{2}{c|}{Exposure 1} & \multicolumn{2}{|c|}{Exposure 2} & \multicolumn{2}{|c|}{Exposure 3} & \multicolumn{2}{|c}{Exposure 4} \\
Mode & RHB1 & RHB2 & RHB1 & RHB2 & RHB1 & RHB2 & RHB1 & RHB2 \\
\hline
Stokes IU & P16 & P2 & P16 & P14 & P4 & P2 & P4 & P14 \\
Stokes IQ & P2 & P14 & P2 & P2 & P14 & P14 & P14 & P2 \\
Stokes IV & P14 & P16 & P2 & P16 & P2 & P4 & P14 & P4 \\
\hline
\end{tabular}
\end{table*}

\subsection{Polarimetry calculations}
\label{sec:polarimetry}

The polarization spectra of \spirou are calculated using the technique introduced by \cite{Donati1997}, which is summarized as follows.  Let $f_{i\parallel}$ and $f_{i\perp}$ be the extracted flux in a given spectral element of fiber A and B channels, where $i=\{1,2,3,4\}$ gives the exposure number in the polarimetric sequence. Note that the extracted flux can be either the extracted spectrum or the extracted telluric corrected spectrum; by default in \APERO, we use the telluric corrected spectrum. The total flux of unpolarized light (Stokes I) is calculated by the sum of fluxes in the two channels and in all exposures, i.e.,

\begin{equation}
F_{I} = \sum_{i=1}^{4}{(f_{i\parallel} + f_{i\perp})}.
\end{equation}

\noindent Let us define the ratio of polarized fluxes as

\begin{equation}
r_{i} = \frac{f_{i\parallel}}{f_{i\perp}},
\label{eq:ratioofpolfluxes}
\end{equation}

\noindent which gives a relative measurement of the flux between the two orthogonal polarization states. In an ideal system, $r=1$ means completely unpolarized light, and other values provide the amount (or the degree) of polarization that can be calculated as in Equation 1 of \cite{Donati1997}, i.e., 

\begin{equation}
P = \frac{f_{\parallel} - f_{\perp}}{f_{\parallel} + f_{\perp}} = \frac{r - 1}{r + 1}.
\label{eq:degreeofpolarization}
\end{equation}

\noindent Therefore, in principle, one could obtain the amount of polarization with a single exposure. However, this measurement is not optimal, since it only records the two states of polarization at the same time but not at the same pixel. To obtain a measurement that records the same state of polarization at the same pixel, it suffices to take a second exposure with one of the quarter-wave analyzers rotated by $90^{\circ}$ with respect to the first exposure, consisting of the 2-exposure mode. One can also use the 4-exposure (2 pairs) mode, where the polarization state in the two channels is swapped between pairs, which better corrects for slight deviations of retarders from nominal characteristics (retardance and orientation) and also corrects for the differences in transmission between the two channels caused, for example, by different throughput of the two fibers, or by a small optical misalignment. For this reason, \spirou only operates in the 4-exposure mode, which is accomplished by rotating the analyzers accordingly in each exposure, as detailed in Table \ref{tab:rhombpositions}. The equation to calculate the degree of polarization for the 4-exposure mode can be obtained in two different ways, by using the ``Difference'' method or by the ``Ratio'' method, as defined in sections 3.3 and 3.4 of \cite{bagnulo2009} and also in Equation 3 of \cite{Donati1997}. The degree of polarization for a given Stokes parameter $X=\{U, Q, V\}$ in the Difference method is calculated by

\begin{equation}
P_{X} =  \frac{1}{4}\sum_{k=1}^{2}{\left(\frac{r_{2k-1}-1}{r_{2k-1}+1} - \frac{r_{2k}-1}{r_{2k}+1}\right)}, 
\label{eq:differencemethod}
\end{equation}

\noindent and for the Ratio method the degree of polarization is given by

\begin{equation}
P_{X} =  \frac{(\prod_{k=1}^{2}{r_{2k-1}/r_{2k}})^{1/4} - 1}{(\prod_{k=1}^{2}{r_{2k-1}/r_{2k}})^{1/4} + 1}.
\label{eq:ratiomethod}
\end{equation}

\noindent Another advantage of using two pairs of exposures is that one can calculate the null polarization (NULL1 and NULL2) as in equations 20 and 26 of \cite{bagnulo2009}, which provides a way to quantify the amount of spurious polarization. The null polarization for the Difference method is given by

\begin{equation}
NULL_{X} =  \frac{1}{4}\sum_{k=1}^{2}{\left[(-1)^{k-1}\left(\frac{r_{2k-1}-1}{r_{2k-1}+1} - \frac{r_{2k}-1}{r_{2k}+1}\right)\right]}, 
\label{eq:nulldifferencemethod}
\end{equation}

\noindent and for the Ratio method the null polarization is given by

\begin{equation}
NULL_{X} =  \frac{\left(\prod_{k=1}^{2}{r_{2k-1}/r_{2k}}\right)^{\frac{(-1)^{k-1}}{4}}  - 1}{\left(\prod_{k=1}^{2}{r_{2k-1}/r_{2k}}\right)^{\frac{(-1)^{k-1}}{4}} + 1}.
\label{eq:nullratiomethod}
\end{equation}

\noindent Finally, the uncertainties of polarimetric measurements can be calculated from the extracted fluxes and their uncertainties (denoted here by $\sigma$) by equations A3 and A10 of \cite{bagnulo2009}. In the Difference method, the variance for each spectral element is given by

\begin{equation}
\sigma_{X}^{2} = \frac{1}{16} \sum_{i=1}^{4}{ \left\{ \left[ \frac{2 f_{i\parallel} f_{i\perp}}{(f_{i\parallel} + f_{i\perp})^{2}} \right]^{2}   \left[ \frac{\sigma_{i\parallel}^{2}}{f_{i\parallel}^{2}} + \frac{\sigma_{i\perp}^{2}}{f_{i\perp}^{2}} \right] \right\} },
\label{eq:errordifferencemethod}
\end{equation}

\noindent and in the Ratio method the variance is given in terms of the flux ratio $r_{i}$ as defined in Equation \ref{eq:ratioofpolfluxes}, i.e.,

\begin{equation}
\sigma_{X}^{2} = \frac{\left( \frac{r_{1}}{r_{2}} \frac{r_{4}}{r_{3}} \right)^{1/2}} { 4 \left[ \left( \frac{r_{1}}{r_{2}} \frac{r_{4}}{r_{3}} \right)^{1/4} + 1\right]^{4}} \sum_{i=1}^{4}{\left[ \frac{\sigma_{i\parallel}^{2}}{f_{i\parallel}^{2}} + \frac{\sigma_{i\perp}^{2}}{f_{i\perp}^{2}} \right]}.
\label{eq:errorratiomethod}
\end{equation}

Applying this formalism to \spirou spectra, we obtain values that vary continuously throughout the spectrum and are systematically above or below zero for each spectrum, which we refer to here as the `continuum polarization'.  For general scientific applications with \spirou, this continuum polarization is actually spurious as it reflects small differences in the injection between  beams, and must therefore be fitted and removed. This step is mandatory before performing measurements in spectral lines. \APERO applies an iterative sigma-clip algorithm to fit either a polynomial or a spline to model the continuum polarization. 

\subsection{Least-Squares Deconvolution} 
\label{sec:lsd}

The least-squares deconvolution method (LSD) is an efficient technique that combines the signal from thousands of spectral lines retaining the same line profile information to obtain a mean velocity profile for the intensity, polarization, and null spectra.  A common application of this technique concerns the measurement of the Zeeman split into Stokes V (circularly polarized) profiles. The Zeeman split is a physical process where electronic transitions occurring in the presence of a magnetic field have their main energy transition level split into two additional levels, forming a double line in the intensity spectrum. An interesting feature of these lines is that they are circularly polarized and their polarizations have opposite signs. Therefore, by observing the circularly polarized spectrum one can obtain a characteristic Stokes V profile that provides a way to detect and characterize the magnetism in stellar photospheres with great sensitivity. 

\APERO implements the LSD calculations using the formalism introduced by \cite{Donati1997}, summarized as follows. Let us first consider the weight of a given spectral line $i$, $w_{i} = g_{i} \lambda_{i} d_{i}$, where $g$ is the Landé factor (magnetic sensitivity), $\lambda$ is the central wavelength, and $d$ is the line depth. Then one can construct the line pattern function,

\begin{equation}
M(v)= \sum_{i=1}^{N_{l}}{w\delta(v - v_{i})}, 
\label{eq:lsdlinepattern}
\end{equation}

\noindent where $N_{l}$ is the number of spectral lines considered in the analysis, $\delta$ is the Dirac function, and $v$ is the velocity. The transformation from wavelength ($\lambda$) to velocity space is performed by the relation $dv/d\lambda = c / \lambda$, where $c$ is the speed of light. The LSD profile is calculated by the following matrix equation:

\begin{equation}
\rm{\bf Z} = \left( \rm{\bf M}^{t}.\rm{\bf S}^{2}.\rm{\bf M} \right)^{-1} \rm{\bf M}^{t} . \rm{\bf S}^{2} . \rm{\bf P}, 
\label{eq:lsdprofile}
\end{equation}

\noindent where $\rm{\bf P}$ is the polarimetric spectrum calculated from Equation \ref{eq:differencemethod} or \ref{eq:ratiomethod}, and $\rm{\bf S}$ is the covariance matrix, a diagonal matrix where each element in the diagonal is given by $S_{jj}=1/\sigma_{j}$, with $\sigma_{j}$ being the uncertainty in the polarimetric spectrum calculated from Equation \ref{eq:errordifferencemethod} or \ref{eq:errorratiomethod}. 

Note that one can also calculate the null polarization LSD profile by substituting the polarimetric spectrum $\rm{\bf P}$ by the null spectrum $\rm{\bf N}$ in Equation \ref{eq:lsdprofile}.  The intensity LSD is also possible, by using the flux spectrum $\rm{\bf F}$, but in this case the line weight in Equation \ref{eq:lsdlinepattern} is simply given by the line depth, i.e, $w_{i} = d_{i}$.

\begin{figure*}
    \centering
    \includegraphics[width=18cm]{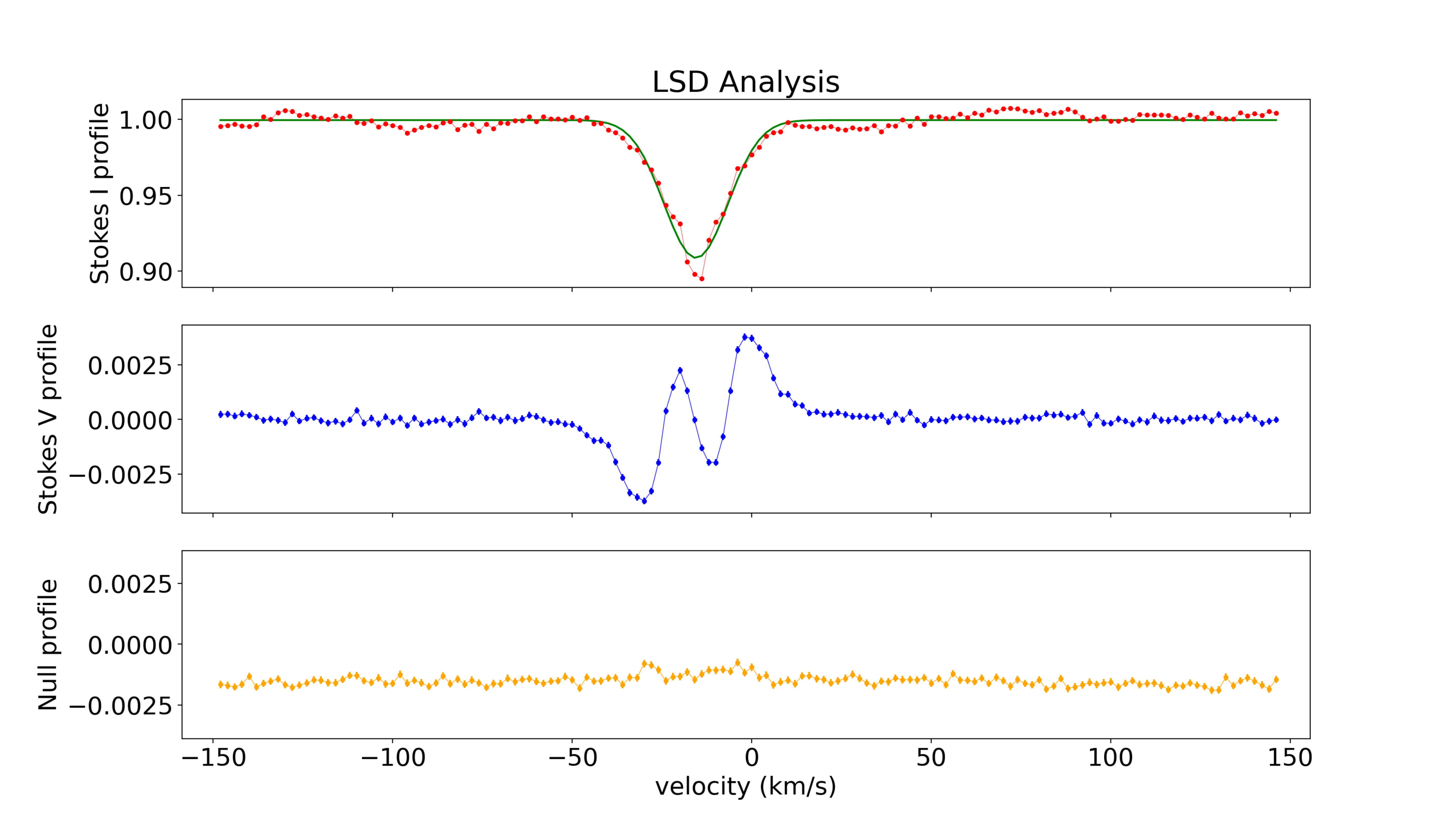}
    \caption{LSD analysis performed on the polarimetric data reduced with APERO and obtained from a 4-exposure Stokes-V sequence of the bright Ap star Gamma Equulei. Panels from top to bottom show Stokes I, Stokes V, and null profiles.}
    \label{fig:polar_lsd_example_gamequ}
\end{figure*}

\begin{figure*}
    \centering
    \includegraphics[width=14cm]{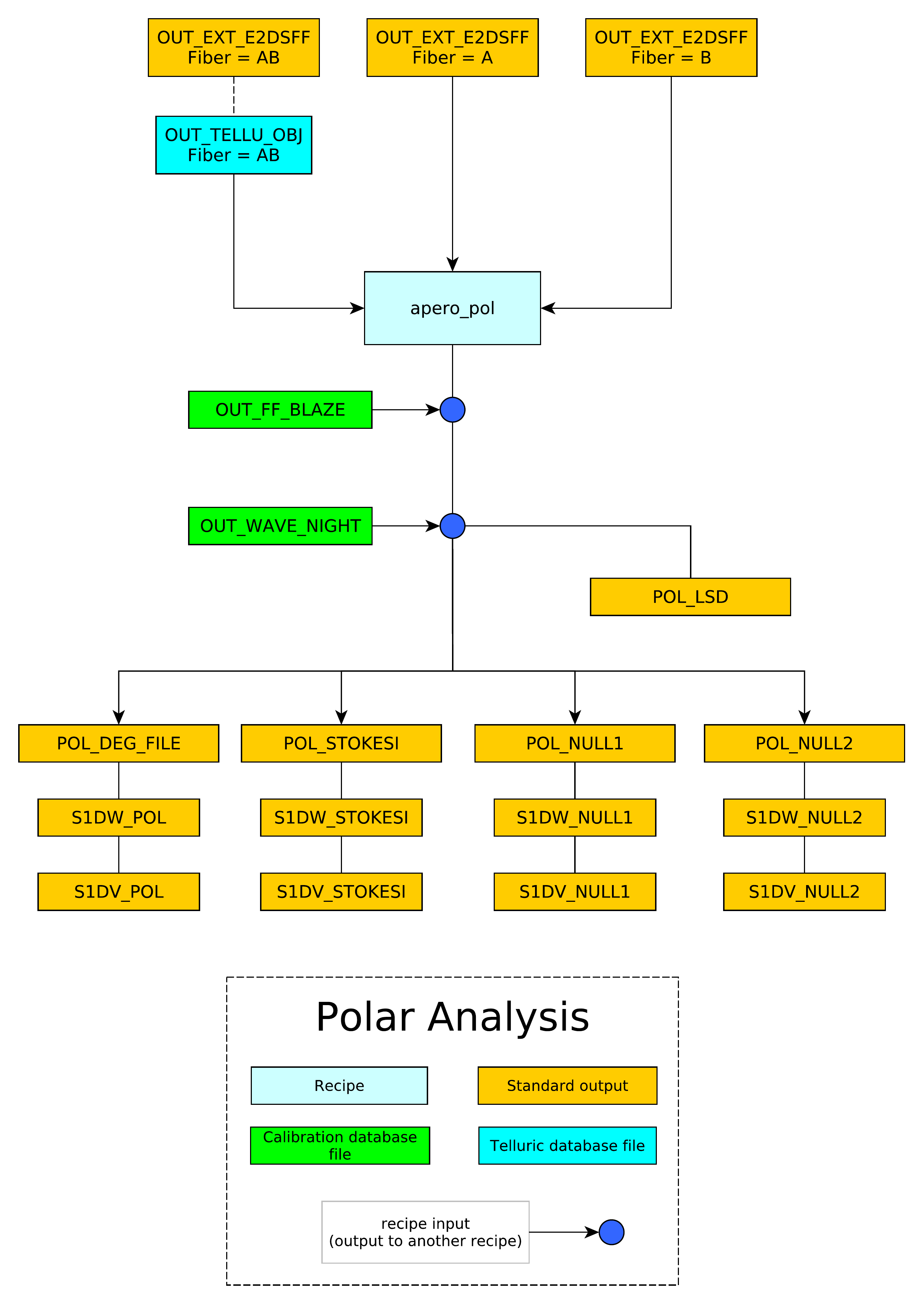}
    \caption{Polarimetric sequence}
    \label{fig:overview_polar}
\end{figure*}

In practice, LSD requires a few important steps to be executed by \APERO. First, each individual spectrum is cleaned using a sigma-clip rejection algorithm to minimize the impact of outliers in the LSD profile. Then we set a grid of velocities to calculate the LSD profile, where the grid is defined by the following parameters: an initial velocity, $v_{0}$, a final velocity, $v_{f}$, and the total number of points in the grid, $N_{v}$. Next, a fast and accurate method is necessary to project the spectral values onto the velocity grid. Finally, an appropriate catalog of spectral lines (line mask) needs to be adopted for the LSD calculations. \APERO selects the line mask from a repository of masks, where the selection is based on the proximity to the effective temperature of the star observed.  The \APERO masks are computed using the VALD catalog \citep{piskunov1995} and a MARCS model atmosphere \citep{Gustafsson2008} with an effective temperature ranging from 2500 to 5000~K in steps of 500~K, and the same surface gravity of $\log g=5.0$~dex. The lines that are effectively used in the LSD analysis are selected with line depths above a given threshold, which is set to 3\% by default and with a Land\'{e} factor of $g_{\rm eff}>0$, resulting in a total of approximately 2500 atomic lines that cover the full spectral range of \spirou. Figure \ref{fig:polar_lsd_example_gamequ} shows an example of an LSD analysis performed on a 4-exposure Stokes-V sequence of the bright Ap star Gamma Equulei, which has a strong magnetic field \citep[e.g.,][]{Bychkov2006} and therefore shows an obvious Zeeman feature in the SPIRou data.

In practice, the LSD analysis is not computed in a standard automated run of \APERO but the module is supplied and can be activated with the use of a single keyword in the \APERO profiles or run after processing.

\newpage
\clearpage

\section{Post processing}
\label{sec:post_processing}

\begin{figure*}
    \centering
    \includegraphics[width=16cm]{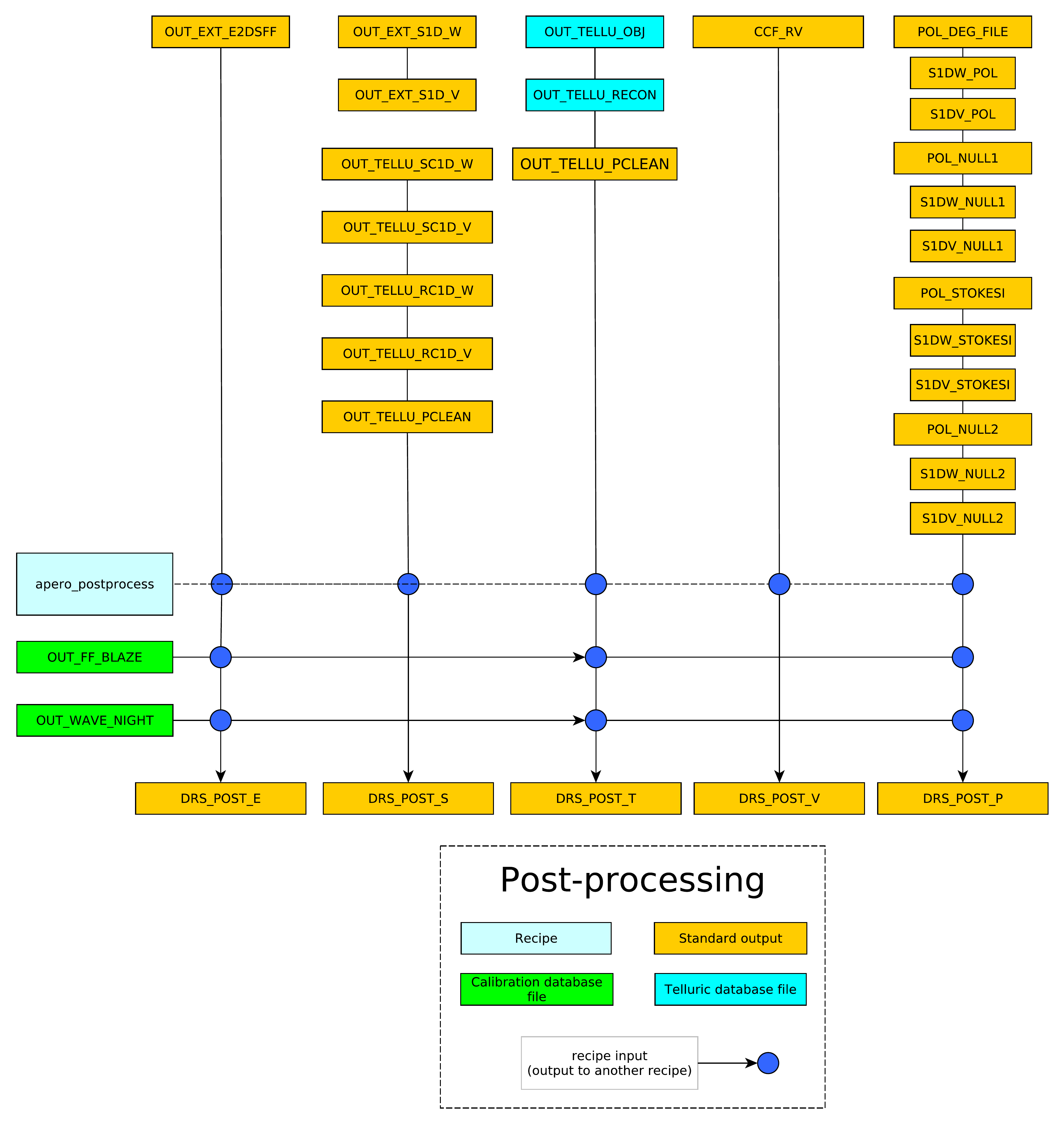}
    \caption{Post-process sequence}
    \label{fig:overview_post}
\end{figure*}

The final data products that go to PIs are composite files of many of the outputs of \APERO. For \spirou, these are sent to the Canadian Data Astronomy Center (CADC)\footnote{accessible from \url{https://www.cadc-ccda.hia-iha.nrc-cnrc.gc.ca/}} but are only produced for science targets and hot stars (i.e., \OBJFP, \OBJDARK, \POLFP, and \POLDARK) and not for calibrations by default. There are currently five post-processing files each linked to a single odometer code. These are the 2D extracted output (e.fits, Section \ref{subsec:e_fits}), the 2D telluric corrected output (t.fits, Section \ref{subsec:t_fits}), the 1D output (s.fits, Section \ref{subsec:s_fits}), the velocity output (v.fits, Section \ref{subsec:v_fits}), and the polarimetric outputs (p.fits \ref{subsec:p_fits}). A summary of the CADC output files is available in table \ref{tab:cadc_outputs} and the post-process sequence is shown in Figure \ref{fig:overview_post}.

\begin{table*}
    \centering
    \begin{tabular}{p{4cm}p{12cm}}
        \hline\hline
         File & Description \\
         \hline 
         (odometer)e.fits & 2D extracted spectrum for fibers AB, A, B, C, wavelength solution, and blaze  \\
         (odometer)s.fits & 1D extracted spectrum for fibers AB, A, B, C, and telluric corrected spectrum if available \\
         (odometer)t.fits & 2D telluric corrected spectrum for fiber AB, A, B, wavelength solution, blaze, and reconstructed atmospheric transmission \\
         (odometer)v.fits & combined and per order CCFs for fitting the radial velocity of the star \\
         (odometer)p.fits & polarimetric products (Polarimetric flux, Stokes I, Null vectors, wavelength solution, and blaze) \\
         \hline
    \end{tabular}
    \caption{Science ready outputs sent to the Canadian Data Astronomy Center, CADC).}
    \label{tab:cadc_outputs}
\end{table*}

\subsection{2D extraction product (e.fits)} \label{subsec:e_fits}

These are the combined extracted products. All extensions are two-dimensional spectra of size $4088\times49$. The `e.fits' file contains the extracted spectrum for each order for each fiber and the matching wavelength and blaze solution for each order and each fiber. The files are identified with a single odometer generated at the time of observation followed by an `e.fits' suffix.

\subsection{2D telluric corrected product (t.fits)} \label{subsec:s_fits}

These are the combined telluric-corrected products. All extensions are two-dimensional spectra of size $4088\times49$. The `t.fits' file contains the telluric corrected spectrum for each order and each fiber and the matching wavelength and blaze solution for each order and each fiber. The files are identified with a single odometer code at the time of observation followed by a `t.fits' suffix.

\subsection{1D extraction and 1D telluric corrected product (s.fits)} \label{subsec:t_fits}

These are the combined 1D spectrum products and consist of two tables. The two tables consist of the 1D spectrum in 1. velocity units and 2. wavelength units. They each consist of the following columns: the wavelength solution, the extracted flux in AB, A, B, and C, the telluric corrected flux in fibers AB, A, and B (if available), and the associated uncertainties for each flux column.  The files are identified with a single odometer code at the time of observation followed by an `s.fits' suffix.

\subsection{Velocity product (v.fits)} \label{subsec:v_fits}

The velocity products are packaged into the `v.fits' file. Currently, only the CCF values (Section \ref{subsec:ccf}) are added as an extension as the LBL products are computed separately. The CCF file consists of the CCF generated for each radial velocity element (by default this is between $\pm$300 \mps in steps of 0.5 \mps) for each order and a combined CCF for the same radial velocity elements. The files are identified with a single odometer code at the time of observation followed by a `v.fits' suffix. Once the LBL module is able to be used with \APERO it will add an extension to the `v.fits' (the `rdb' extension described in the LBL documentation$^{\ref{footnote:lbl_docs}}$).

\subsection{Polarimetric product (p.fits)} \label{subsec:p_fits}

These are the combined polarimetric products. The `p.fits` file consists of eight image extensions and three table extensions. The first two tables are the 1D representations of the 2D polarimetric products (listed in the extensions above) in 1. velocity units and 2. wavelength units. They each consist of the following columns: the wavelength solution, the polarimetric flux, the Stokes I flux, the Null 1 and 2 fluxes, and the associated uncertainties on each flux column. The third table lists the configuration parameters used to run \APERO. Although polarimetric products are the combination of at least 4 odometer codes, files are associated with a single odometer code (the first in the sequence at the time of observation) followed by a `p.fits' suffix.

\section{Discussion}
\label{sec:discussion}

\APERO has been an ongoing effort since its conception in October 2017 (see Appendix \ref{section:version_history} and table \ref{tab:version_history}). With the first light of \spirou in April 2018, it took nearly 2 years for \APERO to start producing precise science results and has been used in publications since 2020 (see Appendix \ref{section:current_science_publications} and table \ref{tab:current_science_pub}). Here we discuss current performances and limitations and planned future work.

\subsection{Performance and limitations}
\label{subsec:issues}

Throughout development, we have tried to optimize the speed of all recipes (e.g., through the use of SQL databases, \textsc{numba}, \textsc{bottleneck}, etc.). With the most current version (\latestversion ), using 35 CPU cores (on a single node), we can reduce all available data (all \spirou legacy survey data and all PI data to which we have access, covering $\sim$430 nights and $\sim$45000 science observations) in 7 days; table \ref{tab:timing} shows a breakdown of the various steps using a 35 core machine. These timings are for all data we have access to, which are equivalent to $\sim$90\% of all data taken with \spirou between April 2018 and June 2022. 

\begin{table*}
    \centering
    \begin{tabular}{J{3cm}J{7cm}lll}
        \hline\hline
        Sequence & Recipes & Number & Time taken & Efficiency \\
        \hline
        Pre-processing & \Apreprocessing & 75761 files & 34.5 hours & 0.89 \\
        &&&&\\ 
        Reference Calibrations\vspace{0.5cm} & \Adarkref, \Abadpixel, \Alocalisation, \Ashaperef, \Ashapenight, \Aflatblaze, \Athermal, \Aleakref, \Awaveref & 1 night & 1.5 hours & - \\
        &&&&\\ 
        Nightly Calibrations\vspace{0.1cm} & \Abadpixel, \Alocalisation, \Ashapenight, \Aflatblaze, \Athermal, \Awavenight  & 432 nights & 7.4 hours &  0.87 \\
        &&&&\\ 
        Extraction & \Aextract  & 46836 files & 63.3 hours & 0.63 \\
        &&&&\\ 
        Telluric (hot star)\vspace{0.1cm} & \Amktellu, \Amkmodel, \Aftellu, \Amtemp & 1043 files & 1.1 hours & 0.70 \\
        &&&&\\ 
        Telluric (science) & \Aftellu, \Amtemp  & 45524 files & 34.6 hours & 0.75 \\
        &&&& \\        
        CCF & \Accf  & 45524 files & 7.0 hours & 0.89 \\
        &&&&\\        
        Polarimetry & \Apolar  & 9880 groups & 2.1 hours & 0.90 \\
        &&&&\\   
        Post-process & \Aprocessing  & 48472 files  & 18.4 hours & 0.93 \\
        \hline
        Total & & & 170.1 hours & 0.76 \\
         & & & 7.1 days & 0.76 \\
        \hline
    \end{tabular}
    \caption{Example full run reducing all SPIRou legacy survey (and some PI) data from April 2018 to April 2022. This was processed on one machine using 35 cores. Note preprocessing is done on both science and calibration observations,  reference calibrations are only run on a single night, and each nightly calibration depends on the availability of specific calibrations thus leading to a range of nights (from 366 to 401). Polarimetric groups consist of 4 individual exposures in different rhomb positions and are only processed for \POLFP and \POLDARK files. The number of specific steps may depend on previous steps (i.e., quality control failures, engineering nights that were excluded, and odometer codes that were present in a list to not be processed).  The efficiency is defined in Equation \ref{equ:efficiency}.
    \label{tab:timing}}
\end{table*}

We find that with 35 CPU cores we reach a point where we start to see input-output bottlenecks, most probably caused by writing to disk and/or writing to the various databases. This manifests as an individual slowdown of each recipe run which limits the efficiency of using more CPU cores (i.e., more recipes run at the same time mean more files being written to disk at the same time and more writing to the various database at the same time causing queuing to occur). We define efficiency in table \ref{tab:timing} as the total CPU time taken divided by the total time taken to run multiplied by the number of cores (Equation \ref{equ:efficiency}). A perfectly efficient code would give a value of 1.

\begin{equation}
    \label{equ:efficiency}
    \text{Efficiency} = \frac{\text{total CPU time}}{\text{total time taken} \times N_{cores}}
\end{equation}

We see that on average we have an efficiency between 0.7 and 0.93. We find that recipes that run quickly but save several files have lower efficiency, i.e., that a bottleneck of writing to disk may be occurring; however, these recipes also run quickly and write frequently to the various databases, and as such it is hard to distinguish between the disk and database bottlenecks. Also, recipes that run slow are rated as more efficient due to the amount of time spent using the CPUs (i.e., science algorithms) against reading/writing to and from disk/databases, thus our metric is far from perfect. Another factor is other processes using the machine at that specific time that cannot be easily taken into account when measuring how efficient we are. We will continue to review the performances, speed up the science algorithms and find ways to make the individual recipes faster.

Currently, \APERO is optimized to run on a single node (i.e., a single machine) with access to many CPU cores on this single node. It is possible to run batches of \APERO in a manual way using multiple runs of \Aprocessing and controlling (manually) when to run the next step (i.e., making sure all pre-processing is done before reference calibrations, that all reference calibrations are done before any night calibrations, etc.). One can also run recipes one-by-one bypassing the use of the \Aprocessing recipe completely. \APERO is optimized to reduce a full data set, this implies many terabytes of raw and reduced data. Currently to reduce a single night of data one requires at the very least a full reference calibration night (i.e., the calibration data from 2020-08-31) and the full set of calibrations for the night to be reduced (and preferably a few nights surrounding it in case any calibrations fail quality control) and the full telluric database of hot stars (from every night processed). This obviously means a large amount of data is required for even a single night or single observation. We plan to release a full calibration database and full telluric database of hot stars with a way to get and install this data, in order to allow any user to reduce a small set of data. We do however always recommend using data reduced with a full data set done in a uniform way. This is currently available at data centers at CFHT, Université de Montréal, and the Laboratoire d'Astrophysique de Marseille.

\subsection{Future work}
\label{subsec:future}

As with most pipelines, improvements are always ongoing. With the LBL RV analysis code, we have seen RV accuracy down to 2\mps with \spirou, which is an indication that \APERO is at least this precise. However, there are several features we plan to add:
\begin{itemize}
    \item The \APERO recipes do not currently propagate uncertainties throughout the data reduction process, which can be problematic when trying to understand the limiting factors in the data analysis. Full error propagation is important as feedback to the engineering team; for example, quantifying the impact of the thermal background from the optical train on the measurement of $K$-band spectroscopic features such as the 2.29\,$\mu$m CO band-head could justify efforts to cool parts of the optical train or not.
    \item In parallel to the propagation of uncertainties, we plan to propagate quality flags on pixel values, for example, whether a given invalid pixel is due to a cosmic ray hit or a hot pixel. In the current framework, pixels are either deemed valid or invalid and flagged as a \NAN, which does not allow one to back-trace the origin of missing data. This is done for JWST data products\footnote{\url{https://jwst-reffiles.stsci.edu/source/data_quality.html}} with a pixel-level data quality encoded with 32-bit integers.
    \item As the \spirou fibers are multi-mode, one expects a certain level of noise to arise from the time-varying weight of modes injected in each of the science fibers. This is minimized through the use of a pupil slicer and fiber scrambler at the injection. As the pupil slicer image provides more information on the flux distribution at the fiber exit than a simple fiber (e.g., the bottom-center panel in Figure \ref{fig:e2ds_grid}), one could decorrelate this spatially-resolved information against the modal noise.
    \item Persistence in infrared arrays is a non-trivial problem for faint-target observations \cite{Artigau2018-H4RG}. For any given image, one sees a decaying remnant image of all previous observations with decreasing amplitude. The amplitude of any given observation falls with an amplitude that is proportional to the inverse of the delay since the last illumination. One notable feature of persistence in infrared arrays is that it is not only the previous frame that matters but the entire history of illumination over the last few hours. A bright target observed for a long time at the beginning of a night (a common example being a multi-hour sequence to monitor a transiting planet) will affect all fainter targets observed later during the night. To add further complexity to the matter, the persistence response varies at the pixel-to-pixel level. Work has begun to construct a persistence model for \spirou. Furthermore, the algorithms need to be run at the observatory level as data obtained earlier in the night may be proprietary and not accessible to all \APERO users.
    \item The main limitation for faint object observations with \spirou or any near-infrared \PRV spectrograph, particularly bluer of the $K$ band, is detector readout noise. As has been demonstrated by \cite{Payeur2022}, machine-learning algorithms can reduce readout noise in long sequences (100 readouts of $\sim$10\,minutes) from $\sim$6\,e$^-$ to $<2$\,e$^-$ (see Table~2 therein). This needs to be performed before the current \APERO steps, as it requires handling the data cubes rather than the 2-D images used here. As long as the output format is maintained, the machine-learning images should be used as inputs to \APERO.
    \item Energetic particles regularly hit infrared arrays and deposit electrons in pixels, leading to spurious signals. These hits happen basically instantaneously (on the time scales relevant to the readout) and manifest themselves as discontinuities in the time series of non-destructive readouts. Efficient algorithms have been proposed to handle cosmic ray hits in ramp-fitting frameworks \citep{Anderson_2011} but have yet to be implemented for \spirou.
    \item The LBL recipes have been designed to use \APERO byproducts, but they have not yet been implemented within the automated \APERO framework. Steps that are currently done manually, such as the association of an appropriate stellar template, will be included within \APERO in the near future.
\end{itemize}

\noindent There are also various planned improvements, they are the optimal extraction (better characterization of extraction weights), the database architecture (currently throttling the maximum number of connections with a large number of cores), some minor memory leaks when parallel processing, handling the thermal contribution at bluer wavelength domains and completing all documentation$^{\ref{footnote:apero_docs}}$.
 
\section{Conclusion}
\label{sec:conclusion}

We present \APERO (A PipelinE to Reduce Observations) and highlight its use as the official pipeline for \spirou. We walk through the steps going from raw data to science-ready products. We detail the pre-processing of raw data to correct detector issues, the production of reference calibrations and nightly calibrations, and the use of these calibrations to correct and extract hot stars and science observations in a consistent, controlled manner. We summarize telluric correction (which will be detailed in a future publication, Artigau et al. in prep), RV analysis, polarimetric analysis, and our post-processing recipes delivering telluric corrected 2D and 1D spectra as well as polarimetry products and enabling precise stable radial velocity calculations (via the LBL algorithm, \citealt{LBL2022}), good to at least $\sim2$ \mps over the timescale of the current lifetime of \spirou (5 years).

\acknowledgments

We would like to thank the anonymous referee for the valuable comments that improved the quality of the paper. The authors wish to thank everyone involved with writing, maintaining, and updating all Python packages. Specifically, \APERO has made extensive use of: 
\begin{tasks}(2)
    \task \modAstropy, \citep{2018AJ....156..123A, 2013A&A...558A..33A} 
    \task \modAstroquery \citep{2019AJ....157...98G} 
    \task \modBarycorrpy \citep{barycorrpy2020_1, barycorrpy2020_2}
    \task \modMatplotlib \citep{Hunter2007} 
    \task \modNumpy \citep{harris2020array} 
    \task \modPandas \citep{McKinney_2010, McKinney_2011} 
    \task \modscipy \citep{Virtanen_2020} 
\end{tasks}

As well as the python packages \textsc{bottleneck} \citep{Goodman2019}, \textsc{gitchangelog} \citep{Lab2018}, \textsc{ipdb} \citep{Chapelle2021}, \textsc{IPython} package \citep{PER-GRA:2007}, \textsc{mysql-connector-python} \citep{Mariz2021}, \textsc{numba} \citep{numba2015}, \textsc{pandastable} \citep{Farrell2016}, \textsc{Pillow} \citep{Murray2021}, \textsc{pyyaml} \citep{Simonov2021}, \textsc{sphinx} \citep{Komiya2021}, \textsc{sqlalchemy} \citep{Bayer2021}, \textsc{Scikit-learn} \citep{scikit-learn}, \textsc{tqdm} \citep{casper_da_costa_luis_2021_5517697}, \textsc{yagmail} \citep{vanKooten2021}, and \textsc{xlrd} \citep{Withers2021}.

This research made use of \textsc{ds9}, a tool for data visualization supported by the Chandra X-ray Science Center (CXC) and the High Energy Astrophysics Science Archive Center (HEASARC) with support from the JWST Mission office at the Space Telescope Science Institute for 3D visualization. 

This research made use of TOPCAT, an interactive graphical viewer and editor for tabular data \citep{2005ASPC..347...29T}. 

\APERO would have been impossible without the use of PyCharm, Git, and GitHub.

This work has made use of data from the European Space Agency (ESA) mission {\it Gaia} (\url{https://www.cosmos.esa.int/gaia}), processed by the {\it Gaia} Data Processing and Analysis Consortium (DPAC, \url{https://www.cosmos.esa.int/web/gaia/dpac/consortium}). Funding for the DPAC has been provided by national institutions, in particular, the institutions participating in the {\it Gaia} Multilateral Agreement.  \APERO has made use of the SIMBAD database, operated at CDS, Strasbourg, France. This research has made use of NASA's Astrophysics Data System. \APERO has made use of the VizieR catalog access tool, CDS, Strasbourg, France. The acknowledgments were compiled using the Astronomy Acknowledgment Generator. 

The authors wish to recognize and acknowledge the very significant cultural role and reverence that the summit of MaunaKea has always had within the indigenous Hawaiian community. We are most fortunate to have the opportunity to conduct observations from this mountain. 

This work was financially supported by the Natural Sciences and Engineering Research Council of Canada and the Fonds Québécois de Recherche - Nature et Technologies. Observatoire du Mont-Mégantic and the Institute for Research on Exoplanets acknowledge funding from Développement Économique Canada, Quebec's Ministère de l'Éducation et de l'Innovation, the Trottier Family Foundation and the Canadian Space Agency. M.H. and I.B acknowledge support from ANID – Millennium Science Initiative – ICN12\_009. C.M., A.C., P.F., X.D., I.B., and J.F.D. acknowledges funding from the French ANR under contract number ANR18CE310019 (SPlaSH) and the Programme National de Planétologie (PNP). This work is supported by the French National Research Agency in the framework of the Investissements d’Avenir program (ANR-15-IDEX-02), through the funding of the “Origin of Life" project of the Grenoble-Alpes University. J.F.D. acknowledges funding from the European Research Council (ERC) under the H2020 research \& innovation programme (grant agreement \#740651 NewWorlds). T.V. would like to acknowledge funding from the Fonds de Recherche du Qu\'ebec - Nature et Technologies (FRQNT, scholarship number 320056), and the Institute for Research on Exoplanets (iREx).

\facilities{CFHT(SPIRou)}
\software{\textsc{python3}: \modAstropy, \modAstroquery, \modBarycorrpy, \textsc{bottleneck}, \textsc{ipdb}, \textsc{ipython}, \modMatplotlib, \textsc{numba}, \modNumpy, \modPandas, \textsc{pandastable}, \textsc{Pillow}, \textsc{pyyaml}, \modscikit, \modscipy, \textsc{sphinx}, \textsc{sqlalchemy}, \textsc{tqdm}, \textsc{xlrd}}

\bibliography{references.bib}{}
\bibliographystyle{aasjournal.bst}

\appendix
\section{Creating a raw \texorpdfstring{\spirou}~ramp image}
\label{section:appendix_fitstoramp}
\setcounter{table}{0}
\renewcommand{\thetable}{A\arabic{table}}

The \spirou detector control software reads the detector continuously every 5.57 s and produces a 2D image ($4096\times4096$) constructed from the linear fit of the pixel value versus time (as well as a slope, intercept, error, and number of frames used for quality checks). The construction of the 2D image from individual readouts being handled at the acquisition step is not included as part of \APERO but as software maintained and used at CFHT. The construction of the 2D frame is performed through the following steps.

Individual detector frames are obtained from the detector control software every 5.57\,s (at time $j$, $t[j]$). A flagging of pixel saturation is performed, and pixels with a non-linearity that would be larger than $\sim10$\,\% are considered unreliable and rejected for all future readouts (binary mask $m[i,j]$ for pixel $i$ at time $j$). A non-linearity correction is applied to pixel fluxes (flux at pixel $i$ and at time $j$ is $f[i,j]$). As individual readouts are in computer memory, intermediate quantities necessary for the computation of the total pixel-level slope are computed. The advantage of preserving these quantities in memory is that one can perform a pixel-level ramp-fitting over an arbitrarily large number of frames without being required to have all pixel values in memory at the same time or without having to access files multiple times.

\begin{itemize}
    \item $\sigma_x[i] = \sum_{j} m[i,j]*f[i,j]$
    \item $\sigma_y[i] = \sum_j m[i,j]*t[j]$
    \item $\sigma_{xy}[i] = \sum_j m[i,j]*f[i,j]*t[j]$
    \item $\sigma_{x^2}[i] = \sum_j m[i,j]*f[i,j]^2$
    \item $n[i] = \sum_j m[i,j]$
\end{itemize}

Among these intermediate quantities, the only one that has a clear signification is $n[i]$: it corresponds to the number of valid (i.e., below the predefined flux level for saturation) readouts obtained over the entire sequence. For normal scientific exposures, $n[i]$ is equal to the total number of readouts for the vast majority of pixels; pixels with a very large dark current being the exception.

From simple linear algebra, one can show that the per-pixel intercept is 
\begin{equation}
 b[i] = \frac{\sigma_x[i]\sigma_{xy}[i] - \sigma_{x^2}[i]*\sigma_y[i]}{\sigma_{x}[i]^2-n\sigma_{x^2}}   
\end{equation}

and correspondingly, the per-pixel slope is:
\begin{equation}
a[i] = \frac{\sigma_{y} - n[i]b[i]}{\sigma_{x}[i]}
\end{equation}

Once the slope image $a[i]$ has been computed, it is corrected for correlated amplifier noise using the side reference pixels (along the fast readout axis), and amplifier offset is corrected using the top and bottom reference pixels (both extremities of the slow readout axis).

As a quality check of pixel-level value, we further compute the  error on the slope to identify pixels having a suspiciously large dispersion of their values around the slope. This is used to flag pixels that have a large slope that is inconsistent with their frame-to-frame accumulation rate. To compute the slope, one has to re-read all frames once (though it is not required to keep them all at once in memory) and compute the following values

\begin{itemize}
    \item $x_p[i] = sx[i]/n[i]$
    \item $y_p[i,j] = b[i]+a[i]*t[j]$
    \item $\varrho_{x^2}[i] = \sum_j (t[j]-x_p[i])m[i,j]$
    \item $\varrho_{y^2}[i] = \sum_j (f[i,j]-y_p[i,j])m[i,j]$
\end{itemize}

From which the slope error is 
\begin{equation}
    \varrho(i) = \sqrt{ \frac{\varrho_{y^2}[i]/(n[i]-2)}{\varrho_{x^2}[i] }}
\end{equation}

\section{Standard image calibration}
\label{section:appendix_standard_image_calibration}
\setcounter{table}{0}
\renewcommand{\thetable}{B\arabic{table}}

After pre-processing (Section \ref{sec:preprocessing}), the reference dark calculation (Section \ref{subsec:ref_dark}) and the bad pixel correction (Section \ref{subsec:bad_pixel}), all images that are used in \APERO need to be calibrated in a standard way (using both the dark reference and bad pixel recipe outputs). This is not a separate recipe but a set of functions that are used in all recipes that use pre-processed files as inputs.

The standard calibration is ordered as follows:

\begin{enumerate}
    \item dark reference correction (Section \ref{subsec:dark_ref_correction}).
    \item flip, resize and re-scale the image  (Section \ref{subsec:flip_resize_rescale}).
    \item flag bad pixels (Section \ref{subsec:flag_badpix}).
    \item correct background flux (Section \ref{subsec:corr_background}).
    \item clean hot pixels (Section \ref{subsec:hotpix_cleaning}).
    \item flag pixels that are out of bounds (Section \ref{subsec:flag_out_of_bounds}).
\end{enumerate}

\subsection{Dark reference correction}
\label{subsec:dark_ref_correction}

The first step of the standard calibration of pre-processed files is to correct the input image for the dark signal.

\begin{equation}
    \label{equ:dark_correction}
    IM_{\text{corr } i,j} = IM_{\text{uncorr }i,j} - N(DARK_{i,j})
\end{equation}

\vspace{0.5cm}
\noindent where $IM_{\text{corr } i,j}$ and $IM_{\text{uncorr }i,j}$ are the flux in \ARRIJ of the corrected image and uncorrected image respectively, $N$ is the number of raw images that went into $IM$ and $DARK_{i,j}$ is flux in \ARRIJ of the reference dark (see Section \ref{subsec:ref_dark}).

The dark reference is taken from the calibration database. If more than one dark reference exists the closest in time to $IM$ is used  (using the header key \MIDEXPOSURE from the header).

\subsection{Flipping, resizing, and re-scaling the image}
\label{subsec:flip_resize_rescale}

For legacy reasons, the image is flipped in the vertical and horizontal directions (see Figure \ref{fig:size_grid}). After this the image is converted from $ADU/s$ to electrons using Equation \ref{equ:adu_to_electons}.

\begin{equation}
    \label{equ:adu_to_electons}
    IM_{\text{electrons } i,j} = IM_{\text{ADU}/\text{s } i,j} \times \text{Gain} \times t_{\rm exp}
\end{equation}

\vspace{0.5cm}
\noindent where $IM_{\text{electons } i,j}$ is the flux in electrons for \ARRIJ, $IM_{\text{ADU}/\text{s } i,j}$ is the flux in $ADU/s$ for \ARRIJ, the gain is taken from the header key \GAIN (although it has remained constant over the lifetime of \spirou), and $exptime$ is the exposure time in seconds (taken from the header key \EXPTIME).

Once the image is in electrons it is then resized. The image is cut in the cross-order direction to start from pixel 250 and end at pixel 3350 (removing a partial blue order and the whole unilluminated dark amplifier region) and in the along-order direction to start from pixel 4 and end at pixel 4092 (removing just the H4RG reference pixels). Thus after this resizing the image is of size 3100$\times$4088 (see Figure \ref{fig:size_grid}).

\subsection{Flagging the bad pixels}
\label{subsec:flag_badpix}

The closest bad pixel map (\BADPIX, see Section \ref{subsec:bad_pixel}) in time to the image (using the header key \MIDEXPOSURE from the header) is loaded from the calibration database and all pixels that are flagged as bad pixels are set to \NAN, this is shown in Equation \ref{equ:badpix_flag}.

\begin{equation}
    \label{equ:badpix_flag}
    IM_{\text{corr } i,j} = \left\{ \begin{array}{cl}
      NaN : & BADPIX_{i,j} \equiv 1 \\
      IM_{i,j} : & \text{otherwise} \\
    \end{array} \right.
\end{equation}

\vspace{0.5cm}
\noindent where $IM_{\text{corr } i,j}$ and $IM_{i,j}$ are the flux in \ARRIJ of the corrected image and the input image respectively. $BADPIX_{i,j}$ is a bad pixel flag (1 or 0) in \ARRIJ from the bad pixel map.

\subsection{Correcting the background flux}
\label{subsec:corr_background}

Within each science image, we take the median of `background' pixels (identified using the \BACKMAP, see Section \ref{subsec:bad_pixel}) within a region and create a map of large-scale background features (middle panel, Figure \ref{fig:backest}). This map is then splined into a $4088\times4088$ image and subtracted from the science frame.

\subsection{Additional cleaning of hot pixels}
\label{subsec:hotpix_cleaning}

Hot pixels are flagged by finding pixels that are 10$\sigma$ (positive or negative) outliers compared to their immediate neighbors. This is in addition to the cosmic ray rejection applied in Section \ref{subsec:cosmic_ray_reject} and the bad pixel flagging (Section \ref{subsec:bad_pixel}) which removes most of the hot pixels. In this additional cleaning of hot pixels, we first construct a flattened image and perform a low-pass filter in the along-order direction, filtering the image so that only pixel-to-pixel structures remain. We then apply median filtering, which removes these big outliers, and then we smooth the image to avoid big regions filled with zeros. We apply a 5-pixel median boxcar and a 5-pixel smoothing in the along-order direction, which blurs along the dispersion over a scale of $\sim$7 pixels. Bad pixels are interpolated with a 2D surface fit by using valid pixels within a $3\times3$ pixel box centered on the bad pixel. 

\subsection{Flagging out of bound pixels}
\label{subsec:flag_out_of_bounds}

Pixel values need to be within reasonable bounds considering the physics of the H4RG detector. If they are not in bounds we set them to \NAN. The upper bound is the saturation per frame time, but as the flux is expressed as a slope (in the \FITSTORAMP), a pixel with a value greater than the saturation point can be recorded by the detector and is nonphysical. The lower bound is set to the negative value of ten times the readout noise, these bounds are shown in Equation \ref{equ:out_of_bounds}.

\begin{equation}
    \label{equ:out_of_bounds}
      IM_{\text{corr } i,j} = \left\{ \begin{array}{cl}
      NaN : & IM_{i,j} > \text{saturation} / t_{\text{frame}}  \\
      NaN : & IM_{i,j} < -10 \times\text{readout noise}  \\
      IM_{i,j} : & \text{otherwise} \\
    \end{array} \right.
\end{equation}

\vspace{0.5cm}
\noindent where $IM_{\text{corr } i,j} $ and $IM_{i,j}$ are the flux in \ARRIJ of the corrected image and the input image respectively, $saturation$ is taken from the header keyword \SATURATION and is converted to electrons via \ref{equ:adu_to_electons}, $t_{\rm frame}$ is the individual frame time (from header keyword \FRAMETIME), and readout noise is taken from the header keyword \RDNOISE.

\section{Shape transformation}
\label{section:affine_transformation}
\setcounter{table}{0}
\renewcommand{\thetable}{C\arabic{table}}

The shape transform algorithm allows three different transformations, that may or may not be used.
Here we define $x$ as the direction along the order, $y$ as the direction across the order.

\begin{enumerate}
    \item a linear transform: defined by $dx$, $dy$, $A$, $B$, $C$, $D$, where $dx$ and $dy$ are shifts $x$ and $y$ respectively and $A$, $B$, $C$, $D$ form the transform matrix:

    \begin{equation}
      \left[ \begin{array}{cl}
      A & B \\
      C & D \\
      \end{array} \right]
    \end{equation}
    
    This combines with $dx$ and $dy$ in order to form a 3$\times$3 matrix:
    
    \begin{equation}
      \left[ \begin{array}{ccc}
      A & B & dx \\
      C & D & dy \\
      0 & 0 & 1 \\
      \end{array} \right]
    \end{equation}
    
    This $3\times3$ linear transformation matrix allows for scaling, rotation, reflection (not used in our case) and shearing \citep{gonzalez2008digital}.
    
    \item a shift in $x$ position, where a shift is defined for each pixel.
    
    \item a shift in $y$ position, where a shift is defined for each pixel.
    
\end{enumerate}

\section{Version history of APERO}
\label{section:version_history}
\setcounter{table}{0}
\renewcommand{\thetable}{D\arabic{table}}

\APERO has been in development since October 2017. Here we list a few of the major versions to give the reader an idea of how long the development of a full pipeline can take.

\begin{table}[ht]
    \centering
    \begin{tabular}{lllp{10cm}}
        \hline
        \hline
        First version & Last version & Date of first version & Main improvements \\
        \hline
        0.0.000 & 0.0.048 & 2017-10-12 & First python version of \APERO \\
        0.1.000 & 0.1.037 & 2018-01-10 & First version to run on \spirou engineering data (H2RG detector) \\
        0.2.000 & 0.2.128 & 2018-04-17 & First version for \spirou commissioning (H4RG upgrade) \\
        0.3.000 & 0.3.077 & 2018-09-06 & First implementation of telluric correction \\
        0.4.000 & 0.4.123 & 2018-12-08 & Re-work wave solution and BERV calculation \\
        0.5.000 & 0.5.124 & 2019-05-10 & Implementation of reference calibrations/recipes \\
        0.6.000 & 0.6.132 & 2019-12-06 & Complete re-ordering of \APERO file structure, first use on NIRPS \\
        0.7.000 & 0.7.255 & 2020-10-16 & Implementation of SQL databases, Telluric pre-cleaning, upgrade calibration recipes, integration of \textsc{spirou-polar} \\
        0.8.000 & active  & 2022-09-11 & Currently in development: uncertainty propagation, \NAN pixel quality flags, optimal extraction weights, database architecture \\
        1.0.000 & -       & - & After 0.8, full documentation, including adding new instruments \\
        \hline
        \hline
    \end{tabular}
    \caption{History of the major versions of \APERO}
    \label{tab:version_history}
\end{table}

\newpage
\clearpage

\section{Current science publications using APERO}
\label{section:current_science_publications}
\setcounter{table}{0}
\renewcommand{\thetable}{E\arabic{table}}

In table \ref{tab:current_science_pub} we list some science publications using \APERO for science. This list is not complete but gives an idea of the range of science enabled by \APERO with \spirou.

\begin{table}[ht]
    \footnotesize
    \centering
    \begin{tabular}{p{13cm}r}
        \hline
        \hline
        Title & Citation \\
        \hline
        Spin-orbit alignment and magnetic activity in the young planetary system AU Mic                                 & \citealt{Martioli2020} \\
        Early science with SPIRou: near-infrared radial velocity and spectropolarimetry of the planet-hosting star HD 189733 	& \citealt{Moutou2020} \\		
        SPIRou: NIR velocimetry and spectropolarimetry at the CFHT														& \citealt{donati_spirou_2020} \\
        Star-disk interaction in the T Tauri star V2129 Ophiuchi: An evolving accretion-ejection structure 				& \citealt{Sousa2021} \\
        Where Is the Water? Jupiter-like C/H Ratio but Strong H2O Depletion Found on Tau Bootis b Using SPIRou			& \citealt{Pelletier2021} \\
        TOI-1278 B: SPIRou Unveils a Rare Brown Dwarf Companion in Close-in Orbit around an M Dwarf 	   				& \citealt{Artigau2021} \\
        Characterizing Exoplanetary Atmospheres at High Resolution with SPIRou: Detection of Water on HD 189733 b 		& \citealt{Boucher2021} \\
        TOI-530b: a giant planet transiting an M-dwarf detected by TESS                                    				& \citealt{Gan2022_530} \\
        TOI-1759 b: A transiting sub-Neptune around a low mass star characterized with SPIRou and TESS     				& \citealt{Martioli2022} \\
        TESS discovery of a sub-Neptune orbiting a mid-M dwarf TOI-2136          						   				& \citealt{Gan2022_2136} \\
        Estimating fundamental parameters of nearby M dwarfs from SPIRou spectra                                        & \citealt{Cristofari2022} \\
        Line-by-line velocity measurements, an outlier-resistant method for precision velocimetry 		   				& \citealt{LBL2022} \\
        TOI-1452 b: SPIRou and TESS reveal a super-Earth in a temperate orbit transiting an M4 dwarf	   				& \citealt{Cadieux2022} \\
        Estimating the atmospheric properties of 44 M dwarfs from SPIRou spectra                                        & \citealt{Cristofari2022b} \\
        CO or no CO? Narrowing the CO abundance constraint and recovering the H2O detection in the atmosphere of WASP-127b using SPIRou &  Boucher et al. (accepted 2022) \\
        Near-IR and optical radial velocities of the active M-dwarf star Gl~388 (AD~Leo)  with SPIRou at CFHT and SOPHIE at OHP & Carmona et al. (submitted 2022) \\
        New insights on the near-infrared veiling of young stars using CFHT/SPIRou data	   				                & Sousa et al. (submitted 2022) \\
        A sub-Neptune planet around TOI-1695 discovered and characterized with TESS and SPIRou                          & Kiefer et al. (submitted 2022) \\
        The rotation period of 43 quiet M dwarfs from spectropolarimetry in the near-infrared: I. The SPIRou APERO analysis & Fouqué et al. (in prep) \\
        Optical and near-infrared stellar activity characterization of the early M dwarf Gl 205 with SOPHIE and SPIRou  & Cortés-Zuleta et al. (in prep) \\
        High-resolution Chemical Spectroscopy of Barnard’s Star with SPIRou                                             & Jahanadar et al. (in prep) \\
        New methods to correct for systematics in near-infrared radial velocity measurements: Application to GL725B with SPIRou data & Ould-Elhkim et al. (in prep) \\
        Characterizing planetary systems with SPIRou: the M-dwarf Planet Search survey and the system of GJ 251        & Moutou et al. (in prep) \\
        \hline
        \hline
    \end{tabular}
    \caption{List of some publications using \APERO for science.}
    \label{tab:current_science_pub}
\end{table}

\clearpage
\newpage

\section{Inputs}
\label{section:appendix_inputs}
\setcounter{table}{0}
\renewcommand{\thetable}{F\arabic{table}}

The currently allowed raw file inputs are listed in table \ref{tab:apero_inputs}. The name becomes the \APERO header key \DPRTYPE. All other columns are header keys found in the raw input files or are added/modified when first processed (in \Apreprocessing and \Aprocessing). 

Although all \spirou raw files have suffixes:
\begin{tasks}[style=itemize](2)
    \task a.fits (\OBSTYPE = \textsc{align})
    \task c.fits (\OBSTYPE = \textsc{comparison})
    \task d.fits (\OBSTYPE = \textsc{dark})
    \task f.fits (\OBSTYPE = \textsc{flat})
    \task o.fits (\OBSTYPE = \textsc{object})
\end{tasks}

\vspace{0.5cm}
\noindent \APERO does not rely on the filenames to assign a \DPRTYPE to a raw input file. Instead, we use header keys to identify file types (Table \ref{tab:apero_inputs}).

\begin{table}
\centering
\footnotesize
\begin{tabular}{llllllll}
    \hline\hline
    name & \textsc{OBSTYPE} & \textsc{SBCCAS\_P} & \textsc{SBCREF\_P} & \textsc{SBCALI\_P} & \textsc{INSTRUME} & \textsc{TRG\_TYPE}$\ast$ & \textsc{DRSMODE}$\ast$ \\
    \hline
    \DARKDARKINT & DARK & pos\_pk & pos\_pk & P4 & SPIRou & - & - \\
    \DARKDARKTEL & DARK & pos\_pk & pos\_pk & P5 & SPIRou & - & - \\
    \textsc{dark\_dark\_sky} & OBJECT & pos\_pk & pos\_pk & - & SPIRou & SKY & - \\
    \textsc{dark\_fp\_sky} & OBJECT & pos\_pk & pos\_fp & - & SPIRou & SKY & - \\
    \DARKFLAT & FLAT & pos\_pk & pos\_wl & - & SPIRou & - & - \\
    \FLATDARK & FLAT & pos\_wl & pos\_pk & - & SPIRou & - & - \\
    \FLATFLAT & FLAT & pos\_wl & pos\_wl & - & SPIRou & - & - \\
    \textsc{flat\_fp} & FLAT & pos\_wl & pos\_fp & - & SPIRou & - & - \\
    \DARKFP & ALIGN & pos\_pk & pos\_fp & - & SPIRou & - & - \\
    \textsc{fp\_dark} & ALIGN & pos\_fp & pos\_pk & - & SPIRou & - & - \\
    \textsc{fp\_flat} & ALIGN & pos\_fp & pos\_wl & - & SPIRou & - & - \\
    \FPFP & ALIGN & pos\_fp & pos\_fp & - & SPIRou & - & - \\
    \LFCLFC & ALIGN & pos\_rs & pos\_rs & - & SPIRou & - & - \\
    \LFCFP & ALIGN & pos\_rs & pos\_fp & - & SPIRou & - & - \\
    \textsc{fp\_lfc} & ALIGN & pos\_fp & pos\_rs & - & SPIRou & - & - \\
    \OBJDARK & OBJECT & pos\_pk & pos\_pk & - & SPIRou & TARGET & SPECTROSCOPY \\
    \OBJFP & OBJECT & pos\_pk & pos\_fp & - & SPIRou & TARGET & SPECTROSCOPY \\
    \textsc{obj\_hcone} & OBJECT & pos\_pk & pos\_hc1 & - & SPIRou & TARGET & - \\
    \textsc{obj\_hctwo} & OBJECT & pos\_pk & pos\_hc2 & - & SPIRou & TARGET & - \\
    \POLDARK & OBJECT & pos\_pk & pos\_pk & - & SPIRou & TARGET & POLAR \\
    \POLFP & OBJECT & pos\_pk & pos\_fp & - & SPIRou & TARGET & POLAR \\
    \textsc{dark\_hcone} & COMPARISON & pos\_pk & pos\_hc1 & - & SPIRou & - & - \\
    \textsc{dark\_hctwo} & COMPARISON & pos\_pk & pos\_hc2 & - & SPIRou & - & - \\
    \FPHC & COMPARISON & pos\_fp & pos\_hc1 & - & SPIRou & - & - \\
    \textsc{fp\_hctwo} & COMPARISON & pos\_fp & pos\_hc2 & - & SPIRou & - & - \\
    \HCFP & COMPARISON & pos\_hc1 & pos\_fp & - & SPIRou & - & - \\
    \textsc{hctwo\_fp} & COMPARISON & pos\_hc2 & pos\_fp & - & SPIRou & - & - \\
    \HCHC & COMPARISON & pos\_hc1 & pos\_hc1 & - & SPIRou & - & - \\
    \textsc{hctwo\_hctwo} & COMPARISON & pos\_hc2 & pos\_hc2 & - & SPIRou & - & - \\
    \textsc{hcone\_dark} & COMPARISON & pos\_hc1 & pos\_pk & - & SPIRou & - & - \\
    \textsc{hctwo\_dark} & COMPARISON & pos\_hc2 & pos\_pk & - & SPIRou & - & - \\
    \hline
\end{tabular}
\caption{All possible inputs currently accepted by \APERO. HDR denotes that a keyword is required from an input file header. $\ast$ denotes header key is added or modified by \APERO before internal use.}
\label{tab:apero_inputs}
\end{table}

\section{APERO Products}
\label{section:appendix_outputs}
\setcounter{table}{0}
\renewcommand{\thetable}{G\arabic{table}}

\APERO produces outputs after every recipe. These are saved to the reduced directory (except for the \Apreprocessing recipe that saves outputs into the working directory).
These are intermediary products and are used to create the CADC outputs in post-processing (see Section \ref{sec:post_processing}).

\begin{table}[ht]
\centering
\footnotesize
    \begin{tabular}{p{6cm}p{3cm}p{2cm}p{6cm}}
         \hline\hline
         File & Recipe & Frequency & Description \\ 
         \hline
(id)\_pp.fits & \Apreprocessing & every file & preprocessed file \\
(HASH)\_pp\_dark\_ref.fits & \Adarkref & ref night & reference dark file \\
(HASH)\_pp\_badpixel.fits & \Abadpixel & every night &  bad pixel map file \\
(HASH)\_pp\_order\_profile\_\{AB,C\}.fits & \Alocalisation & every night & order profile file \\
         (HASH)\_pp\_loco\_\{AB,C\}.fits & \Alocalisation & every night & localization center map file \\
(HASH)\_pp\_shapex.fits & \Ashaperef & ref night & dx shape map file \\
         (HASH)\_pp\_shapey.fits & \Ashaperef & ref night & dy shape map file \\
         (HASH)\_pp\_fpref.fits & \Ashaperef & ref night & FP reference file \\
(HASH)\_pp\_shapel.fits & \Ashapenight & every night & local shape map file \\
(HASH)\_pp\_blaze\_\{AB,A,B,C\}.fits & \Aflatblaze & every night & blaze correction file \\
         (HASH)\_pp\_flat\_\{AB,A,B,C\}.fits & \Aflatblaze & every night & flat correction file \\
(HASH)\_pp\_e2ds\_\{AB,A,B,C\}.fits & \Athermal & every night & 2D Extracted \DARKDARKINT and/or \DARKDARKTEL file [49x4088] \\
         (HASH)\_pp\_e2dsff\_\{AB,A,B,C\}.fits & \Athermal & every night & 2D extracted + flat fielded \DARKDARKINT and/or \DARKDARKTEL file [49x4088]\\
         (HASH)\_pp\_s1d\_v\_\{AB,A,B,C\}.fits & \Athermal & every night & 1D extracted + flat fielded \DARKDARKINT and/or \DARKDARKTEL file with constants velocity bins \\
         (HASH)\_pp\_s1d\_w\_\{AB,A,B,C\}.fits & \Athermal & every night & 1D extracted + flat fielded \DARKDARKINT and/or \DARKDARKTEL with constants wavelength bins \\
         (HASH)\_pp\_thermal\_e2ds\_int\_\{AB,A,B,C\}.fits & \Athermal & every night & extracted thermal internal dark calibration file\\
         (HASH)\_pp\_thermal\_e2ds\_tel\_\{AB,A,B,C\}.fits & \Athermal & every night & extracted thermal telescope dark calibration file\\
(id)\_pp\_e2ds\_\{AB,A,B,C\}.fits & \Awaveref & ref night & 2D Extracted \DARKFP file [49x4088] \\
         (id)\_pp\_e2dsff\_\{AB,A,B,C\}.fits & \Awaveref & ref night & 2D extracted + flat fielded \DARKDARKINT and/or \DARKDARKTEL file [49x4088]\\
         (id)\_pp\_s1d\_v\_\{AB,A,B,C\}.fits & \Awaveref & ref night & 1D extracted + flat fielded \DARKDARKINT and/or \DARKDARKTEL file with constants velocity bins \\
         (id)\_pp\_s1d\_w\_\{AB,A,B,C\}.fits & \Awaveref & ref night & 1D extracted + flat fielded \DARKDARKINT and/or \DARKDARKTEL with constants wavelength bins \\
         (HASH)\_pp\_leak\_ref\_\{AB,A,B,C\}.fits & \Aleakref & ref night & leak correction reference file \\
(HASH)\_pp\_e2ds\_\{AB,A,B,C\}.fits & \Awaveref & ref night & 2D extracted \FPFP and \HCHC file [49x4088] \\
         (HASH)\_pp\_e2dsff\_\{AB,A,B,C\}.fits & \Awaveref & ref night & 2D extracted + flat fielded \FPFP and \HCHC file [49x4088]\\
         (HASH)\_pp\_s1d\_v\_\{AB,A,B,C\}.fits & \Awaveref & ref night & 1D extracted + flat fielded \FPFP and \HCHC file] with constants velocity bins \\
         (HASH)\_pp\_s1d\_w\_\{AB,A,B,C\}.fits & \Awaveref & ref night & 1D extracted + flat fielded \FPFP and \HCHC file] with constants wavelength bins \\
         (HASH)\_pp\_wavesol\_ref\_\{AB,A,B,C\}.fits & \Awaveref & ref night & reference 2D wavelength solution [49x4088] \\
         (HASH)\_pp\_wavecav\_AB.fits & \Awaveref & ref night & Cavity width measurement file \\ 
(HASH)\_pp\_e2ds\_\{AB,A,B,C\}.fits & \Awavenight & every night & 2D extracted \FPFP and \HCHC file [49x4088] \\
         (HASH)\_pp\_e2dsff\_\{AB,A,B,C\}.fits & \Awavenight & every night & 2D extracted + flat fielded \FPFP and \HCHC file [49x4088]\\
         (HASH)\_pp\_s1d\_v\_\{AB,A,B,C\}.fits & \Awavenight & every night & 1D extracted + flat fielded \FPFP and \HCHC file] with constants velocity bins \\
         (HASH)\_pp\_s1d\_w\_\{AB,A,B,C\}.fits & \Awavenight & every night & 1D extracted + flat fielded \FPFP and \HCHC file] with constants wavelength bins \\
         (HASH)\_pp\_wave\_night\_ref\_\{AB,A,B,C\}.fits & \Awavenight & every night & reference 2D wavelength solution [49x4088] \\
         \hline
    \end{tabular}
    \caption{Main \APERO products, where id for SPIRou is the odometer code and the HASH is a checksum created by multiple inputs (of the same data type) to the given recipe. A full list of data products for each recipe can be found in the documentation. Continued in table \ref{tab:apero_outputs2}.}
    \label{tab:apero_outputs1}
\end{table}

\begin{table}
\centering
\footnotesize
    \begin{tabular}{p{6cm}p{3cm}p{2cm}p{6cm}}
         \hline\hline
         File & Recipe & Frequency & Description \\ 
         \hline
(id)\_pp\_e2ds\_\{AB,A,B,C\}.fits & \Aextract & every file & 2D extracted science/hot star file [49x4088] \\
         (id)\_pp\_e2dsff\_\{AB,A,B,C\}.fits & \Aextract & every file & 2D extracted + flat fielded science/hot star file [49x4088]\\
         (id)\_pp\_s1d\_v\_\{AB,A,B,C\}.fits & \Aextract & every file & 1D extracted + flat fielded science/hot star file with constants velocity bins \\
         (id)\_pp\_s1d\_w\_\{AB,A,B,C\}.fits & \Aextract & every file & 1D extracted + flat fielded science/hot star file with constants wavelength bins \\
(id)\_pp\_tellu\_trans\{AB,A,B\}.fits & \Amktellu & every hot star file & Measured telluric transmission file [49x4088] \\
         (id)\_pp\_tellu\_pclean\{AB,A,B\}.fits & \Amktellu & every hot star file & Telluric pre-cleaning (corrected, transmission mask, measured absorption, sky model) [49x4088] \\
trans\_model\_AB.fits & \Amkmodel & one & Model of all telluric transmission files (residuals in water, dry and a dc level). \\
(id)\_pp\_e2dsff\_tcorr\{AB,A,B\}.fits & \Aftellu & every hot star/science file & 2D telluric corrected extracted flat fielded file [49x4088] \\
         (id)\_pp\_s1d\_w\_tcorr\_\{AB,A,B\}.fits & \Aftellu & every hot star/science file & 1D Telluric corrected extracted flat fielded file with constants velocity bins \\
         (id)\_pp\_s1d\_v\_tcorr\_\{AB,A,B\}.fits & \Aftellu & every hot star/science file & 1D telluric corrected extracted flat fielded file with constants velocity bins \\
         (id)\_pp\_e2dsff\_recon\{AB,A,B\}.fits & \Aftellu & every hot star/science file & 2D telluric reconstructed absorption file [49x4088] \\
         (id)\_pp\_s1d\_w\_recon\_\{AB,A,B\}.fits & \Aftellu & every hot star/science file & 1D telluric reconstructed absorption file with constants wavelength bins \\
         (id)\_pp\_s1d\_v\_recon\_\{AB,A,B\}.fits & \Aftellu & every hot star/science file & 1D telluric reconstructed absorption file with constants velocity bins \\
         (id)\_pp\_tellu\_pclean\{AB,A,B\}.fits & \Amktellu & every hot star/science file & Telluric pre-cleaning (corrected, transmission mask, measured absorption, sky model) [49x4088] \\
Template\_\{object\}\_tellu\_obj\_AB.fits & \Amtemp & once per object & 2D telluric corrected template of a hot star or science object [49x4088] \\
         Template\_s1d\_\{object\}\_sc1d\_w\_file\_AB.fits & \Amtemp & once per object & 1D telluric corrected template of a hot star or science object with constants wavelength bins \\
         Template\_s1d\_\{object\}\_sc1d\_v\_file\_AB.fits & \Amtemp & once per object & 1D telluric corrected template of a hot star or science object with constants velocity bins \\
(id)\_pp\_e2dsff\_tcorr\{AB,A,B\}\_ccf \_\{mask\}\_\{AB,A,B,C\}.fits & \Accf & every hot star/science file & The CCF output file (CCFs per order and fitted parameters) \\
(HASH)\_pp\_e2dsff\_tcorr\_pol\_deg.fits & \Apolar & every polarimetric group & 2D polar file [49x4088] \\
         (HASH)\_pp\_e2dsff\_tcorr\_null1\_pol.fits & \Apolar & every polarimetric group & 2D Null 1 file [49x4088] \\
         (HASH)\_pp\_e2dsff\_tcorr\_null2\_pol.fits & \Apolar & every polarimetric group & 2D Null 2 file [49x4088] \\
         (HASH)\_pp\_e2dsff\_tcorr\_StokesI.fits & \Apolar & every polarimetric group & 2D Stokes I file [49x4088] \\
         (HASH)\_pp\_e2dsff\_tcorr\_s1d\_w\_pol.fits & \Apolar & every polarimetric group & 1D polarimetry, null 1, null 2 and stokes I file with constants wavelength bins \\
         (HASH)\_pp\_e2dsff\_tcorr\_s1d\_v\_pol.fits & \Apolar & every polarimetric group & 1D polarimetry, null 1, null 2 and stokes I file with constants velocity bins \\
         \hline
    \end{tabular}
    \caption{Main \APERO products, where id for SPIRou is the odometer code and the HASH is a checksum created by multiple inputs (of the same data type) to the given recipe. A full list of data products for each recipe can be found in the documentation. Continued from table \ref{tab:apero_outputs1}.}
    \label{tab:apero_outputs2}
\end{table}

\newpage
\clearpage
\section{Preliminary usage with NIRPS}
\label{section:appendix_use_with_nirps}
\setcounter{figure}{0}
\renewcommand{\thefigure}{H\arabic{figure}}

One of our main goals with \APERO was to keep the code generic enough that adding new instruments is possible. To document this we detail here the changes required to add \nirpshe and \nirpsha modes to \APERO. This work is preliminary as commissioning of \nirps is currently underway and we expect additional changes will be required when larger data set over longer periods of time exist (including long science sequences); however \APERO with \nirps has already demonstrated precision equivalent to \spirou. The specific details of all changes are not in the scope of this paper and will be part of a future publication. Currently, after extraction, there are no code differences between \nirps and \spirou reductions.

\subsection{\nirps: A comparison with \spirou}
\label{subsec:appendix_nirps_overview}

\nirps is very similar to \spirou but differs in several ways. We list key differences that \APERO must handle:
\begin{itemize}
    \item there are two modes, high efficiency (\nirpshe) and high accuracy (\nirpsha).
    \item there is only one science fiber and one calibration fiber.
    \item wavelength domain of 980\,nm to 1800\,nm (negligible thermal emission)
    \item missing order(s) around 1400\,nm
    \item the resolution is higher $\sim$100 000 and $\sim$ 80 000 for \nirpshe and \nirpsha modes respectively.
    \item there is no slicer and \nirpshe and \nirpsha have differing fiber geometry.
    \item there are 73 echelle orders extracted by \APERO.
    \item there are no dark unilluminated amplifiers.
\end{itemize}

\subsection{APERO changes for NIRPS}
\label{subsec:appendix_changes_for_nirps}

We use Figure \ref{fig:apero_design} as our reference to changes within \APERO sub-packages. It is worth noting that adapting \APERO for use with \nirps did change some code used for \spirou, as having a second instrument with some unique characteristics informed us of code that could be improved for both instruments. These changes are not mentioned in this section. \APERO is designed to have each of these instruments in the same code base thus there is no separate installation, nor download of additional code is required for \APERO for usage with \nirps.

No code was changed in the following \APERO sub-packages: \texttt{apero.core}: with the exception of \texttt{apero.core.instruments}, \texttt{apero.documentation}, \texttt{apero.io}, \texttt{apero.lang}, \texttt{apero.plotting}, \texttt{apero.setup}, and \texttt{apero.tools}. Minimal code (adding 3 or fewer changes) was changed in the following \APERO sub-package: \texttt{apero.base}: adding of \nirpshe and \nirpsha to the supported instruments list. Some code changes were added to the following \APERO sub-packages:
\begin{itemize}
    \item \texttt{apero.data}: data files were copied from \spirou and updated for \nirpshe and \nirpsha. A few FITS files had to be updated using external scripts. We plan that these scripts will be ingested into \APERO as tools to be used on other instruments directly with sufficient documentation to do so.
    \item \texttt{apero.recipes}: recipe scripts were copied from \spirou (mostly unchanged other than filename). One recipe was added that did not exist before (a reference flat done before the preprocessing recipe), which was added to the preprocessing sequence for \nirpshe and \nirpsha.
    \item \texttt{apero.science}: some science algorithms were added (and called from the recipes) in addition to or to replace \spirou science algorithms. For example, as \nirps does not have an unilluminated region similar to \spirou we have to handle the detector corrections in preprocessing slightly differently, building a background image from the between-order regions.
\end{itemize}

\noindent Substantial code was added to the following \APERO sub-package: \texttt{apero.core.instruments}: configuration, constants, keywords, file definitions, \texttt{pseudo constants} and recipe definitions were copied from \spirou and updated for \nirpshe and \nirpsha. We also removed the polarimetry recipes as there is no polarimetry mode for \nirps.

Note that \texttt{pseudo constants} are constants and variables that cannot be described by a single number or string of characters, such as a decision between science and reference fibers for a specific step of \APERO, or a specific fix to a FITS header key for a specific instrument. These \texttt{pseudo constants} are python functions designed to keep these instrument-specific options separate from the rest of the code.

\section{Glossary}
\label{section:appendix_glossary}
\setcounter{table}{0}
\renewcommand{\thetable}{I\arabic{table}}

In Table \ref{tab:glossary} we present a list of terms used throughout this paper.

\begin{table}[ht!]
\centering
\footnotesize
    \begin{tabular}{p{4cm}p{14cm}}
         \hline\hline
         Term & Description \\ 
         \hline
         \APERO & A PipelinE to Reduce Observations. \\
         \APERO profile & A specific setup of \APERO (i.e. with a certain set of constants, reduction directories, database setups, etc). \\
         amplifier & Independent electronic readout circuits operating in parallel and used to minimize the total readout time. \\
         CCF & Cross-correlation function \\
         CADC & Canadian Data Astronomy Center. \\
         CFHT & The Canada, France, Hawaii Telescope, situated on Maunakea, Hawaii, US. \\
         \DPRTYPE & Data product type - this describes what is in the science and reference fibers and distinguishes different calibrations and observations from each other. \\
         \DRSMODE & Data product mode - for \spirou this is spectroscopy data or polarimetry data. \\
         e2ds & An extracted order-by-order spectrum. \\
         e2dsff & An extracted order-by-order spectrum that has been flat fielded. \\
         fast-axis (long-axis) & Axis parallel to the amplifier direction on the detector. For \spirou this is 4096 pixels per amplifier. \\
         FITS file & A Flexible Image Transport System file to hold images, tables, and metadata in the form of a FITS header. \\
         FITS header & A Flexible Image Transport System metadata holder. Consists of 8 character `keys`, a value, and a comment. \\
         FP & A Fabry Perot etalon used for calibration.  \\
         hash & A short unique hexadecimal string of characters generated from a long string of characters. \\
         HC & A hollow cathode lamp used for calibration. \\
         hot star & Bright, fast rotators of B or A spectral type that are spectrally featureless under $\sim$100\kmps. \\
         LBL & Line-by-line method for measuring radial velocity, presented in \cite{LBL2022}. \\
         \nirps & Near Infra Red Planet Searcher, a spectrograph on the 3.6m telescope in La Silla, Chile. \\ 
         odometer & A unique sequential number used by CFHT to identify individual observations. \\
         order & A domain on the detector at specific wavelengths generated by light passing through a diffraction grating.\\ 
         PI & Principal investigator. \\
         pipeline & Software that takes data from the origin to a destination. \\
         post-processed data & Data that is given to PIs after \APERO has been completed. \\
         pre-processed data & Data that has been corrected for detector issues - the first step for handling raw data. \\
         \PRV & Precision radial velocity, measurements at the order of \mps accuracy. \\
         raw data & For the purposes of \APERO this is the RAMPS from CFHT. \\
         recipe & A top-level script similar to a cookbook recipe where simple steps follow one another; most calculations and algorithms are hidden from these recipes. \\
         reduced data & Data that has been created from the raw data using a pipeline. \\
         rhomb & An ensemble of prisms used to rotate polarization states. \\ 
         run.ini file & A configuration file used for a specific reduction sequence i.e. a science sequence or a calibration sequence. \\
         reduction sequence & A set of recipes run in a certain order with specific filters on which files to reduce for a specific purpose. \\
         reference calibration & A calibration done once (not on a nightly basis). \\
         science observation & Any observation that is taken by the telescope specifically for the purposes of science i.e. SLS data and PI data. \\
         slicer & Device that is used to split an image into narrower images to increase spectral resolution. \\
         slow-axis (short-axis) & Axis perpendicular to the amplifier direction on the detector. For \spirou this is 128 pixels per amplifier. \\
         \spirou & Spectro Polarimètre Infra ROUge, a spectrograph for CFHT. \\
         SLS & The SPIRou Legacy Survey. \\
         Wollaston prism & A device that allows the incoming beam (either from the telescope or the calibration unit) to be split into two lyorthogonal polarized beams. \\
         1/f noise & A noise component arising from the detector readout electronics that has a low-frequency component that is common to all amplifiers and sampled by the reference pixels.\\
         \hline
    \end{tabular}
    \caption{Glossary of terms}
    \label{tab:glossary}
\end{table}

\end{document}